\documentclass[10pt,preprint]{aastex}
\usepackage{epsfig}

\begin{document}

\title{The Co-Evolution of Galaxies and Supermassive Black Holes: Insights
  from Surveys of the Contemporary Universe}

\author{Timothy M. Heckman}

\affil{Center for Astrophysical Sciences, Department of Physics \&
  Astronomy, The Johns Hopkins University, Baltimore, MD 21218, USA}

\author{Philip N.  Best}

\affil{Institute for Astronomy, Royal Observatory Edinburgh, Blackford
  Hill, Edinburgh EH9 3HJ, UK}

{\large{\bf Keywords}}\\[3mm]
Active Galactic Nuclei, Seyfert Galaxies, QSOs, Radio Galaxies\\ 

{\large{\bf Abstract}}

We summarize what large surveys of the contemporary (low-redshift)
universe have taught us about the physics and phenomenology of the
processes that link the formation and evolution of galaxies and their
central supermassive black holes. The surveys are powerful because their
large size and high data quality enable the careful control of systematics
needed to untangle the complex inter-relationships in a robust way. We
emphasize the particular importance of high-quality optical spectra from
the Sloan Digital Sky Survey. We present a picture in the contemporary
universe in which the population of Active Galactic Nuclei (AGN) can be
divided into two distinct populations.  The Radiative-Mode AGN are
associated with black holes that produce radiant energy powered by
accretion at rates in excess of $\sim$1\% the Eddington Limit. This
population today is primarily associated with less massive black holes
growing in high-density pseudo-bulges at a rate sufficient to produce the
total mass budget in these black holes in about 10 Gyr. The circum-nuclear
environment contains high density cold gas and associated on-going
star-formation. Major mergers are not the primary mechanism for
transporting this gas inward; secular processes appear dominant. The
association between black hole growth and star-formation means that
stellar feedback will be generic in these objects. Strong AGN feedback in
the form of global outflows is seen only in the most powerful AGN. In the
Jet-Mode AGN the bulk of energy extracted from the accretion process takes
the form of collimated outflows (jets). This population is associated with
the more massive black holes in the more massive (classical) bulges and
elliptical galaxies. Neither the accretion onto these black holes nor
star-formation in their host bulge is significant today. These AGN are
probably fueled by the accretion of slowly cooling hot gas that is limited
by the feedback/heating provided by AGN radio sources. Surveys of the
high-redshift universe are painting a very similar picture for the
characteristic population of AGN that are responsible for the bulk of the
mass in relic black holes in today's fossil record: they live in normal
galaxies with the (high) star-formation rates characteristic of those
epochs. We speculate that a mode in which black holes are formed in major
mergers and generate galaxy-scale winds may be responsible for the
minority population lying above the knee in the AGN luminosity function.
Noting that the volume-averaged ratio of star formation to black hole
growth has remained broadly similar over the past 10 Gyrs, we argue that
the processes that linked the cosmic evolution of galaxies and
supermassive black holes are still at play today.

\section{INTRODUCTION}
\subsection{A Brief History}

It has been nearly a century since Hubble (1925) decisively demonstrated
that the Andromeda nebula is a vast island universe of stars similar to
our own Milky Way galaxy. His discovery soon thereafter of an expanding
cosmos filled with galaxies opened up a vista of an immense and seemingly
serene universe.  However, this picture of serenity was deceptive.  As we
write this review, it is the 50 year anniversary of Maarten Schmidt's
(1963) realization that the radio source 3C273 was associated with an
optically unresolved object with the (then) enormous redshift of
0.158. Subsequent identifications of a whole population of quasi-stellar
radio sources at even higher redshifts were soon followed by the discovery
of a more numerous population of otherwise-similar but radio-quiet
quasi-stellar galaxies (Sandage 1965).  [It is fascinating in hindsight that the discovery of the cosmic X-ray background (Giacconi et~al. 1962) and of the Kerr metric (Kerr 1963) occurred at the same time.] The concept of the Violent
Universe was born.

It is beyond the scope of this article to review the twists and turns that
finally led to the consensus that all these quasi-stellar objects (QSOs)
were the extremely luminous active nuclei of distant galaxies.
Qualitatively similar objects had been found much earlier in the nuclei of
some relatively nearby galaxies by Seyfert (1943). It is sobering that the
significance of this discovery was long unrecognized: the first
investigation into the nature of Seyfert galaxies was published 16 years
later by Burbidge et~al.\ (1959) who concluded that NGC 1068 was
explosively ejecting ionized gas from its nucleus. Likewise, although
other radio sources had been identified with distant galaxies before the
discovery of QSOs -- e.g.\ Cygnus A by Baade \& Minkowski (1954) -- there
was great confusion as to the origin of this very large-scale
radio-emission which had no obvious connection to the galaxy nucleus.
 
For most of the past five decades the communities that studied galaxies
and active galactic nuclei (AGNs) remained largely disconnected. AGNs were
studied primarily as laboratories in which to probe exotic high-energy
processes. There was some effort to understand the role that the
environment might play in triggering or fueling the AGN -- for example,
see the ancient review in this journal by Balick \& Heckman (1982) -- but
there was almost no idea that AGNs played any significant role in the
evolution of typical galaxies.  Of course, things could hardly be more
different today. The notion of the co-evolution of galaxies and AGN has
become inextricably engrained in our current cosmogony. Indeed, our review
represents the third article in this journal in consecutive years that
deals with some aspect of this co-evolution (Fabian 2012, Kormendy \& Ho
2013).

The reasons for this change are easy to see.  First came the realization
that powerful AGNs (as represented by QSOs) were only the tip of the
iceberg.  Extensive surveys in the radio, optical, and X-ray domains
revealed local populations of Seyfert and radio galaxies and established
that signs of lower-level activity were commonplace in the nuclei of
early-type galaxies.  This strongly suggested that the AGN phenomenon --
rather than being simply a rare spectacle -- was a part of the lifecycle of
typical galaxies. The demography and physical properties of these
low-power AGN has been comprehensively reviewed in this journal by Ho
(2008).

\begin{figure}[!t]
\begin{center} 
\psfig{file=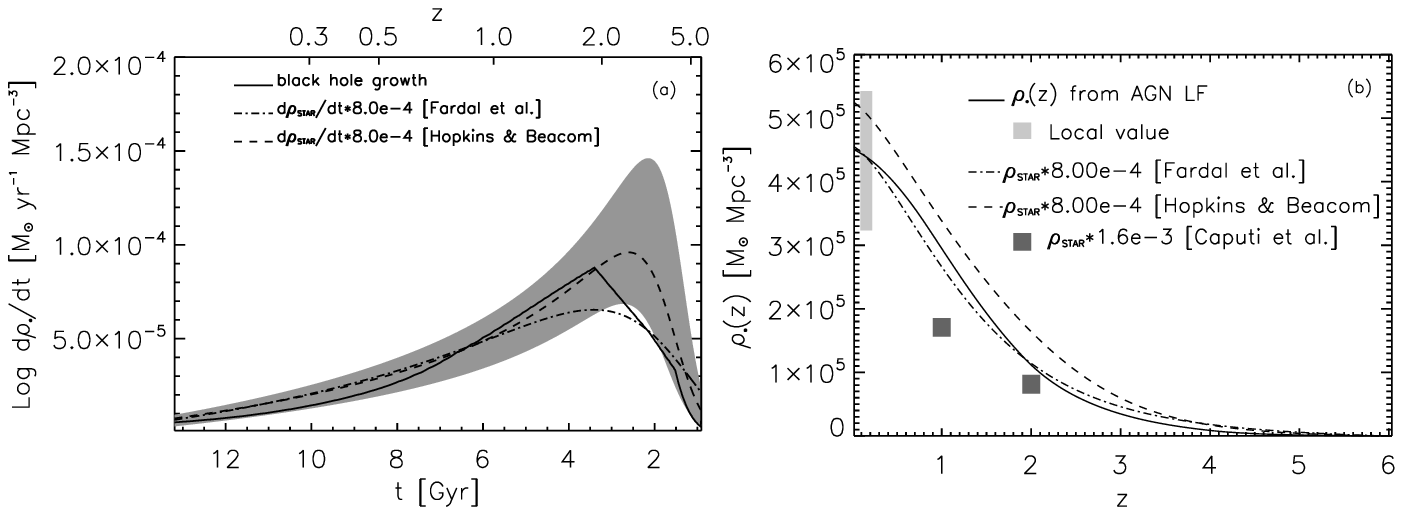,width=8.8cm,clip=} 
\end{center} 
\caption{\label{shankarhist} The cosmic history of black hole growth and
  stellar mass growth.  The average black hole accretion rate is compared
  to the SFR as a function of redshift, where the latter is given by
  Hopkins \& Beacom (2006) and Fardal et~al.\ (2007), scaled by the factor
  0.8$\times$10$^{-3}$. The shaded grey area shows the 3$\sigma$
  uncertainty region from Hopkins \& Beacom (2006). Figure from Shankar et
  al.\ (2009). }
\end{figure}

A second reason followed from the documentation of the overall build-up
over most of cosmic time of the populations of galaxies (as traced via
star-formation) and of supermassive black holes (SMBHs, whose growth is
traced by AGN). The evolution of the two populations is strikingly
similar: a steep rise in both the star-formation rate (SFR) and SMBH
growth rate by about a factor of 10 from redshift z = 0 to 1, a broad
maximum in both rates at $z \sim$ 2 to 3 and then a relatively steep
decline at higher redshifts (see Fig.~\ref{shankarhist} -- Shankar
et~al.\ 2009 and references therein). For at least the last $\sim$11 Gyr
of cosmic history the ratio of these two growth rates has remained roughly
constant with a value of-order 10$^3$.  Thus, at least in a
volume-averaged sense, the growth of galaxies and SMBHs has been
synchronized somehow.

Finally, and even more remarkably, observations over the past decade have
not only established that SMBH exist in the nuclei of (probably) all
galactic bulges, they have shown that the properties of these present-day
SMBHs and the galaxies in which they live are linked on a
galaxy-by-galaxy-basis.  This impressive fossil record of co-evolution has
been exhaustively reviewed by Kormendy \& Ho (2013).  In addition to the
accumulating observational evidence for the co-evolution of galaxies and
SMBHs, a considerable theoretical motivation to invoke this linkage has
developed as well.  Without some form of feedback from AGN, neither
current semi-analytic models nor numerical simulations can successfully
reproduce the properties of massive galaxies. This fascinating subject has
been reviewed in this journal by Fabian (2012).

In conclusion, while the evidence to date remains indirect, it is hard not
to infer that the cosmic evolution of galaxies and of SMBHs have seemingly
been driven by a suite of inter-linked physical processes.

\subsection{Our Perspective: Large Surveys of the Contemporary Universe}

Our goal here is to review the evidence for the co-evolution of galaxies
and SMBHs as derived from large surveys of the contemporary (low redshift)
universe. The majority of the growth of SMBH and of the stellar components
in galaxies occurred between redshifts of roughly 0.5 and 2, and the
present-day growth rates of both populations are over an
order-of-magnitude smaller than during the peak epoch. A reasonable
question is then whether one can learn anything very useful by studying
the contemporary universe. We hope that by the end of the review the
reader will agree with us that one can actually learn a great deal.

We believe that there are a number of reasons why this is the case. The
contemporary universe contains a rich fossil record that (once
successfully decoded) reveals the processes that produced this
record. Moreover, the basic physical processes that operated during the
peak epoch of SMBH/galaxy growth are still in place and can be studied in
greater detail due to their relative proximity.  However, the primary
reason to study the contemporary universe is that it is the only place
where has it been possible to carry out the surveys of galaxies and SMBHs
that are of both sufficiently large size and whose data are of
sufficiently high quality to have finally allowed us to be able to fully
explore the complex inter-relationship between these two populations in a
statistically robust way. This has led to a much clearer picture of both
the fossil record of galaxy/SMBH co-evolution and of the processes by
which the co-evolution continues to play out.

The particular importance of high-quality optical spectroscopy cannot be
over-stated: these data form the interpretational backbone of the whole
structure defined by extensive multi-band imaging and photometric
surveys. Of most relevance to the specific topic of our review, studies of
the lower-luminosity objects that dominate the AGN population in the local
universe benefitted immensely from galaxy spectra obtained as part of the
main galaxy sample of the Sloan Digital Sky Survey (SDSS -- Strauss
et~al.\ 2002).  The high quality of the spectra enabled them to be used to
characterize both the AGN and the stellar populations in the host
galaxies.  Finally, the uniformity and completeness of the SDSS main
galaxy sample rendered it ideal for statistical studies of the
multi-parameter characteristics of the galaxy population in the
contemporary universe, including their AGN properties.

The uniformity and wide sky coverage of the 2dF Galaxy Redshift Survey
(2dFGRS; Colless et~al.\ 2001) and SDSS galaxy redshift survey make them
an ideal starting point for studies of AGN and their host galaxies in
regions of the electromagnetic spectrum other than the optical.  This
encompasses the study of radio-selected AGN using surveys such as the 1.4
GHz National Radio Astronomy Observatory (NRAO) VLA Sky Survey (NVSS;
Condon et~al.\ 1998) by Sadler et~al.\ (2002) and Mauch \& Sadler (2007)
and the Faint Images of the Radio Sky at Twenty centimeters (FIRST; Becker
et~al.\ 1995) survey by Best et~al.\ (2005) and Best \& Heckman (2012).
Recently, samples of tens of thousands of nearby mid-infrared-detected AGN
covering large areas of the sky have been constructed by cross-correlating
sources detected by the Wide-field Infrared Survey Explorer (WISE; Wright
et~al.\ 2010) with the SDSS (Donoso et~al.\ 2012, Shao et~al.\ 2013).
X-ray data having the depth and wide-field sky coverage to fully exploit
the SDSS galaxy sample do not exist.  However, to date the SWIFT/BAT
survey has detected over 700 (mostly local) AGN at energies above 15 keV
(Baumgartner et~al.\ 2013). Near-IR data from WISE and vacuum-ultraviolet
data from the Galaxy Evolution Explorer (GALEX; Martin et~al.\ 2005)
provide important additional information about AGN host galaxies.  In
subsequent sections, we will summarize the new scientific insights that
resulted from all these large surveys.

\subsection{The Landscape of the Galaxy Population and How It Got That Way}

To help frame the main issues addressed in this review it is helpful to
briefly summarize the basic properties of the population of galaxies in
the contemporary universe and the current thinking about how these
galaxies were built. The reader is referred to the review by Madau \&
Dickinson in this volume for all the details.

\begin{figure}[!t]
\begin{center} 
\psfig{file=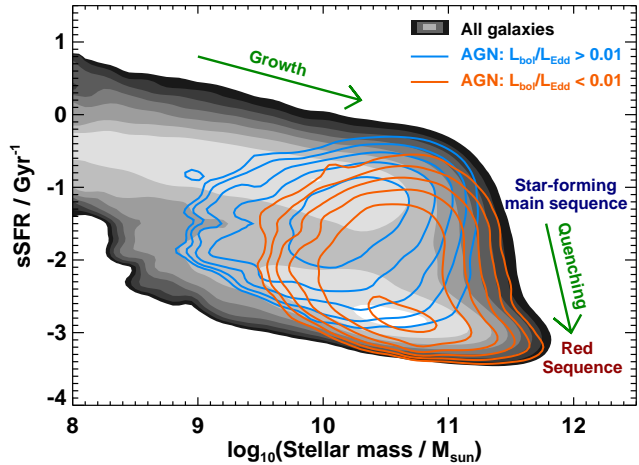,width=9.5cm,clip=} 
\end{center} 
\caption{\label{mass_ssfr} The distribution of galaxies in the SDSS main
  galaxy sample on the plane of stellar mass {\it vs.}\ specific star
  formation rate (sSFR $=$ SFR/M$_*$). The greyscale indicates the
  volume-weighted distribution of all galaxies, with each lighter color
  band indicating a factor of two increase. Galaxies predominantly fall
  within two regions: a `main sequence' of star-forming galaxies, and a
  red sequence of `quenched' galaxies. The blue and red contours show the
  volume-weighted distributions of high ($>$1\%; mostly radiative-mode)
  and low ($<1$\%; mostly jet-mode) Eddington-fraction AGN, with contours
  spaced by a factor of two.}
\end{figure}

Results from the SDSS have shown that the galaxy population in the
contemporary universe occupies a very small part of the parameter space
defined by the structure, stellar content, and chemical composition of a
galaxy. In particular, the existence of clear bimodality in the galaxy
population was revealed (Kauffmann et~al.\ 2003b, Blanton et~al.\ 2003,
Baldry et~al.\ 2004). One population (blue, for short) consists of
galaxies with significant on-going star-formation, small stellar masses
(M$_*$), low stellar surface mass densities ($\mu_* = 0.5M_*/(\pi
R_{50}^2)$, where $R_{50}$ is the radius containing 50\% of the light; see
Section~2.4), and small concentrations ($C = R_{90}/R_{50}$) of their light
(late Hubble type). The other (red) consists of galaxies with little
on-going star formation, large M$_*$, high $\mu_*$, and large $C$ (early
Hubble type). The characteristic parameter values that mark the
transition between populations are M$_* \sim 10^{10.5}$\,M$_{\odot}$,
$\mu_* \sim 10^{8.5}$\,M$_{\odot}$kpc$^{-2}$, and $C \sim 2.6$. Subsequent
work showed that the blue population is characterized by a tight almost
linear relationship between the star-formation rate (SFR) and M$_*$
(Brinchmann et~al.\ 2004, Schiminovich et~al.\ 2007). This has come to be
called the star-forming main sequence. Fig.~\ref{mass_ssfr} shows the
relationship between M$_*$ and the specific star formation rate (sSFR $=$
SFR/M$_*$) in the contemporary universe.

Over the past few years, deep surveys have established that the
qualitative distribution shown in Fig.~\ref{mass_ssfr} is characteristic
of the galaxy population out to at least a redshift of 2 (e.g.\ Whitaker
et~al.\ 2013). The most significant difference is that the actual values
of sSFR on the star forming main sequence evolve extremely rapidly with
time: Elbaz et~al.\ (2011) find as the age of the universe (t$_{cos}$)
increased from 2.2 to 13.6 Gyr ($z = 3$ to 0) the characteristic value of
the sSFR declined with time as $t_{cos}^{-2.2}$.

The simplest picture for the evolution of a typical galaxy (e.g.\ Lilly
et~al.\ 2013) is that it evolves along the (strongly evolving) blue
star-forming main sequence, increasing in mass through the accretion of
cold gas from the cosmic web and (secondarily) through mergers with other
galaxies. As it approaches a critical mass, its supply of cold gas is shut
off, the star formation is quenched, and the galaxy then evolves into the
red population. It can continue to increase in mass through subsequent
mergers with other galaxies. It is presently not clear whether the
mass-scale at which quenching occurs pertains to the stellar mass or the
dark matter halo mass.

The physical process(es) that quench the galaxy are unclear. In part,
quenching may be due to a change in the nature of accretion: rapid accretion
of cold streams of infalling gas at low mass transitioning to slow
accretion of hot gas in hydrostatic equilibrium at high mass (e.g.\ Dekel
et~al.\ 2009). In principle these processes are included in numerical and
semi-analytic models of galaxy evolution. Nevertheless, these models
require some additional process to be at play in order to reproduce the
observed properties of massive galaxies. Heating and/or the ejection of
surrounding gas by an AGN-driven outflow to suppress the cold accretion is
a popular idea. In addition the models also require AGN feedback to keep
galaxies that arrive in the red/dead population from forming too many
stars from the slow accretion of their hot halo gas. Thus, in the current
paradigm, AGN play a crucial role in the evolution of massive galaxies.

\section{BASIC METHODOLOGY}

\subsection{ Overview of the Local AGN Population}

The fundamental property that we consider to define an AGN is that its
power source involves extracting energy from the relativistically-deep
potential well of a SMBH at or near the center of a galaxy.

In this paper we will present strong empirical evidence that the
low-redshift population of AGN can be divided into two main
categories. The first category consists of objects whose dominant
energetic output is in the form of electromagnetic radiation produced by
the efficient conversion of the potential energy of the gas accreted by
the SMBH.  Historically, these objects have been called either Seyfert
galaxies or QSOs depending upon rather vague and arbitrary criteria
involving luminosity and/or redshift. In the rest of this review we will
refer to these as radiative-mode AGN. The second category consists of
objects that produce relatively little radiation, and whose primary
energetic output takes the form of the bulk kinetic energy transported in
two-sided collimated outflows (jets). These jets may be ultimately powered
by the accretion of gas or by tapping the spin energy of the SMBH.
Historically, this AGN population has been called (low-excitation) radio
galaxies. In this review we will refer to them as jet-mode AGN. As we will
show in this review, these two populations are virtually disjoint in terms
of the basic properties of their SMBHs and host galaxies.

\begin{figure}[!t]
\centerline{\psfig{file=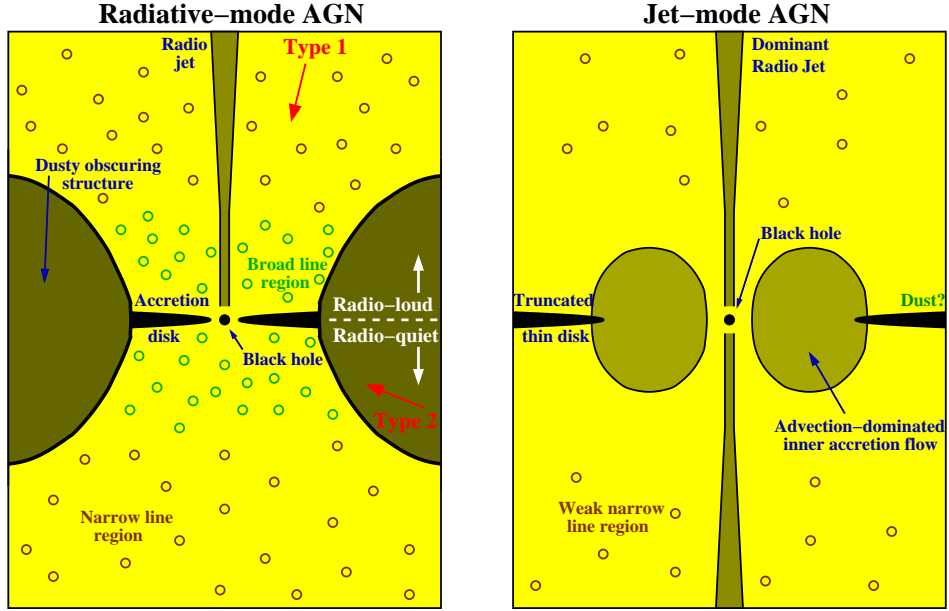,angle=-90,width=13cm,clip=}}
\vspace*{0.5cm}
 
\caption{\label{agnschematic} Schematic drawings of the central engines of
  radiative-mode and jet-mode AGN (not to scale). Radiative-mode AGN (left
  panel) possess a geometrically-thin, optically-thick accretion disk,
  reaching in to the radius of the innermost stable orbit around the
  central supermassive black hole. Luminous ultraviolet radiation from
  this accretion disk illuminates the broad-line and narrow-line emission
  regions. An obscuring structure of dusty molecular gas prohibits direct
  view of the accretion disk and broad-line regions from certain lines of
  sight (Type 2 AGN), whereas they are visible from others (Type 1
  AGN). In a small proportion of sources (predominantly towards the high
  end of the range of black hole masses) powerful radio jets can also be
  produced. In jet-mode AGN (right panel) the thin accretion disk is
  replaced in the inner regions by a geometrically-thick
  advection-dominated accretion flow. At larger radii (beyond a few tens
  of Schwarzschild radii, the precise value depending upon properties of
  the accretion flow, such as the Eddington-scaled accretion rate), there
  may be a transition to an outer (truncated) thin disk. The majority of
  the energetic output of these sources is released in bulk kinetic form
  through radio jets. Radiative emission is less powerful, but can ionize
  weak, low-ionization narrow-line regions, especially where the
  truncation radius of the thin disk is relatively low.}
\end{figure}

Let us then begin by describing the basic building blocks for these two
types of AGN (see Fig.~\ref{agnschematic} for schematic diagrams). We
refer readers to the texts of Krolik (1999), Netzer (2013), Osterbrock \&
Ferland (2005), Peterson (1997), and the review by Yuan \& Narayan in this
volume, for the gory details.  In the first category (radiative-mode AGN),
the SMBH is surrounded by a 
geometrically-thin, optically-thick accretion disk through which an inflow
occurs.  The accretion disk has a radial temperature gradient and the
resulting total thermal continuum emission emerges in the extreme
ultraviolet through visible portion of the electromagnetic spectrum. The
accretion disk is surrounded by a hot corona which Compton-up-scatters the
soft seed photons from the disk into the X-ray regime. As the X-rays
impact the accretion disk their spectral energy distribution is modified
through fluorescence and reflection off the accretion disk.  The ionizing
radiation from the disk and corona heats and photo-ionizes a population of
dense gas clouds located on scales of light-days to light-years from the
SMBH leading to the production of UV, optical, and near-IR permitted
emission-lines. The velocity dispersion of the population of clouds is
typically several thousand km s$^{-1}$, leading to its designation as the
Broad Line Region based on the resulting emission-line spectrum.

On larger scales, the SMBH and accretion disk are surrounded by a region
of dusty molecular gas (which we will refer to as the obscuring
structure). Its inner radius is set by the sublimation temperature of the
most refractory dust grains and is hence larger in more luminous AGN. In
this region some of the incident UV/visible photons from the accretion
disk and the soft X-rays from the corona are absorbed by the dust and this
absorbed energy emerges as thermal infrared emission. The total column
density of the obscuring structure spans a range in inferred column
densities from roughly $10^{23}$ to $10^{25}$cm$^{-2}$. The highest column
densities are sufficient to absorb even hard X-rays (these cases are
Compton-thick).  As ionizing radiation escapes along the polar axis of the
obscuring structure it photo-ionizes gas on circum-nuclear scales (few
hundred to few thousand pc). This more quiescent and lower density
population of clouds produces UV, optical, and infrared forbidden and
permitted emission-lines, Doppler-broadened by several hundred
km\,s$^{-1}$, and is hence called the Narrow Line Region.

Observing an AGN from a sight-line nearer the polar axis of the obscuring
structure yields a clear direct view of the SMBH, the disk/corona, and
Broad Line Region. These are called Type 1 (or unobscured) AGN. When
observing an AGN from a sight-line nearer the equatorial plane of the
obscuring structure, this central region is hidden and these are called
Type 2 (or obscured) AGN. This is the basis for the standard Unified Model
for radiative-mode AGN (e.g.\ Antonucci 1993) which asserts that the Type
1 and 2 populations differ only in the viewing angle from which the AGN is
observed. The presence of AGN can still be inferred in the Type 2 objects
from the thermal infrared emission from the obscuring structure, from hard
X-rays transmitted through the structure (when it is Compton-thin), and
from the emission-lines with tell-tale line ratios from the Narrow Line
Region.

In some cases, the obscuring material can be the larger-scale dusty
interstellar medium of the host galaxy. This is particularly relevant when
the host galaxy's disk is viewed at a large inclination or the galaxy is
in the throes of an on-going major merger with a strong central
concentration of dusty gas. This material will sometimes be sufficient to obscure
the optical, UV, and soft X-ray emission from the AGN accretion disk and
Broad Line Region, but insufficient to attenuate the hard X-rays
(e.g.\ Gelbord 2003).  The effect of this is nicely demonstrated by Lagos
et~al.\ (2011), who use SDSS data to show that the host galaxies of
optically-obscured AGN are skewed toward edge-on orientations, while those
optically-classified as Type 1 have a much higher probability of having a
face-on orientation.

In the second category (jet-mode AGN) a distinct mode of accretion onto
the SMBH exists that is apparently associated with low accretion rates and
which is radiatively inefficient. The geometrically-thin accretion disk is
either absent, or is truncated in the inner regions, and is replaced by a
geometrically-thick structure in which the inflow time is much shorter
than the radiative cooling time (e.g.\ Narayan \& Yi 1994, 1996, Quataert
2001, Narayan 2005, Ho 2008).  These are called advection-dominated or
radiatively-inefficient accretion flows (ADAFs/RIAFs). A characteristic
property of these flows is that they are capable of launching two-sided
jets. Note that powerful jets are also launched by a small fraction of
radiative-mode AGN (e.g.\ radio-loud QSOs). The jets are (far and away)
most easily detected via the synchrotron emission they produce at radio
wavelengths. This can extend from optically-thick (synchrotron
self-absorbed) emission on pc-scales all the way out to regions far beyond
the stellar body of the galaxy (reaching Mpc-scales in extreme cases). In
typical local radio galaxies the jets travel at relativistic velocities
(Lorentz gammas of several) when launched, but appear to rapidly
decelerate and destabilize as they interact with the gaseous halo of the
host galaxy and transition to sub-sonic turbulent plumes. At the highest
radio luminosities (most commonly found in the radio-loud radiative-mode
AGN) the jets survive as highly-collimated structures until they terminate
as bright shocks (hot spots) at the interface with the circum-galactic or
inter-galactic medium.  The survival or disruption of the jets leads to
the Fanaroff \& Riley (1974) morphological classification of radio
galaxies (see Urry \& Padovani 1995 for further discussion).

Missing from the above description are the objects that may trace the
lowest luminosity portion of the AGN population. This population has been
reviewed in this journal by Ho (2008). Since his review, evidence has
grown that at least some of the objects previously classified as the
lowest-luminosity AGN (the Low-Ionization Nuclear Emission-line Regions,
or LINERs, with the very weakest optical emission-lines) are not bona fide
AGN (e.g.\ Cid Fernandes et~al.\ 2012, Yan \& Blanton 2013). However, the
remaining more powerful LINERs are very likely to be members of the
jet-mode AGN population (e.g.\ Nagar et~al.\ 2005, Ho 2008, Best \& Heckman
in preparation), albeit with rather modest radio luminosities.

\begin{figure}[!t]
\centerline{\psfig{file=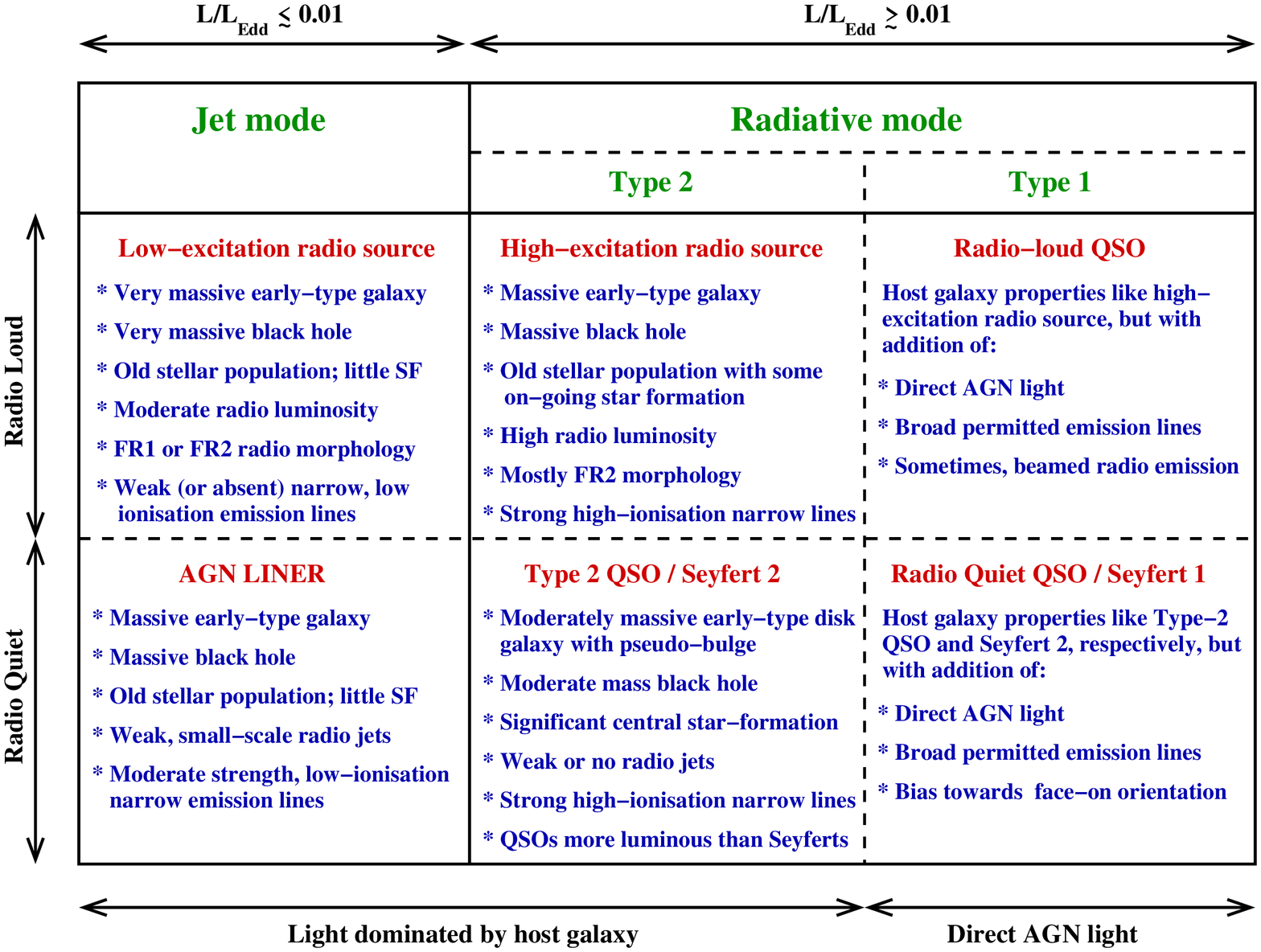,angle=-90,width=\textwidth,clip=}}
\vspace*{0.5cm}
 \caption{\label{props_tab} The categorisation of the local AGN population
   adopted throughout this review. The blue text describes {\it typical}
   properties of each AGN class. These, together with the spread of 
   properties for each class, will be justified throughout the review.}
\end{figure}

Fig.~\ref{props_tab} summarizes this description of the local AGN
population, and how the common AGN classes fit within in. The blue text
within that figure describes the properties of a {\it typical} example of 
each category of AGN. As the review develops, we will describe the 
evidence for this, and discuss the range of properties seen for each AGN
class. 

\subsection{Finding AGN}

This review is focused on insights into the co-evolution of SMBHs and
galaxies that have been derived from large surveys of the local
universe. For such investigations of the radiative-mode AGN it is the
obscured (Type 2) AGN that are far and away the more valuable. In these
objects the blinding glare of the UV and optical continuum emission from
the central accretion disk has been blocked by the natural coronagraph
created by the dusty obscuring structure. The remaining UV and optical
continuum is generally dominated by the galaxy's stellar component
(Kauffmann et~al.\ 2003a) which can then be readily characterized. In the
sections to follow we will therefore restrict our discussion of
radiative-mode AGN to techniques that can recognize Type 2 AGN. For the
jet-mode AGN the intrinsic UV and optical emission from the AGN is
generally weak or absent unless the observer is looking directly down the
jet axis (e.g.\ Urry \& Padovani 1995). Thus, the host galaxy properties
can be easily studied without contamination.

The relevant techniques that have been used for very large surveys of the
local universe are based on: 1)~optical spectroscopy to identify emission
from the AGN-powered narrow-line region, 2)~mid-IR photometry to identify
the warm dust emission from the obscuring structure, 3)~hard X-ray imaging
spectroscopy to detect emission from the accretion disk corona, and
4)~non-thermal radio continuum emission to locate jets and jet-powered
structures.

For the radiative-mode population, one important caveat to keep in mind is
the degree to which Type 2 AGN and their hosts are the same as those of
Type 1. In the most simplistic version of the AGN Unified Model described
above, this complete overlap would be the case.  However, in reality the
fraction of the sky covered by obscuring material (as seen from the SMBH)
will vary significantly from AGN-to-AGN. In this case, the AGN observed as
Type 1 (2) will be systematically drawn from the portion of the AGN parent
population with a smaller (larger) covering-factor for the obscuring
material. It would therefore not be surprising if the properties of
Type 1 and Type 2 AGN and their hosts differed in some systematic fashion.

\subsubsection {Optical Emission-Line Surveys} 

\begin{figure}[!t]
\begin{center} 
\psfig{file=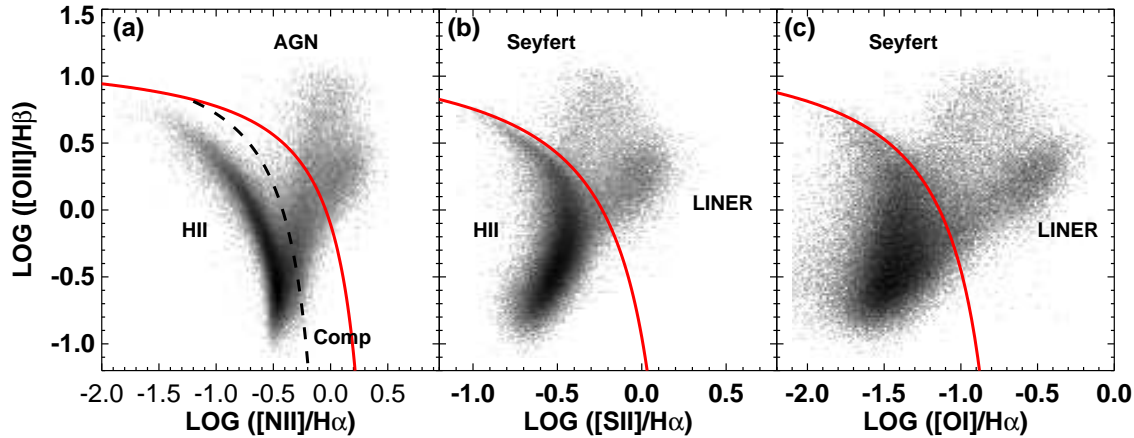,width=16cm,clip=} 
\end{center} 
\caption{\label{bptdiagrams} A set of BPT diagnostic diagrams for the SDSS
  main galaxy sample taken from Kewley et~al.\ (2006). The red line is
  the maximal starburst line from Kewley et~al.\ (2001) while the dashed
  line in the left panel shows the more stringent dividing line between
  pure star-forming galaxies and AGN adopted by Kauffmann
  et~al.\ (2003a). The separation between star-forming galaxies and AGN is
  most cleanly done using the left panel while the center and right panels
  can be used to separate Seyferts from LINERs.}
\end{figure}

Large-scale spectroscopic surveys of the contemporary universe
(particularly 2dFGRS and SDSS) have revolutionized the study of
emission-line AGN, by providing high-quality spectra for over a million
galaxies, with well-understood selection functions. The broad
emission-lines and strong non-stellar continuum emission in Type 1
radiative-mode AGN offer easily detectable and unambiguous evidence of the
presence of an AGN origin (detailed discussions of the identification of
Type 1 AGN and of the criteria used to discriminate between Type 1 and
Type 2 AGN in the SDSS can be found in Richards et~al.\ 2002 and Hao
et~al.\ 2005a). In contrast, the optical signature of Type 2
radiative-mode AGN, and of jet-mode AGN, is the presence of narrow
emission lines. In general, such narrow emission-lines can also be
produced by ordinary populations of O stars associated with on-going star
formation.  However, the relative intensities of the strong forbidden and
permitted emission-lines resulting from photoionization by the relatively
hard continuum from an AGN differ systematically from those produced by
the much softer continuum of O stars.  This approach to recognizing AGN
was first systematized by Baldwin et~al.\ (BPT -- 1981) and later refined
by Veilleux \& Osterbrock (1987). The diagnostic diagrams (hereafter BPT
diagrams) compare pairs of emission line flux ratios. Primary among these
is the flux ratio pair [OIII]\,5007/H$\beta$ vs.\ [NII]\,6584/H$\alpha$,
but the [OI]\,6300/H$\alpha$ and [SII]\,6716,6731/H$\alpha$ flux ratios
provide diagnostics that are complementary to [NII]/H$\alpha$
(e.g.\ Kewley et~al.\ 2006; see Fig.~\ref{bptdiagrams}). The same
diagnostic diagrams work for Type 1 AGN provided that only the narrow
components of the Balmer lines are considered (e.g.\ Stern \& Laor 2012a).

We show a set of BPT diagrams for the SDSS main galaxy sample in
Fig.~\ref{bptdiagrams} (taken from Kewley et~al.\ 2006).  The basic
morphology of the BPT diagrams reveals the presence of two populations.
The first is that of purely star-forming galaxies whose locus runs from the
upper left to lower middle of the plots in a sequence of downwardly increasing
metallicity (e.g.\ Pettini \& Pagel 2004). The second is the AGN
population, which is joined to the star forming sequence at the
high-metallicity end and then extends upward and to the right.

Physically, the observed difference in the emission from high-metallicity
gas photoionized by an AGN reflects the increased rate of photoelectric
heating by the hard AGN radiation field.  This leads to a higher gas
temperature and an increased strength of the collisionally-excited
forbidden lines relative to the recombination lines of Hydrogen. Since
real galaxies contain both AGN and regions of star-formation within the
observed region, the extension of the AGN locus away from the star-forming
sequence can be viewed as a mixing line along which the contribution to
the total emission-line spectrum by the AGN increases relative to that
from the star-forming regions as the points move further away from the
star-forming locus. This inference is confirmed in a statistical sense by
the result that the relative distribution of points along this sequence
migrates downward toward the star-forming sequence as the redshift of the
galaxy population at fixed stellar mass increases and the SDSS fiber
encompasses a progressively larger region of the host galaxy (Kauffmann
et~al.\ 2003a, Kewley et~al.\ 2006, LaMassa et~al.\ 2013).

Kewley et~al.\ (2001) used photo-ionization and stellar population
synthesis models to define an upper boundary to the possible location of
star-forming galaxies in the BPT diagrams, above which a clean sample of
AGN can be selected. Kauffmann et~al.\ (2003a) used empirical methods,
based on the tight locus of star-forming galaxies within the
[NII]/H$\alpha$ BPT diagram, to produce an alternative demarcation which
should provide a more complete census of the AGN population that is less
biased against AGN in strongly star-forming galaxies. These demarcation
lines are shown in Fig.~\ref{bptdiagrams}. Other authors have proposed
further refinements of this division (e.g.\ Stasinska et~al.\ 2006, Kewley
et~al.\ 2006, Cid Fernandes et~al.\ 2011).  This diagram has often been
used to separate emission-line galaxies into three classes:
star-formation-dominated (below and to the left of the Kauffmann
et~al.\ line), AGN-dominated (lying above and to the right of the Kewley
line), and Composite (between the Kewley and Kauffmann lines). In this
paper we will not distinguish between the Composite and AGN-dominated
objects (we regard them all as AGN). 

It is important to emphasize here that while the optical emission-lines used to classify AGN and star-formation-dominated galaxies can be substantially affected by dust absorption, these optical classifications agree with mid-IR based classifications in the great majority of cases (LaMassa et al.\ 2012).

In 1980 a separate class of optical emission line AGN was proposed, known
as LINERs (Heckman 1980). LINERs are distinguished from Seyfert nuclei by
the relative strength of their low-ionization emission lines; the original
definition was based on the forbidden lines of oxygen, with LINERs having
emission-line flux ratios [OII]\,3727/[OIII]\,5007 $>$ 1 and
[OI]\,6300/[OIII]\,5007 $>$ 0.32. Subsequent classifications have built
upon the BPT diagrams separating the AGN population into Seyferts and
LINERs (see Fig.~\ref{bptdiagrams}). Although LINERs typically have lower
nuclear luminosities than Seyferts, many of the fundamental
characteristics of the objects close to the demarcation boundary are very
similar to those of the low-luminosity end of the distribution of Seyferts
(e.g.\ Kewley et~al.\ 2006, Ho 2008, Netzer 2009).

Stasinska et~al.\ (2008), Sarzi et~al.\ (2010), Cid Fernandes
et~al.\ (2011), Capetti \& Baldi (2011), and Yan \& Blanton (2012) have
all presented evidence that those LINERs with the weakest emission-lines
(which are located in galaxies with predominantly old stars) are not
powered by an AGN. Instead most of these papers argue that they are
produced by photo-ionization of neutral atomic gas by a population of
Post-Asymptotic Giant Branch stars (PAGB). A problem with this idea is
that HST ultraviolet images of the bulge of M31 (the site of a LINER) do
not reveal the predicted population of these PAGB stars (Brown
et~al.\ 1998, 2008, Rosenfield et~al.\ 2012).  For these weak LINERs to be
photo-ionized by old stars, a population that is less luminous and more
numerous than the PAGB stars would be needed.

In this review we will be conservative and consider the possibility that
weak LINERs are not members of the AGN population. Here we consider weak
LINERs to be objects with [OIII]\,5007 equivalent widths smaller than
$\sim$1\AA\ (e.g.\ Capetti \& Baldi 2011). It is important to note that
although these objects constitute the majority of LINERs in the SDSS, they
contribute a negligible amount to the overall AGN emissivity of the
contemporary universe.  It is also important to note that objects
classified as LINERs in the SDSS span over two-orders-of-magnitude in
[OIII] luminosity (L$_{\rm [OIII}$) and roughly three-orders-of-magnitude
  in $L_{\rm [OIII]}/M_{\rm BH}$ (Kewley et~al.\ 2006, Netzer 2009; where
  $M_{\rm BH}$ is the black hole mass in solar units).  While a stellar
  origin for the LINER emission is likely at the lowest luminosities, the
  more powerful LINERs are most likely to be AGN (Best \& Heckman, in
  preparation).

\subsubsection{ X-Ray and Infrared Surveys}

X-ray surveys have played a critical role in defining the radiative-mode
AGN population and delineating its evolution over cosmic time (see Brandt
\& Hasinger 2005 for a thorough review).  In our review we are most
interested in the population of Type 2 obscured AGN in the contemporary
universe. The ROSAT all-sky survey (Voges 1999) has the requisite
wide-field coverage, but the survey is very shallow compared to the SDSS
in terms of finding AGN, and its 0.1-2.4 keV band is too soft to recover
the bulk of the obscured AGN population (e.g.\ Anderson et~al.\ 2003, Shen
et~al.\ 2006). Hard X-ray observations are much more useful, and the
current state-of-the-art is represented by the survey undertaken with the
Burst Alert Telescope (BAT) onboard the Swift gamma-ray burst
observatory. As of late 2012, BAT observations have yielded a catalog of
1171 X-ray sources detected in the 14-195 keV band, drawn from a survey
with nearly uniform sensitivity over the entire sky based on 70 months of
observations (Baumgartner et~al.\ 2013). Of these over 700 are AGN.  At
these high X-ray energies only Compton-thick obscured AGN will be
under-represented, and indeed the 70 month catalog contains roughly equal
numbers of Type 1 and Type 2 Seyfert galaxies.

The signature of an obscured AGN in the infrared is unusually strong
mid-IR ($\sim$ 3 to 30 $\mu$m) emission produced by the dusty obscuring
structure.  IRAS pioneered the detection and characterization of the local
AGN populations in the mid and far-IR (e.g.\ Miley et~al.\ 1985, Spinaglio
\& Malkan 1989). Unfortunately, the sensitivity of IRAS was not well
matched to the bulk of the optically-selected local AGN in the SDSS
(e.g.\ Pasquali et~al.\ 2005). The Spitzer Space Telescope was
considerably more sensitive than IRAS but did not undertake wide-field
surveys optimized to the local AGN population. Nonetheless, a number of
different techniques based on near- and mid-IR color selection were
developed to find AGN using Spitzer data (e.g.\ Lacy et~al.\ 2004, Stern
et~al.\ 2005, Richards et~al.\ 2006, Donley et~al.\ 2008). The situation
has been revolutionized with the completion of the mission of NASA's
Wide-field Infrared Survey Explorer (WISE) which has detected near- and
mid-IR emission from roughly 30,000 optically-classified AGN in the SDSS
main galaxy sample (Donoso et~al.\ 2012, Shao et~al.\ 2013) and many other
AGN as well (e.g.\ Stern et~al.\ 2012, Rosario et~al.\ 2013).

It is important to note here that while samples of AGN selected by their
optical emission-lines, their mid-IR emission, or their hard X-ray
emission are roughly similar to one another in terms of both the
properties of the AGNs and the host galaxies (e.g.\ La Massa et~al.\ 2010,
Kauffmann et~al.\ 2003a, Winter et~al.\ 2010, Shao et~al.\ 2013), they may
not be identical (e.g.\ Koss et~al.\ 2011, Juneau et~al.\ 2013, Hickox
et~al.\ 2009).

\subsubsection{Radio Continuum Surveys}

Radio continuum surveys offer an important means to find AGN, particularly
the jet-mode AGN for which jets represent the bulk of the AGN's energetic
output. For radiative-mode AGN, the dusty obscuring structure is optically
thin to radio continuum emission, so the detection of radio-emitting
Type-2 AGN is also reasonably straightforward. Deep radio sky surveys such
as NVSS, FIRST, the Westerbork Northern Sky Survey (WENSS; Rengelink
et~al.\ 1997) and the Sydney University Molonglo Sky Survey (SUMSS; Bock
et~al.\ 1999) cover large sky areas with sufficient sensitivity to probe
the majority of the radio-AGN in the contemporary universe, reaching down
to at least $P_{\rm 1.4GHz} \sim 10^{23}$W\,Hz$^{-1}$ out to $z \sim
0.1$. More problematic issues than simple detection, however, are the
accurate association of observed radio sources with their optical host
galaxies, and the confirmation that the observed radio emission arises
from AGN activity rather than being associated with star formation within
the host galaxy.

The NVSS was the first large-area radio survey of sufficiently high
angular resolution (45 arcsec) to permit automated cross--correlation with
optical surveys (cf.\ Machalski \& Condon 1999). However, the positional
accuracy of NVSS sources is only 1-2 arcsec at 10 mJy flux levels,
increasing to 4-5 arcsec or more at the 2.5mJy flux limit of the survey
(Condon et~al.\ 1998). This leads to significant uncertainties in
cross-identifying the radio sources with their optical host galaxies,
which are further exacerbated by the extended and multi-component nature
of some radio sources: even though $>95$\% of $z < 0.1$ radio sources are
contained within a single NVSS component, the radio centroid of asymmetric
extended sources may not match the host galaxy position. This leads to a
trade-off between the reliability and the completeness of the matched
sample. For example, Sadler et~al.\ (2002) cross-matched the NVSS with
galaxies from the 2dFGRS, accepting matches within 10 arcsec of the
optical galaxy; they calculated that the catalog would be 90\% complete
at this level, but that 5-10\% of the matches were expected to be false
identifications.

Higher reliability samples can be derived using radio surveys with better
angular resolution, such as the 5 arcsec resolution of FIRST (cf.\ the SDSS
cross-matching by Ivezi{\'c} et~al.\ 2002). However, these surveys can miss
extended radio emission, leading to the radio luminosity of sources larger
than a few arcsec being systematically low (Becker et~al.\ 1995), and also
to a much larger fraction of extended radio sources being split into
multiple components. These multi-component sources can usually be
associated with host galaxies by visual inspection, which is practical for
moderate sample sizes (e.g.\ Mauch \& Sadler 2007), but not for the very
large samples produced from SDSS.  Automated cross-matching routines
therefore need to be used, which take account of the possible
multi-component nature of radio sources.  So-called collapsing algorithms
aim to identify multi--component radio sources by using the angular
separations, flux densities and flux-density ratios of close source pairs
to determine whether these are likely to be associated (e.g.\ Magliocchetti
et~al.\ 1998); for optical cross-matching, the relative separations and
position angles from the host galaxy position offer further information
(e.g.\ McMahon et~al.\ 2002, de Vries et~al.\ 2006). Best et~al.\ (2005a)
developed a combined approach, using data from both NVSS (for reliable
luminosities) and FIRST (for accurate cross-matching), along with all of
the above information and incorporating visual analysis for the most
extended radio sources. They were able to develop samples with $\sim95$\%
completeness and $\sim99$\% reliability. Adaptations of this method have
since been widely used (e.g.\ Sadler et~al.\ 2007, Donoso et~al.\ 2009, Lin
et~al.\ 2010, Best \& Heckman 2012).

The derived radio source populations comprise a mixture of AGN and
star-forming galaxies. A variety of different methods have been adopted in
the literature for separating radio-loud AGN from star-forming
galaxies. Perhaps the cleanest of these uses the ratio of far-infrared to
radio luminosity: star-forming galaxies show a very tight correlation
between these properties (see review by Condon et~al.\ 1992), whereas
radio-loud AGN are offset to brighter radio luminosities. This technique
has been widely used at higher redshifts (e.g.\ Ibar et~al.\ 2008, Sargent
et~al.\ 2010, Padovani et~al.\ 2011, Simpson et~al.\ 2012), but the
shallow depth of wide-area far-IR surveys has limited its use locally
(cf.\ Sadler et~al.\ 2002). For local galaxies, the information provided
by the large spectroscopic surveys can be used instead. BPT
emission-line-ratio diagnostic diagrams offer an approach consistent with
that used to classify emission-line selected AGN, although a large
fraction of the radio population lack the required emission line
detections (Sadler et~al.\ 1999, Best et~al.\ 2005a). Radio-loud AGN can
also be identified as being offset from the tight correlation that
star-forming galaxies display between H$\alpha$ emission line luminosity
and radio luminosity (Kauffmann et~al.\ 2008, Vitale et~al.\ 2012). An
alternative method developed by Best et~al.\ (2005a) considers the
4000\AA\ break strength (a measure of stellar population age; see Section
2.4.2) against the ratio of radio power per stellar mass, on the basis
that star-forming galaxies with a wide range of star formation histories
occupy a similar locus in this plane.

\begin{figure}[!t]
\begin{tabular}{cc}
\psfig{file=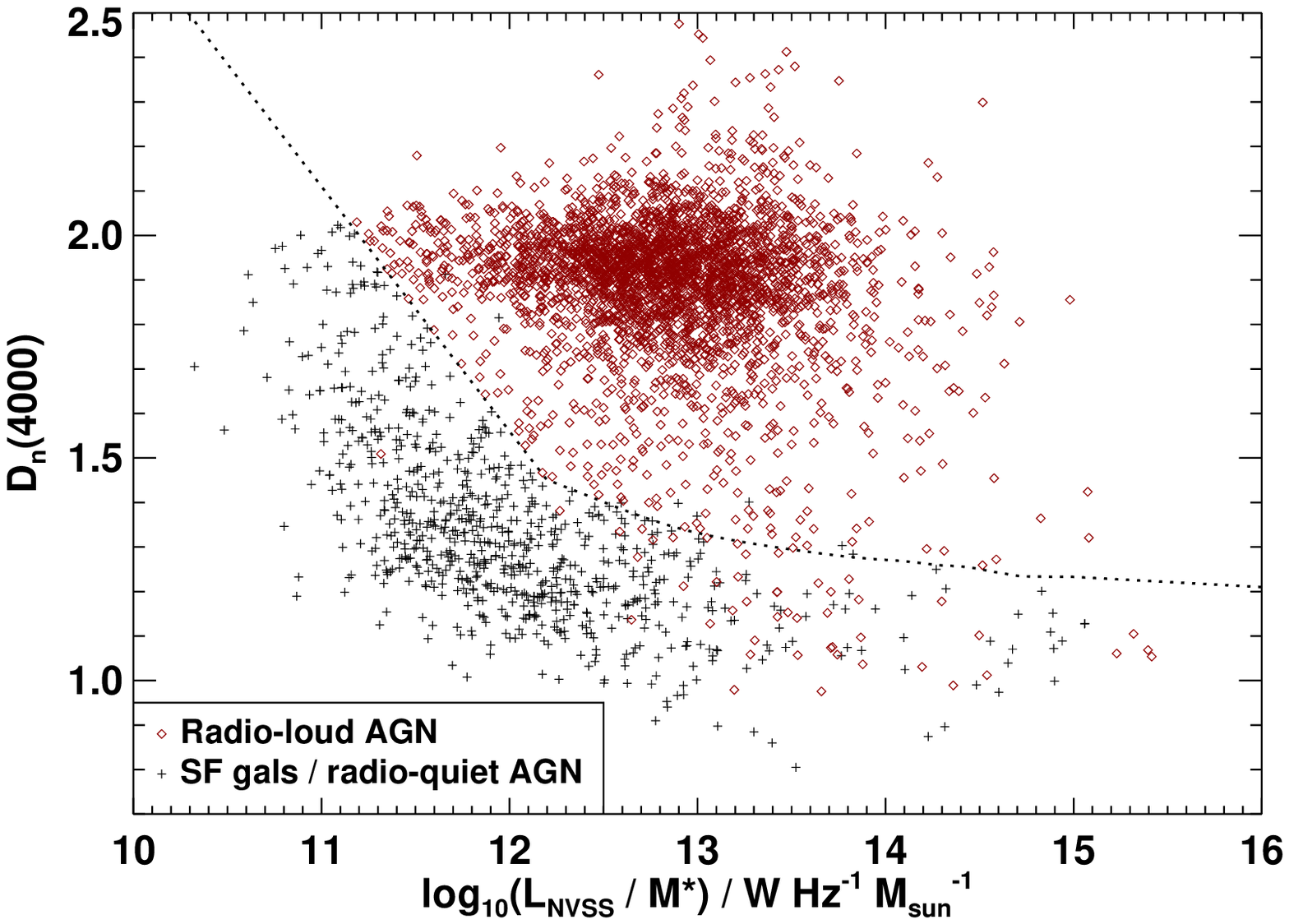,width=7.5cm,clip=} 
\psfig{file=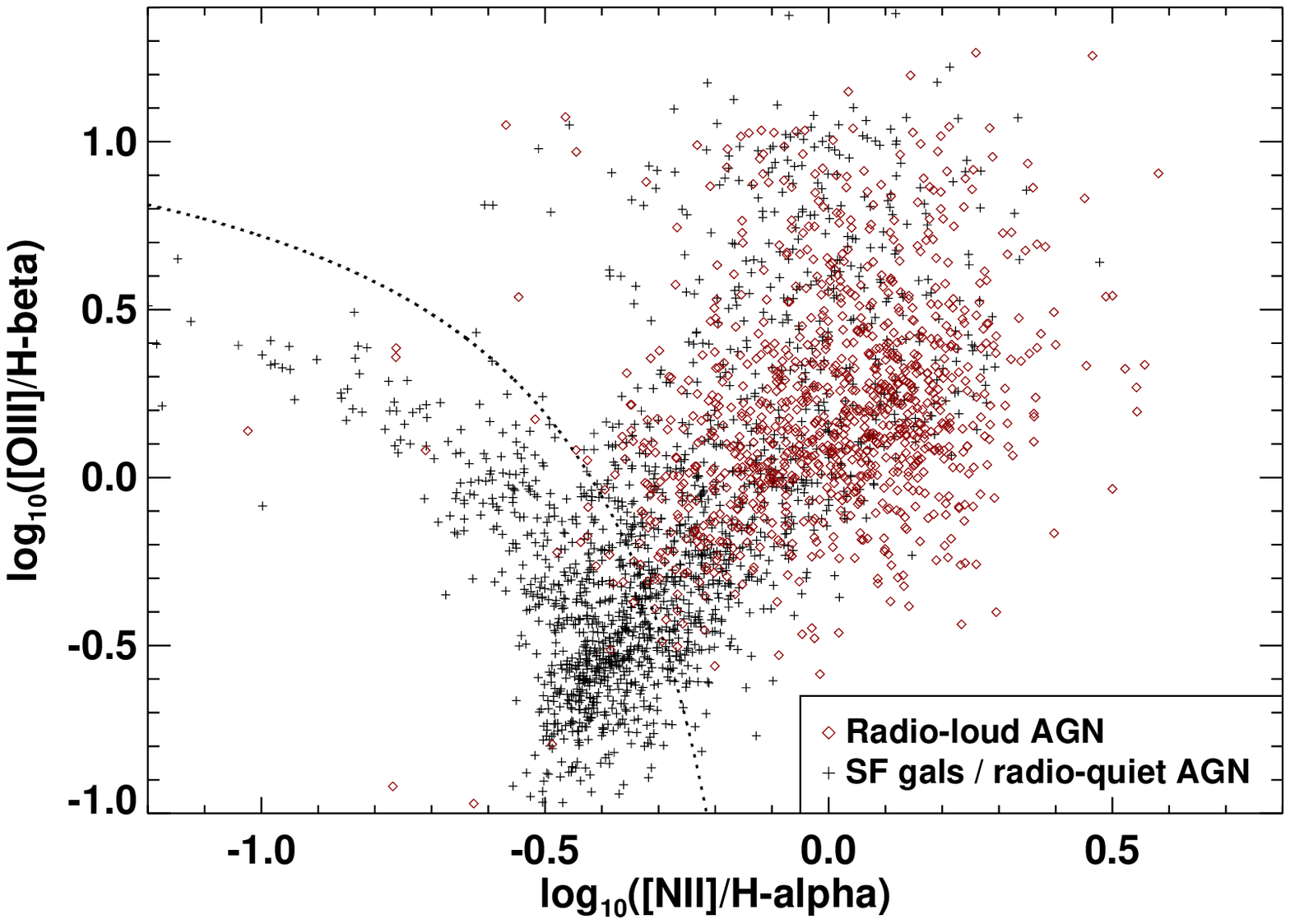,width=7.5cm,clip=} 
\end{tabular}
\centerline{\psfig{file=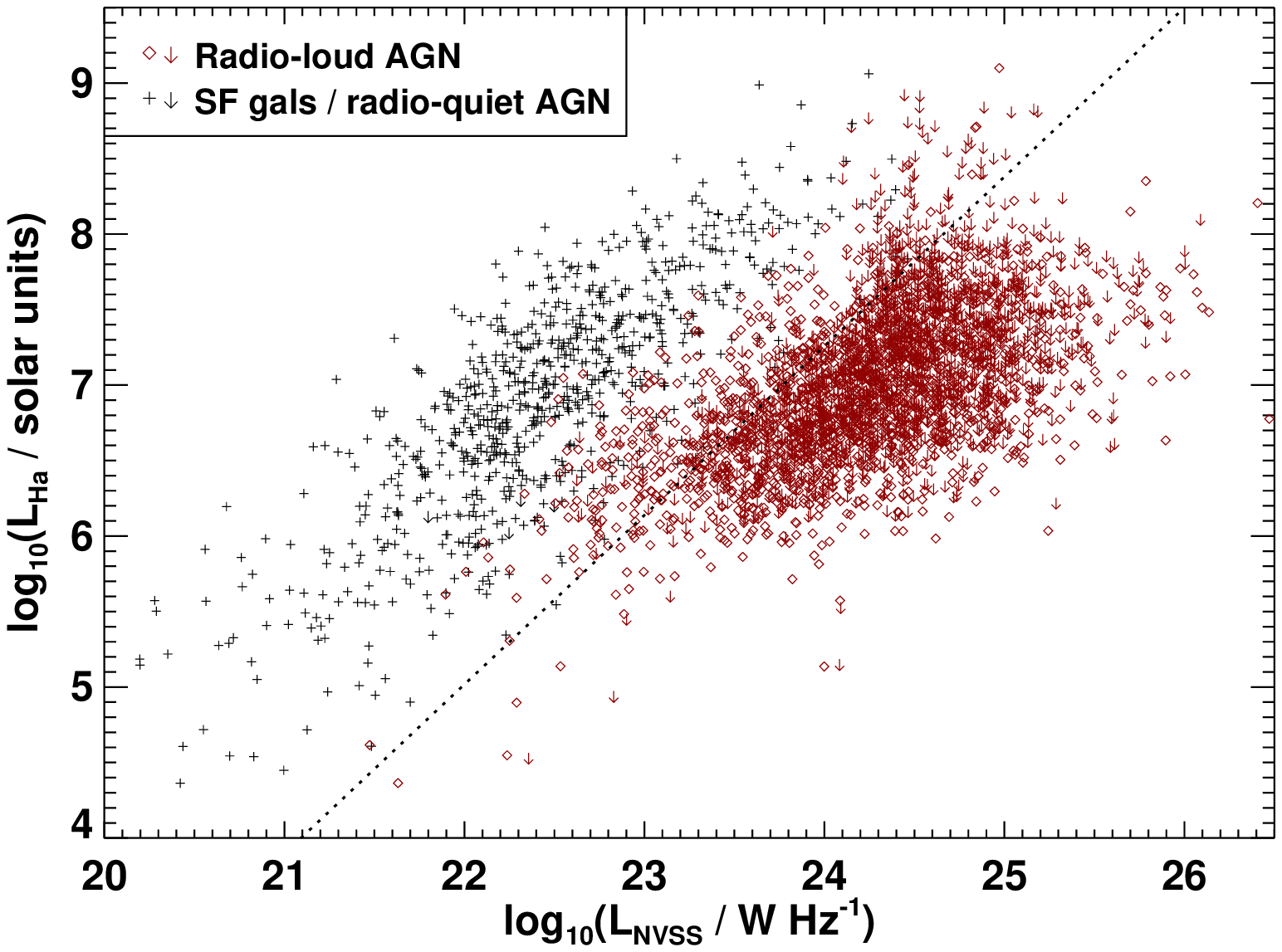,width=7.5cm,clip=}} 
\caption{\label{sfagn_rad} Diagnostic diagrams showing the galaxies within
  the SDSS main galaxy sample that are detected in the radio waveband
  above $S_{\rm 1.4GHz} = 5$\,mJy. These diagrams are used to separate
  genuine radio-loud AGN from other radio-detected galaxies in which the
  radio emission is powered by star-formation (this includes both
  radio-quiet AGN and star-forming galaxies). The upper-left plot shows
  the `D(4000) {\it vs.}\ $L_{\rm 1.4GHz}$/M' method, originally developed by
  Best et al.\ (2005a). The upper-right plot shows the widely-used `BPT'
  emission line ratio diagnostic. The lower plot shows the relationship
  between H$\alpha$ and radio luminosity (arrows indicate upper limits to
  the H$\alpha$ luminosity). In all plots, the dotted lines indicate the
  division used by Best \& Heckman (2012) for that classification
  method. For the `BPT' diagnostic diagram, this division separates AGN
  from star-forming galaxies in terms of what powers the emission lines,
  rather than what powers the radio emission. The division line in the
  $L_{{\rm H}\alpha}$ {\it vs.}\ $L_{\rm 1.4GHz}$ plot was defined to ensure a
  clean sample of genuine radio-loud AGN below the division. Through
  combinations of the locations of galaxies in the three diagrams (see
  Best \& Heckman for details) sources can be given an overall
  classification as radio-loud AGN (red diamonds) or star-forming galaxies
  / radio quiet AGN (black crosses). Adapted from Best \& Heckman (2012).}
\end{figure}

These different AGN/SF classification schemes agree on the classification
of most sources, but do produce significant differences (see the
comparison by Best \& Heckman 2012). Primary amongst these differences is
that way that `radio-quiet' AGN are classified. These AGN often display
compact radio cores and weak jet-like structures, but those which lack
these features lie on the same far-IR to radio correlation as star-forming
galaxies (Roy et~al.\ 1998). This suggests that the bulk of their radio
emission is powered by star formation. BPT diagnostics classify these
sources as AGN, since the AGN drives the emission line ratios away from
the star-forming locus, but in contrast the sources are generally
classified as star-forming using the far-IR to radio or emission line to
radio luminosity ratio diagnostics, or the 4000\AA\ break strength. Best
\& Heckman (2012) combined these different classification methods to
provide a sample of 18,286 radio-loud AGN from the SDSS, excluding
radio-quiet AGN (see Fig.~\ref{sfagn_rad}). This is the sample used to
construct relevant plots throughout this review.

\subsection{Determining AGN Physical Properties}

\subsubsection{Proxies for Bolometric Luminosity in Radiative-Mode AGN}

In this section we will discuss how best to estimate the bolometric
luminosity of radiative-mode AGN. Type 2 AGN are (by definition) obscured
and this complicates the estimation of perhaps the most basic physical
property of an AGN: its total (bolometric) luminosity. We cannot directly
observe the great bulk of this radiant energy, and must therefore rely on
attempting to calibrate various proxies for the bolometric luminosity
based on the radiation that does escape.  Let us then assess these various
proxies and begin by briefly listing their pros and cons.

{\it Proxy 1: The [OIII]5007 emission-line.} The proxy that we will
primarily use later in this review is the luminosity of the [OIII]5007
narrow emission-line.  The [OIII] line is produced in AGN by ionizing
radiation that escapes along the polar axis of the dusty obscuring
structure where it photo-ionizes and heats ambient gas clouds on radial
scales of-order 0.1 to 1 kpc from the SMBH (in the Narrow-Line Region, or
NLR).

Pros: 1) This line is easily observed and can be measured (if present) in
the existing SDSS spectra of a few million galaxies. 2) Since the NLR lies
outside the very high column density in the obscuring structure its
emission suffers far less attenuation due to dust near the AGN than the optical
emission from the accretion disk.  Modest amounts of dust extinction can
be corrected using the ratio of the fluxes of the narrow H$\alpha$ and
H$\beta$ emission-lines (the Balmer decrement).  Typical implied
corrections are 1 to 2 magnitudes (e.g.\ Kewley et~al.\ 2006).

Cons: 1) Significant amounts of more wide-spread dust extinction due to
the galaxy interstellar medium may be present and it may not be possible
to reliably correct for this extinction if it is too large. 2) The [OIII]
line can be contaminated by a contribution due to star-formation (which
will be significant in the composite objects shown in
Fig.~\ref{bptdiagrams}).  Examples of the empirical techniques for
correcting for contamination by stellar ionization are described in
Kauffmann \& Heckman (2009) and Wild et~al.\ (2010). 3) The luminosity of
the line will depend on the uncertain fraction of the AGN ionizing
luminosity that is intercepted by clouds in the NLR.

{\it Proxy 2: The [OIV]25.9 micron and [NeV]14.3,24.3 micron
  emission-lines.} These proxies have the same physical origin as the
[OIII]5007 line, arising from photoionized gas in the NLR.

Pros: 1) They arise well outside the region of the heaviest
obscuration. 2) As mid-IR lines they are much less affected by dust
extinction than the [OIII]5007 line. 3) They have ionization potentials
above the HeII edge at 54.4 eV (where hot stars produce very little
radiation). They therefore suffer much less contamination from stellar
photoionization than [OIII]. This is especially true for [NeV]. This
advantage relative to [OIII] has been exploited by Satyapal et~al.\ (2008,
2009) and Goulding \& Alexander (2009) to identify examples of
low-luminosity AGN in the nuclei of nearby late-type galaxies that were
not recognized as AGN based on optical spectra that were dominated by
regions of star formation.

Cons: 1) They are expensive to observe (having been detected in hundreds
of objects rather than hundreds of thousands). 2) The luminosity will
depend on the uncertain fraction of the AGN ionizing luminosity that is
intercepted by clouds in the NLR.

{\it Proxy 3: The Mid-Infrared Continuum.}  This is produced by warm dust
in the obscuring structure and as such represents the reprocessing of the
primary ultraviolet/visible emission from the accretion disk and the soft
X-rays from its corona.

Pros: 1) Thanks to WISE, we now have measurements of the near- to mid-IR
continuum in a very large sample of galaxies in the nearby universe
(comparable in size to the SDSS main galaxy sample). 2) It represents a
substantial fraction of the bolometric luminosity (roughly 25\% in Type 1
AGN where we directly observe nearly all the emission).

Cons: 1) The mid-infrared light from the obscuring structure will not be
emitted in an isotropic way and correction for this effect is uncertain at
the level of factors of several. This effect is largest at the shortest
mid-infrared wavelengths. 2) The net observed mid-infrared emission will
be the sum of the emission from the AGN and dust heated by hot stars. A
decomposition of these two emission processes is uncertain and
model-dependent. This effect is worse at longer infrared wavelengths.

{\it Proxy 4: The Hard X-ray Continuum.}  In a Type 2 AGN the observed
hard X-ray emission will be a combination of whatever radiation succeeds in
traversing the obscuring structure plus emission from the central AGN that
has been scattered or reprocessed into our line-of-sight (e.g.\ Murphy \&
Yaqoob 2009).

Pros: 1) The hard X-ray emission probes regions very close to the SMBH
(the accretion disk corona), but suffers far less obscuration than optical
or UV observations of the accretion disk. 2) This hard X-ray emission is a
rather unambiguous indicator of an AGN (rather than a starburst) at
luminosities greater than about $10^{42}$ erg s$^{-1}$.

Cons: 1) Hard X-rays are poor probes of the AGN emission at the high
column densities often inferred for the obscuring structure. This is true
for column densities above $10^{24}$ cm$^{-2}$ in the 2-10 keV band and
above $10^{25}$ cm$^{-2}$ even in the Swift BAT band. These latter are the
Compton-thick AGN. 2) The SWIFT/BAT wide all-sky survey is quite shallow
relative to SDSS in its ability to detect AGN.

\begin{figure}[!t]
\begin{center} 
\psfig{file=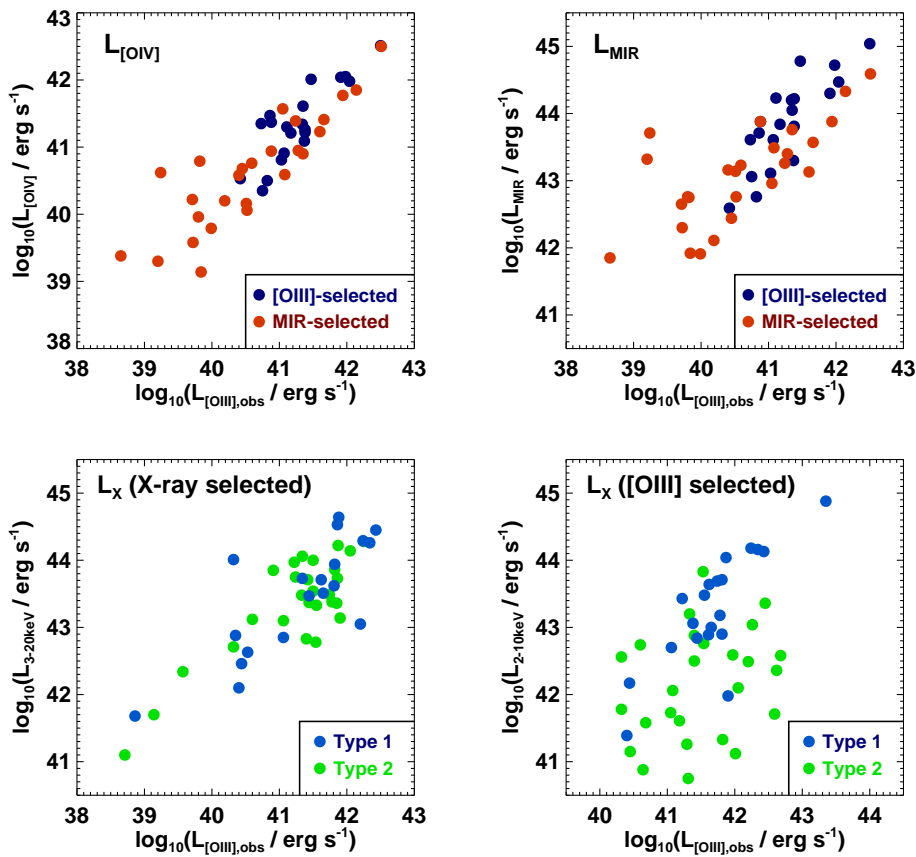,width=13cm,clip=} 
\end{center} 
\caption{\label{lamassafig} Relationships between different estimators of
  bolometric luminosity for radiative-mode AGN. Top panels: a comparison
  of the [OIII] emission line luminosity with the [OIV] line luminosity
  (left) and mid-infrared continuum luminosity (right), for samples of galaxies
  selected from their [OIII] emission line and their 12$\mu$m mid-IR flux
  (data from LaMassa et~al.\ 2010). Bottom panels: the hard X-ray {\it vs.}
  [OIII] luminosity for X-ray selected (left) and emission-line selected
  (right) samples of AGN (data from Heckman et~al.\ 2005). Type 2 AGN in
  the emission line selected samples do not show a good correlation due to
  X-ray absorption.}
\end{figure}

Having briefly described the various proxies, we now summarize the
evidence as to how well they actually do in estimating the AGN bolometric
luminosity.  The problem with this is that we do not know the ground truth
needed to make this evaluation. The best approach is to inter-compare
these different proxies among the Type 2 AGN and also to evaluate their
behavior relative to the Type 1 AGN (where obscuration is minimized); see
Fig.~\ref{lamassafig}.

LaMassa et~al.\ (2010) examined a wide range of proxies in two complete
samples of Type 2 AGN. One was selected on the basis of the flux of the
[OIII]5007 emission-line from SDSS and the other was selected on the basis
of the mid-infrared (12$\mu$m) flux measured with IRAS (hereafter the
SDSS-OIII and IRAS-MIR samples). The proxies they considered for the
bolometric luminosity were the luminosities of the [OIII]5007 line, the
[OIV] 25.9 micron line, the mid-infrared (12$\mu$m) continuum, the hard
X-ray Swift BAT continuum, and the 8.4 GHz nuclear radio continuum. They
examined the correlations between all possible pairs of the proxies and
compared the distributions of the flux ratios of all proxy-pairs to those
of samples of unobscured Type 1 AGN (cf.\ Fig.~\ref{lamassafig}). They
concluded that the luminosity of the [OIV] line is the single best proxy
and that the radio and hard X-ray luminosities are the worst. The problem
with the hard X-ray luminosity is that many of the Type 2 AGN in the
SDSS-OIII and IRAS-MIR samples are Compton-thick and significantly
obscured even in the Swift BAT band (14-195 keV) -- see also Heckman
et~al.\ (2005), Winter et~al.\ (2010). Similar conclusions were drawn by
Rigby et~al.\ (2009) based on comparison of the flux ratio of the [OIV]
25.9 micron line and hard X-ray continuum (above 10 keV) in
optically-selected Type 1 and Type 2 AGN drawn from the Revised
Shapley-Ames Catalog (hereafter the RSA sample).

An analysis of multiple proxies in the RSA sample was undertaken by
Diamond-Stanic et~al.\ (2009). They found that the ratio of the flux of
the 2-10 keV continuum and the [OIV] emission-line was smaller on-average
in the Type 2 AGN than in Type 1 AGN and varied by about three
orders-of-magnitude.  This implies a wide range in the amount of
attenuation of the 2-10 keV X-ray emission in these Type 2 AGN (see
Fig.~\ref{lamassafig}).  They also found that the flux ratio of the
[OIII]5007/[OIV]25.9 micron lines was systematically smaller in Type 2
than in Type 1 AGN and showed more scatter in the RSA sample.  They argued
that this was a result of a greater amount of dust-extinction in Type 2
AGN (by typically one to two magnitudes).  LaMassa et~al.\ (2010) found a
much smaller offset in this ratio in the SDSS-OIII and IRAS-MIR samples
(implying a mean excess extinction of [OIII] by only about 0.5 magnitudes
in the Type 2 AGN).  These results do underscore the importance of
correcting the [OIII] flux for extinction (when possible).

Shao et~al.\ (2013) have analyzed WISE observations of a sample of 30,000
Type 2 AGN identified in SDSS spectra. They find an excellent overall
correlation between the luminosities of the (extinction-corrected)
[OIII]5007 emission-line and the (starlight-subtracted) 4.6 micron
emission from hot AGN-heated dust (rms scatter of about 0.3
dex). Moreover, the partitioning of the AGN emission as a function of host
galaxy parameters is the same based on the [OIII] and 4.6 micron emission
(once LINERs were excluded).  Finally they show that the existing
color-color techniques that have been used to identify AGN from Spitzer
and WISE data recover only the most luminous AGN (objects in which the AGN
significantly outshines its host galaxy in the near- and mid-IR). This
misses most of the optically identified Type 2 AGN in SDSS, which have
lower luminosities.

A simple summary of the above is that there no single perfect proxy for
the bolometric luminosity in obscured AGN.  Even relatively good proxies
will produce an estimate of the bolometric luminosity of any individual
obscured AGN that will be uncertain at the level of a factor of several
(in an rms sense).  Given the huge number of AGN with measured [OIII]5007
luminosities (e.g.\ Hao et~al.\ 2005a, Kauffmann \& Heckman 2009), we will
adopt this as our default proxy for the optically-selected Type 2 AGN.
Based on an analysis of multi-waveband data for two samples of Type 1 AGN,
Heckman et~al.\ (2004) calculated a mean bolometric correction to the
[OIII]5007 luminosity of a factor of 3500. This refers to [OIII]
luminosities derived from the raw [OIII] flux with no correction for dust
extinction and assumes that Type 1 and Type 2 AGN have the same bolometric
correction.  Kauffmann \& Heckman (2009) adopted a mean bolometric
correction of 600 for the extinction-corrected [OIII]5007 luminosity. This
was based on the multi-waveband data subsequently presented in LaMassa
et~al.\ (2010) and is consistent with the average value for the extinction
of the [OIII] emission as measured in Type 2 Seyferts by Kewley
et~al.\ (2006). We adopt this factor 600 bolometric correction for
calculations within this review.

Netzer (2009) finds similar mean bolometric corrections for the observed
and extinction-corrected [OIII]5007 luminosities. He argues, however, that
the bolometric correction for a given AGN will depend on the ionization
state of the gas in the NLR and that a combination of the luminosities of
the [OIII]5007 and [OI]6300 lines is a more robust estimator of the
bolometric luminosity. We agree with this point, but note that the
[OI]6300 line is significantly weaker than the [OIII]5007 line (by a
factor of typically 3 in LINERs and 10 in Seyfert nuclei).  Requiring the
[OI] line to be well-detected diminishes the ability to probe AGN at lower
luminosities and/or over larger volumes in the SDSS main galaxy sample.
The difference between Netzer's method and ours is most significant in the
case of LINERs.  As we have discussed above, a number of recent papers
have concluded that the majority of LINERs in the SDSS (those with the
weakest emission-lines) are not bona fide AGN. However, for the minority
population of LINERs that are clearly AGN we agree that Netzer's
methodology is better.

Finally, it is important to note that these bolometric corrections can be
luminosity dependent. This dependence is well-established in the hard
X-ray regime where the bolometric correction increases systematically with
luminosity (e.g.\ Marconi et~al.\ 2004). There have been claims and
counter-claims about a luminosity dependence for the bolometric correction
to the [OIII] luminosity.  Most recently, Stern \& Laor (2012b) have
analyzed SDSS spectra of a large sample of Type 1 AGN and find that in
log-log plots the [OIII] luminosity has a sub-linear dependence on the UV
continuum luminosity and on the broad H$\alpha$ luminosity, but a linear
dependence on the hard X-ray luminosity. They derive a bolometric
correction to the [OIII] luminosity that increases by an
order-of-magnitude over the range between $L_{\rm [OIII]} \sim 10^{6.5}$
and $10^9$ L$_{\odot}$.  However, Shao et~al.\ (2013) find a linear
dependence of the [OIII] luminosity on the mid-IR luminosity (from WISE)
for Type 2 SDSS AGN over the same range in [OIII] luminosity. LaMassa
et~al.\ (2010) found a super-linear dependence of $L_{\rm [OIII]}$ on the
mid-IR continuum luminosity for Type 2 AGN.  Hainline et~al.\ (2013)
confirm this relationship, extending it to the yet higher luminosities of
Type 2 QSOs.  The implied bolometric correction to $L_{\rm [OIII]}$
decreases by about an order-of-magnitude between $L_{\rm [OIII]} \sim
10^{6}$ and $10^{9.5}$ L$_{\odot}$ (i.e. the opposite of the result Stern
\& Laor found for Type 1 AGN). The reason for this disagreement is not
clear. Hainline et~al.\ also find a possible saturation in $L_{\rm
  [OIII]}$ at MIR luminosities above about $10^{12}$ L$_{\odot}$ and
attribute this to a transition to matter-bounded conditions where the AGN
is effectively photo-ionizing the entire galaxy ISM (see also Netzer
et~al.\ 2006). We emphasize in closing that most of the results reviewed
below (and our own new analyses) have assumed a luminosity-independent
bolometric correction for $L_{\rm [OIII]}$ in Type 2 AGN.

\subsubsection{Mechanical Energy Outflow Rates in Radio Jets}

We now turn our attention to estimating the energy outflow rate in radio
jets. In jet-mode AGN, this can greatly exceed the radiative bolometric
luminosity. It can also make a significant contribution to the total
energetics in the small population of radio-loud radiative-mode AGN.

The AGN jets are observable through their radio synchrotron
emission. Monochromatic radio luminosity represents only a small fraction
of the energy transport: the mechanical (kinetic) power of the jets is
estimated to be about 2 orders of magnitude larger (Scheuer 1974). The jet
mechanical power ($P_{\rm mech}$) can be estimated from the synchrotron
emission using the minimum energy condition to estimate the energy stored
in the radio lobes. This assumes that the magnetic field has the strength
that minimizes the combined energy content of particles and magnetic
fields needed to produce the observed synchrotron emission (see Miley
1980). This is then combined with the age of the radio source (from
synchrotron spectral ageing techniques, or radio source growth models;
cf.\ Willott et~al.\ 1999) and an efficiency factor that accounts for
losses due to work done by the expanding radio source (Rawlings \&
Saunders 1991).

Our incomplete knowledge of the physics of radio sources leads to
significant uncertainties in this estimation. Primary amongst these are
the composition of the radio jet plasma and the low energy cutoff of the
electron energy distribution. The observed synchrotron emission traces
only the highly relativistic electron population, and there are
indications that the energy density in protons may be up to an order of
magnitude higher (e.g.\ Bell 1978). Similarly, extrapolating the electron
energy distribution down from the observed limit of $\gamma > 100$ to
$\gamma \sim 1$ can increase the energy estimates by a factor of a few
(depending upon the spectral index). Willott et~al.\ (1999) combined all
of the uncertainties into a single factor, $f_{\rm W}$, which through
observational constraints they determined to lie between $\sim$ 1 and
20. Subsequent work favored values towards the upper end of that range
(e.g.\ Blundell \& Rawlings 2000, Hardcastle et~al.\ 2007). The mechanical
jet power derived by Willott et~al., expressed in terms of the 1.4 GHz
radio luminosity, corresponds to:

$P_{\rm mech,sync} = 4 \times 10^{35} (f_{\rm W})^{3/2} ({\rm L}_{\rm
  1.4GHz} / 10^{25}{\rm W Hz}^{-1})^{0.86}$ W (Eq 1)

\noindent Note that weak dependencies on other radio source properties
such as environment and source size also exist (see also Shabala \&
Godfrey 2013), but can be ignored to first-order. O'Dea et~al.\ (2009)
used deep multi-frequency radio data for 31 powerful radio galaxies to
observationally estimate the pressure, volume and spectral age of the
lobes, and hence the derive jet power. Daly et~al.\ (2012) combined these
with the radio luminosities to derive a relation between $P_{\rm mech}$
and $L_{\rm radio}$ and found a slope of 0.84, almost exactly the same as
the Willott et~al.\ theoretical prediction. Their normalization is
consistent with Eq 1 for $f_{\rm W} \sim 5$.

\begin{figure}[!t]
\begin{center} 
\psfig{file=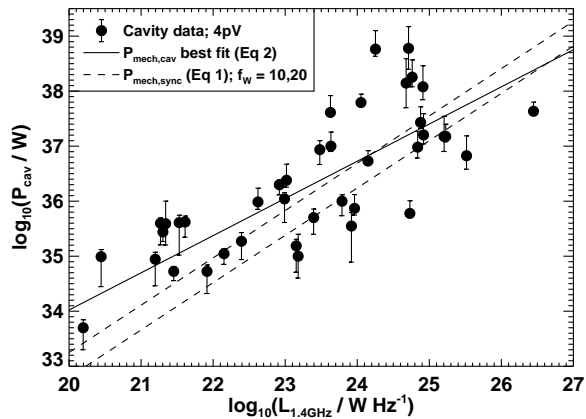,width=8.5cm,clip=} 
\end{center} 
\caption{\label{pcavlrad} The jet mechanical energy of radio sources,
estimated as 4$pV$ from cavities and bubbles in the X--ray gas, versus 
the monochromatic 1.4\,GHz radio luminosity. Data are primarily sourced
from Cavagnolo et~al.\ (2010), B{\^ i}rzan et~al.\ (2008) and Rafferty
et~al.\ (2006). The solid line indicates the best-fit linear relation, 
given in Eq 2. For comparison, the dotted lines show the synchrotron
estimate of Eq 1, for values of $f_{\rm W} = 10$ and 20.}
\end{figure}

The expanding radio sources inflate lobes of relativistic plasma, which in
X-rays can be observed as bubbles, or cavities, in the surrounding hot gas
(B{\"o}hringer et~al.\ 1993). As X-ray facilities developed the
sensitivity and angular resolution to accurately measure the extent of
these cavities and the pressure of the surrounding intergalactic medium,
an alternative method of estimating the radio jet power became
available. B{\^ i}rzan et~al.\ (2004) derived the $pV$ energy associated
with the cavities in a sample of groups and clusters, and used the
buoyancy timescale (e.g.\ Churazov et~al.\ 2001) to estimate their
ages. They combined these cavity powers with the monochromatic 1.4 GHz
radio luminosities to show that the two were well-correlated, with a
scatter of $\sim$0.7 dex. Further cavity power measurements in systems
ranging from giant elliptical galaxies to the most massive clusters have
refined this correlation (e.g.\ Dunn et~al.\ 2005, Rafferty et~al.\ 2006,
B{\^ i}rzan et~al.\ 2008, Cavagnolo et~al.\ 2010; see
Fig.\ref{pcavlrad}). The largest uncertainty in this method is the
determination of the cavity energy from the measured pressure and volume:
E$_{\rm cav} = f_{\rm cav} pV$. The work done inflating the cavity implies
$f_{\rm cav} > 1$, while for the relativistic plasma of the radio lobes
the enthalpy of the cavity is 4$pV$. Additional heating arising directly
from the radio jet shocks (e.g.\ Forman et~al.\ 2007) could imply higher
values of $f_{\rm cav}$, with some authors arguing that $f_{\rm cav}$ can
be as high as $\sim 10$ (e.g.\ Nusser et~al.\ 2006). The value of $f_{\rm
  cav}=4$ is the one usually adopted (e.g.\ Cavagnolo et~al.\ 2010) and is
consistent with a broad balance between AGN heating and radiative cooling
in massive clusters (see Section 6).

Fig.~\ref{pcavlrad} shows a compilation of jet mechanical energy (assuming
4$pV$) and radio luminosity measurements, primarily drawn from the
analysis of Cavagnolo et~al.\ (2010), recast into the same format as Eq
1. From this we derive a best-fit linear relation of

$P_{\rm mech,cav} = 7 \times 10^{36} f_{\rm cav} (L_{\rm 1.4 GHz} /
10^{25}{\rm W Hz}^{-1})^{0.68}$ W (Eq 2)

Eqs 1 and 2 show very similar slopes, and for factors of $f_{\rm cav}=4$
and $f_{\rm W}=15$, the normalizations also agree for typical sources of
$L_{\rm 1.4 GHz} \sim 10^{25}$W\,Hz$^{-1}$ (see Fig.~\ref{pcavlrad}).
Although discrepancies would exist for lower values of $f_{\rm W}$
(cf.\ Daly et~al.\ 2012), the degree of agreement between these two
independent estimators is encouraging. Hereafter in this review, Eq 2 with
$f_{\rm cav}=4$ is adopted to calculate jet mechanical power from the
radio luminosity.

\subsubsection{Black Hole Masses, Eddington Ratios, \& Accretion Rates}

The masses for supermassive black holes (SMBH) have been determined in two
types of situations. In the case of Type 1 AGN, the time-dependent
response of the gas in the Broad Line Region (BLR) to changes in the
ionizing continuum flux can be used to infer a size for the BLR (from
light-travel-time arguments). This size, together with the measured
velocity dispersion in the BLR can be used to estimate $M_{\rm BH}$ under
the assumption that gas is acting solely under the dynamical influence of
the SMBH.  This technique is called reverberation mapping and has been
recently reviewed by Peterson (2013).  The sub-sample of SMBH with masses
determined by reverberation mapping can be used to calibrate a relation
between $M_{\rm BH}$ and a combination of the widths of the broad
emission-lines and the optical continuum luminosity in Type 1 AGN
(e.g.\ Kaspi et~al.\ 2000, Bentz et~al.\ 2013).

\begin{figure}[!t]
\begin{center} 
\begin{tabular}{cc}
\psfig{file=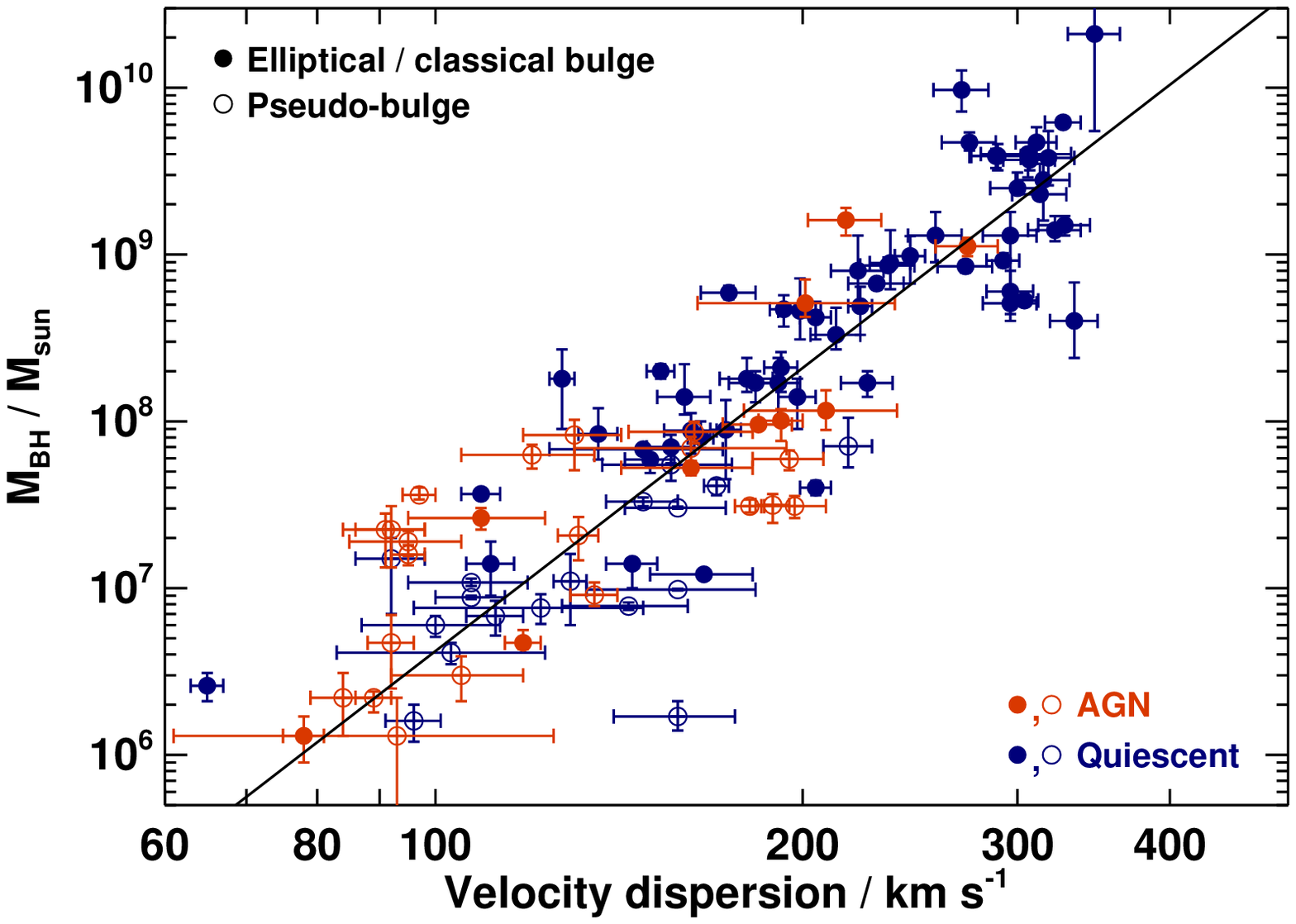,width=7.8cm,clip=} 
&
\psfig{file=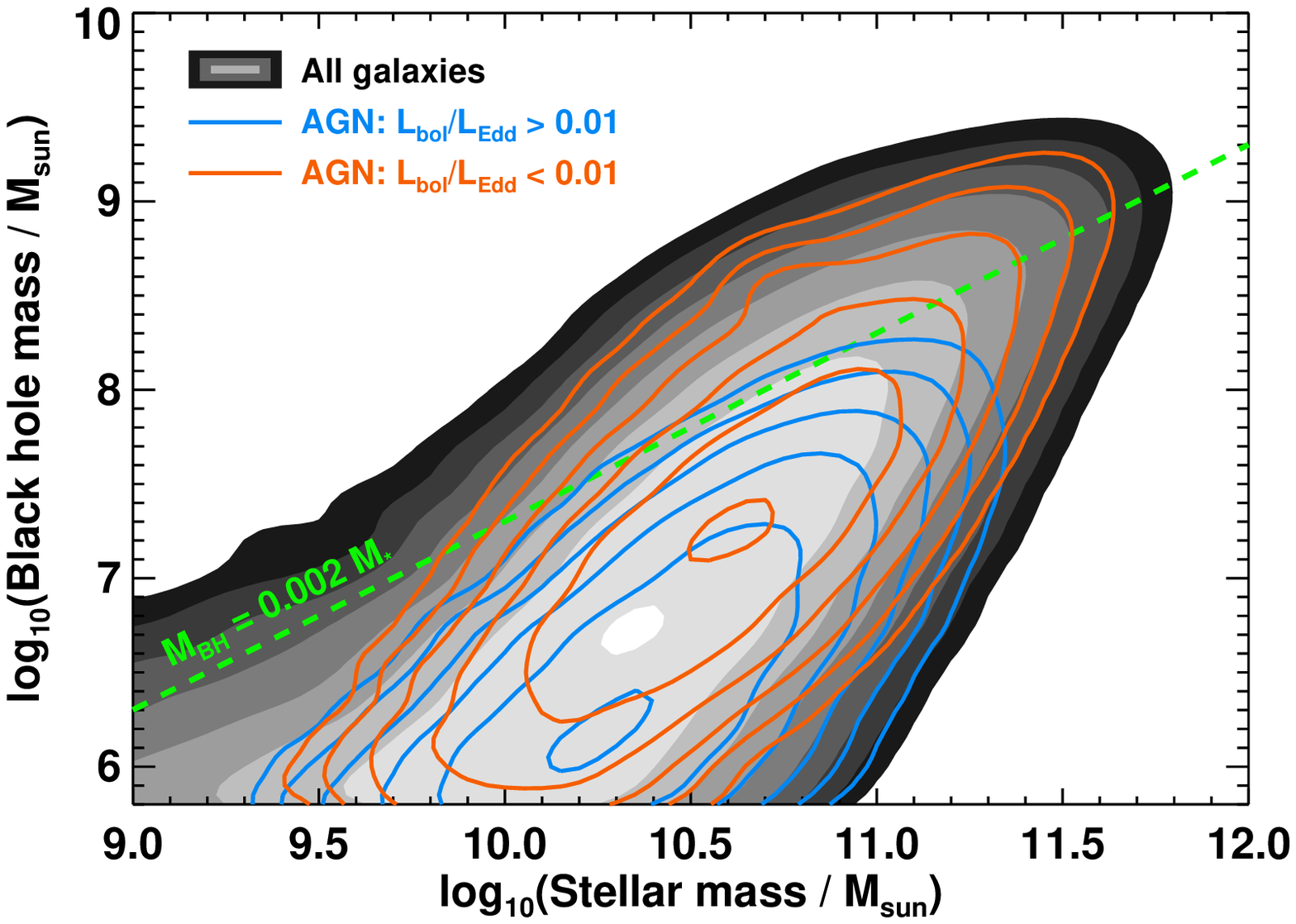,width=7.8cm,clip=} 
\end{tabular} 
\end{center} 
\caption{\label{mass_bhmass} Left: The M$_{\rm BH}$-$\sigma$ relation of
  local galaxies with direct black hole mass measurements (data from Woo
  et~al.\ 2013 and references therein). Both AGN (the color-coding
  includes both radiative-mode and jet-mode AGN) and quiescent galaxies
  are consistent with the McConnell \& Ma (2013) relation, shown by the
  solid line. Right: The distribution of galaxies in the SDSS main galaxy
  sample on the stellar mass {\it vs.}\ black hole mass plane, using black
  hole masses derived from the M$_{\rm BH}$-$\sigma$ relation.  The
  greyscale indicates the volume-weighted distribution of all galaxies,
  with each lighter color band indicating a factor of two increase. It is
  clear that the black hole mass is not a fixed fraction of the {\it
    total} stellar mass. This is also true for AGN: the blue and red
  contours show the volume-weighted distributions of high ($>$1\%; mostly
  radiative-mode) and low ($<1$\%; mostly jet-mode) Eddington-fraction
  AGN, with contours spaced by a factor of two.}
\end{figure}

In this review we are focused on Type 2 and jet-mode AGN and their
relationship to the more numerous population of inactive galaxies. In
these cases, $M_{\rm BH}$ has been measured directly through the dynamical
modeling of the spatially resolved kinematics of stars and/or gas. This
subject has been recently reviewed by Kormendy \& Ho (2013).  There are
still fewer than 100 galaxies in which $M_{\rm BH}$ has been determined
directly.  In order to estimate $M_{\rm BH}$ in large surveys like SDSS,
indirect (secondary) techniques must be used.  The most widely used
secondary technique is based on a fit to the relationship between $M_{\rm
  BH}$ and the stellar velocity dispersion ($\sigma$) of the surrounding
galactic bulge (the M-$\sigma$ relation -- Ferrarese \& Merritt 2000,
Gebhardt et~al.\ 2000; see Fig.~\ref{mass_bhmass}).  This technique is
particularly useful for the SDSS main galaxy sample. The SDSS fiber
typically covers the region within a radius of a few kpc of the
nucleus. The typical hosts for the SDSS AGN are early-type galaxies (see
Section 4 below) and thus the measured stellar velocity dispersion is
representative of the bulge.

Most of the results in the literature that used the M-$\sigma$ relation
were based on the fit given in Tremaine et~al.\ (2002): ${\rm log} (M_{\rm
  BH} / M_{\odot}) = 8.13 + 4.02 {\rm log} [\sigma/(200{\rm km
    s}^{-1})]$. More recent work has found a steeper slope and a different
normalization than the Tremaine et al.\ fit (see the Kormendy \& Ho 2013
review). More specifically, the recent analyses by Graham \& Scott (2013),
McConnell \& Ma (2013), and Woo et~al.\ (2013) find respective slopes of
(6.18,5.64,5.48) and zero-points of (8.15, 8.32, 8.33).  Woo
et~al.\ (2013) also conclude that the relationship between black hole mass
and $\sigma$ is consistent between samples of AGN and quiescent galaxies
(see Fig.~\ref{mass_bhmass}), and this conclusion was reinforced by an
investigation of powerful AGN with more massive black holes by Grier
et~al.\ (2013).  In terms of applicability to SDSS, note that the SDSS
spectra can only reliably measure stellar velocity dispersions above
$\sigma \sim 70$\,km\,s$^{-1}$, corresponding to black hole masses larger
than of-order $10^6$ M$_{\odot}$ (depending on which version of M-$\sigma$
is adopted).  A complementary probe of lower-mass SMBH in SDSS is provided
by using mass estimates based on the broad emission-line width for Type 1
AGN (Greene \& Ho 2007b, Dong et~al.\ 2012, Kormendy \& Ho 2013).

Another secondary technique uses the correlation between $M_{\rm BH}$ and
the stellar mass of the bulge (e.g.\ Marconi \& Hunt 2003, Haring \& Rix
2004). While this is straightforward to apply to elliptical galaxies, the
typical AGN host galaxies in the SDSS sample have significant bulge and
disk components (Section 4). The SDSS images are shallow and do not
well-resolve the bulge component in typical cases.  This makes the task of
robustly determining the bulge/disk mass ratio rather expensive in terms
of computational time and the need for human intervention (e.g.\ Gadotti
2009, Lackner \& Gunn 2012).  Thus, this technique for estimating $M_{\rm
  BH}$ is less readily applicable to the SDSS AGN sample.  Moreover, the
new data compiled by Kormendy \& Ho (2013) and the analysis in Graham \&
Scott (2013) both imply that the correlation of $M_{\rm BH}$ is
significantly poorer with stellar bulge mass than with $\sigma$.

In this review we derive values for $M_{\rm BH}$ using the simple
power-law M-$\sigma$ relation in McConnell \& Ma (2013), specifically:

${\rm log} (M_{\rm BH} / M_{\odot}) = 8.32 + 5.64 {\rm log} [\sigma/(200{\rm km
    s}^{-1})]$ (Eq 3)
 
It is important to emphasize here that it is {\it not} a reasonable
approximation to assume that the black hole mass is some fixed fraction of
the {\it total} stellar mass. To illustrate this we show in the right
panel of Fig.~\ref{mass_bhmass} a plot of the total galaxy stellar mass
{\it vs.}\ the black hole mass as estimated using our adopted M-$\sigma$ relation
for the SDSS main galaxy sample. The ratio of estimated black hole to
galaxy mass drops from an average of about $10^{-2.5}$ at M$_* \sim
10^{11.5}$ M$_{\odot}$ to about $10^{-4}$ at M$_* \sim 10^{10}$
M$_{\odot}$. As also shown in this figure, this is true for galaxies in
general and for the hosts of both radiative-mode and jet-mode AGN.

With an estimate of the AGN bolometric luminosity and supermassive black
hole mass in hand it is useful to then recast things in terms of the
Eddington ratio for the AGN ($L_{\rm bol}/L_{\rm Edd}$).  Here $L_{\rm
  Edd}$ is the luminosity of the classical Eddington limit:

$L_{\rm Edd} = (4 \pi G m_p c/\sigma_T) M_{\rm BH} = 3.3 \times 10^4
M_{\rm BH}$ (Eq4)

\noindent where for the numerical form of the relation the luminosity and
mass are in solar units.  More uncertain still is an estimate of the
accretion rate onto the SMBH. This is typically computed assuming a fixed
efficiency $\epsilon$ for the conversion of accreted mass into radiant
energy:

$L_{\rm bol} = \epsilon \dot{M} c^2$                        (Eq 5)

It is typically assumed that $\epsilon \sim 0.1$. This relatively high
efficiency is well-motivated theoretically for a black hole radiating at a
significant fraction of the Eddington limit (i.e. the radiative-mode
AGN), but may substantially under-estimate the accretion rate of AGN with
very small Eddington ratios (the jet-mode AGN).  We will discuss this more
quantitatively in Section 5 below.

\subsection{ Determining Basic Galaxy Parameters}

\subsubsection{From Imaging and Photometry}

With the advent of digital surveys of large regions of the sky,
astronomers were motivated to design automated algorithms for measuring
the principle properties of galaxies (e.g.\ Lupton et~al.\ 2002). The size
of a galaxy is most often parameterized by the radius that encloses a
fixed fraction (often 50\%; written as $R_{50}$) of the total light.  The
shape of the radial surface-brightness profile provides a simple way to
estimate the prominence of the bulge with respect to the disk: galaxies
with concentration index $C = R_{90}/R_{50} > 2.6$ are predominantly
bulge-dominated systems, with Hubble classification of Type Sa or earlier,
whereas galaxies with $C < 2.6$ are mostly disk-dominated systems
(Shimasaku et~al.\ 2001).  Gadotti (2009) performed 2-dimensional
bulge/bar/disk decompositions of a representative sample of galaxies from
the SDSS and found that the concentration index correlated more tightly
with bulge-to-disk ratio than the Sersic index $n$ obtained from
1-dimensional profile fitting. Lackner \& Gunn (2012) reached a similar
conclusion from analysis of a larger SDSS sample.  Recently, the power of
citizen science has been exploited to create visual classifications of
roughly 900,000 SDSS galaxies (Lintott et~al.\ 2011). These
classifications have been used to document the nature of the host galaxies
of AGN.

The stellar mass-to-light ratio (and hence the stellar mass) of a galaxy
can be measured using multi-band optical and near-IR photometry to an
accuracy of $\sim 10\%$. While this spectral energy distribution will be
sensitive to both the (luminosity-weighted) age of the stellar population
and the amount of dust reddening, these two effects operate in such a way
that the relation between optical color and $M/L$ is linear with little
scatter if galaxies have undergone relatively smooth recent star formation
histories (Bell \& De Jong 2001, Salim et~al.\ 2005). These estimates
assume a fixed stellar initial mass function (IMF).  The effective surface
mass density $\mu_*$ is then defined as $0.5M_*/(\pi R_{50}^2$).

The star formation rate of a galaxy can also be estimated from fits to the
ultraviolet through near-IR multiband photometry (e.g.\ Salim
et~al.\ 2005). Ultraviolet photometry from GALEX is particularly important
as it probes the young stellar population in star-forming galaxies. We
note, however, that in the absence of far-infrared photometry, which
allows the thermal emission from dust to be measured directly, star
formation rates estimated from UV/optical photometry can be significantly
biased if the range of dust extinction values in the SED model library is
incorrect (Saintonge et~al.\ 2011, Wuyts et~al.\ 2011a).  Most of this
dust emission powered by the reprocessing of ultraviolet light from
massive stars emerges in the far-infrared.  Unfortunately, the depth of
the all-sky IRAS survey is not well matched to the SDSS main galaxy
sample, with only about 2\% of these galaxies being reliably detected by
IRAS at 60 and 100 $\mu$m (Pasquali et~al.\ 2005).  The situation has been
improved in some respects by the WISE survey which detects about 25\% of
the SDSS main galaxies in all four WISE bands (3.4, 4.6, 12, and 22
$\mu$m).  The disadvantage of the WISE data for estimating the star
formation rate is that larger and more uncertain corrections need to be
made to estimate the total infrared luminosity than in the case of
IRAS. In the case of AGN, the WISE data suffer from the further problem of
potential contamination by AGN-powered emission from warm dust in the
obscuring structure, even at the longest-wavelength WISE band (22 $\mu$m).

Finally, radio continuum emission can also be used to estimate the
star-formation rate since it shows a strong correlation with the
far-infrared emission in star-forming galaxies (Condon et~al.\ 1992).
Given the depth of the FIRST survey, the sensitivity of this technique
over the SDSS volume is rather similar to that of the IRAS Faint Source
Catalog, but -- for our purposes -- it has the disadvantage of being more
susceptible to contamination by AGN emission than the far-IR.

\subsubsection{From Spectra}
\label{d4000sec}

The spectra available for galaxies from the SDSS are taken through 3
arcsecond diameter fibers.  At the median redshift of the main galaxy
sample ($z \sim 0.1$), only about 25-30\% of the total light from the
galaxy is sampled. As a result, estimates of the global star formation
rates of SDSS galaxies cannot be made using only the spectra (Brinchmann
et~al.\ 2004).  The spectra do, however, provide a very useful probe of
the central stellar populations of normal galaxies and AGN hosts.

The break occurring at 4000\AA\ is the strongest discontinuity in the
optical spectrum of a galaxy.  The main contribution to the opacity comes
from low-ionization metals. In hot stars, the elements are multiply
ionized and the opacity decreases, so the 4000\AA\ break will be small
for young stellar populations and large for old, metal-rich galaxies.  A
break index, D(4000), was defined initially by Bruzual (1983).  A
definition using narrower and more closely-spaced continuum bands was
introduced by Balogh et~al.\ (1999), and has the advantage of being
considerably less sensitive to reddening effects.

Strong Balmer absorption lines arise in galaxies that experienced a burst
of star formation that ended $\sim 10^8$ to $10^9$ years ago, when A stars
are the dominant contributors to the optical continuum (e.g.\ Dressler \&
Gunn 1983).  Worthey \& Ottaviani (1997) defined an H$\delta_A$ index
using a central bandpass bracketed by two pseudo-continuum bandpasses.
Kauffmann et~al.\ (2003b) used stellar population models (Bruzual \&
Charlot 2003) to show that the locus of galaxies in the
D(4000)-H$\delta_A$ plane is a powerful diagnostic of whether galaxies
have been forming stars continuously or in bursts over the past 1-2
Gyr. In addition, these two stellar indices are largely insensitive to the
dust attenuation effects that complicate the interpretation of broad-band
colors. These indices are very useful diagnostics of the recent star
formation history of Type 2 and jet-mode AGN.

Measurements of the H$\delta_A$ index that have a precision that is good
enough to provide useful constraints require relatively high
signal-to-noise data.  However, the H$\delta$ line itself is only one of a
handful of other high-order Balmer lines in the blue/near-UV portion of
the spectrum. The situation can be significantly improved by utilizing the
combined information in these lines. One technique for doing so uses
Principal Component Analysis (PCA). This approach has been successfully
applied to SDSS spectra by Wild et~al.\ (2007) and Chen
et~al.\ (2012). More examples of the use of PCA and other sophisticated
techniques for extraction of information about the stellar content and
star-formation history from the SDSS galaxy spectra include Panter
et~al.\ (2003, 2007), Cid Fernandes et~al.\ (2005), and Yip
et~al.\ (2004).

\section{BLACK HOLE MASS \& AGN LUMINOSITY FUNCTIONS}

\subsection{The Radiative-Mode Population}

The luminosity functions for the radiative-mode AGN (Seyferts and QSOs)
can be constructed using a variety of different indicators. Some
indicators, such as broad-band optical luminosities, are effective at
probing unobscured black holes accreting near their Eddington limit at
high redshifts. However, when the accretion rate is low, these broad-band
optical luminosities are dominated by stellar emission from the host
galaxy. Moreover, the optical/UV continuum will not recover the obscured
(Type 2) AGN population.  Other indicators, such as high-ionization
nebular emission line luminosities, mid-infrared continuum emission, and
X-ray emission have been used to probe lower accretion rate and obscured
systems. Nice syntheses of these multi-waveband data have been undertaken
by Hopkins et~al.\ (2007) and Shankar et~al.\ (2009) who use them to
produce bolometric luminosity functions for the radiative-mode AGN that
span the redshift range from $\sim$ 0 to 6.

At a given redshift, the AGN bolometric luminosity function can be
approximated as a double power-law:

$d\Phi/d{\rm log}(L) = \phi_* / ((L/L_*)^{\gamma_1} + (L/L_*)^{\gamma_2})$         (Eq6)

\noindent This is characterized by a faint end slope ($\gamma_1$), a
bright end slope ($\gamma_2$), the characteristic luminosity of the break
in the function ($L_*$) and the co-moving space density of AGN with that
luminosity ($\phi_*$).  The evolution of the bright end of the luminosity
function from redshifts $\sim 0$ to 3 can be well-described by a so-called
pure luminosity evolution (PLE) model in which $L_*$ evolves strongly with
redshift, while $\gamma_2$ and $\phi_*$ do not (Hopkins et~al.\ 2007).
Fiore et~al.\ (2012) show that this behavior of the bright AGN population
may extend to even higher redshifts.  The value for $L_*$ (bolometric)
increases from about $10^{11.5}$ L$_{\odot}$ at $ z\sim 0$ to 10$^{13}$
L$_{\odot}$ at $z \sim 2$ to 3, and then declines at higher redshift
(e.g.\ Hopkins et~al.\ 2007, Shankar et~al.\ 2009).

The first clear indications for a departure from the PLE model came from
X-ray surveys of AGN using Chandra, which generated complete samples of
Type 1 AGN down to significantly lower luminosities than was possible with
optical surveys (see Brandt \& Hasinger 2005).  These surveys demonstrated
that the space density of lower luminosity AGN peaks at lower redshift
than that of higher luminosity AGN, a phenomenon that is often referred to
as AGN down-sizing. Similar results were later obtained by optical QSO
surveys that pushed down to fainter flux levels. The QSO luminosity
function of the completed 2dF-SDSS optically-selected QSO survey (Croom
et~al.\ 2009) agrees extremely well in both amplitude and shape with the
X-ray selected QSO luminosity functions of Hasinger et~al.\ (2005).

In terms of the luminosity function, this departure from a pure PLE model
is primarily manifested as an evolution in the faint end slope with
$\gamma_1$ increasing from $\sim$0.3 at $z = 3$ to $\sim$0.9 at $z = 0$
(Hopkins et~al.\ 2007). The steep faint-end slope at low-z is seen for
both Type 1 and Type 2 AGN and in both the [OIII]5007 (Hao et~al.\ 2005a)
and hard X-ray (Ueda et~al.\ 2011 and references therein) luminosity
functions.  Note that a faint end slope of 1 corresponds to equal radiant
output per bin in log L.  This has an important implication: while at
high-redshifts the bulk of the radiant output occurs in AGN near $L_*$,
this is not the case in the contemporary universe where it is spread rather
evenly across a broad range in luminosity (and only about 20\% is produced
by AGN with $L_{\rm bol} > L_*$).

\begin{figure}[!t]
\begin{center} 
\psfig{file=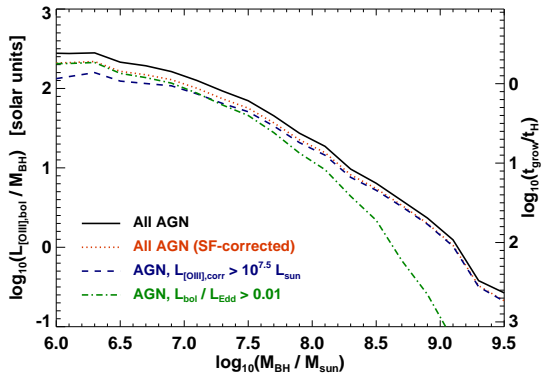,width=8cm,clip=} 
\end{center} 
\caption{\label{heck04a} Logarithm of the ratio of the total bolometric
  radiated luminosity (as calculated from the [OIII] emission line; see
  Section 2.3.1) per unit volume due to emission-line AGN of a given black
  hole mass to the total mass per unit volume in black holes at that black
  hole mass (both in solar units) in the contemporary universe. The right
  axis shows the corresponding growth time (mass-doubling time) for the
  population of black holes in units of the Hubble time. The solid black
  line shows the result obtained by integrating over all AGNs in the
  sample. The dotted red line shows the effect of correcting the [OIII]
  luminosities for the contribution from star formation in composite
  galaxies. The dashed blue line shows the result obtained if only AGN
  with (reddening-corrected and star-formation corrected) [OIII]
  luminosities above $10^{7.5} L_{\odot}$ are included. The dash-dot green
  line includes only AGN with bolometric luminosities above 1\% of the
  Eddington rate (this cut effectively reduces the sample to only
  radiative-mode AGN). Adapted from Heckman et~al.\ (2004).}
\end{figure}

Heckman et~al.\ (2004) used data from the SDSS to compute the integrated
[OIII] luminosity emitted by (Type 2 and jet-mode) emission-line AGN per
unit black hole mass for nearby black holes spanning the mass range from
$3 \times 10^6 -10^9$ $M_{\odot}$.  They adopted a bolometric correction
$L_{\rm bol}/L_{\rm [OIII]}=3500$ (see Section 2.3.1) and a mass-to-energy
conversion efficiency of $\epsilon = 0.1$ to transform these ratios into
estimates of the time taken for the entire population of black holes to
have grown to their present-day mass, given their present-day
volume-averaged accretion rate.  They defined the mass-dependent
volume-averaged growth time for the local population of black holes to be
the total mass of the BH population per unit volume in a given BH mass bin
divided by the total growth rate of the BH population per unit volume in
the same BH mass bin. They then found that this volume-averaged growth
time of the population was of-order the Hubble time for black holes with
masses less than $10^{7.5} M_{\odot}$ , but that this growth time
increased sharply for higher mass black holes, implying that most of their
mass was accreted at high redshifts (Fig.~\ref{heck04a}).  Greene \& Ho
(2007a, 2009) reached similar conclusions for the local Type 1 AGN
population in the SDSS.

Heckman et~al.\ (2004) also studied the fraction of black holes of mass
$M_{\rm BH}$ with accretion rates above a given value. They showed that
these cumulative accretion rate functions cut off sharply at $\dot{M}_{\rm
  BH}/\dot{M}_{\rm Edd} \simeq 1$. Many more low mass black holes than
high mass black holes had accretion rates near the Eddington limit (0.5\%
of $10^7 M_{\odot}$ black holes are accreting above a tenth the Eddington
limit, as compared to 0.01\% of black holes with masses $3 \times 10^8
M_{\odot}$). In a later paper, Kauffmann \& Heckman (2009) analyzed
Eddington ratio distribution functions in narrow bins of black hole mass.
They found that the distribution functions can be decomposed into a
log-normal distribution of accretion rates, plus a power-law function that
dominates at low Eddington ratio. We will describe the dependence of this
distribution function on the age of the central stellar population in
Section 4.2 below (see also Fig.~\ref{eddfracfuncs}).  

\subsection{The Jet-Mode AGN}

The broad nature of the radio luminosity function of local galaxies has
been well-established for many years (e.g.\ Toffolatti et~al.\ 1987,
Condon 1989), but the large spectroscopic surveys have enabled both
improved statistics and an accurate separation of the star-forming and
radio-AGN populations. Radio luminosity functions derived from the Las
Campanas redshift survey (Machalski \& Godlowski 2000), the 2dFGRS (Sadler
et~al.\ 2002), the SDSS (Best et~al.\ 2005a, Best \& Heckman 2012) and the
6-degree-field galaxy survey (6dFGS; Mauch \& Sadler 2007) are all in
excellent agreement, and show that the local radio luminosity function of
AGN is well-fitted by a double power-law model similar in form to Eq 6:

$\rho = \rho_0 / ((P/P_0)^{\alpha} + (P/P_0)^{\beta})$                  (Eq7)

At 1.4 GHz, the best-fitting values from combining all four spectroscopic
surveys above give $\rho_0 = 10^{-(5.33 \pm 0.12)}$ Mpc$^{-3}$
log$_{10}(P)^{-1}$, $P_0 = 10^{24.95 \pm 0.14}$ W Hz$^{-1}$, $\alpha=0.42
\pm 0.04$ and $\beta = 1.66 \pm 0.21$. Mauch \& Sadler (2007) show that,
at the faint end, the radio luminosity function of radio-loud AGN
continues with the same slope down to at least $P_{\rm 1.4GHz} \sim
10^{20}$W\,Hz$^{-1}$. They argue that the luminosity function must begin
to turn-over at around $P_{\rm 1.4GHz} \sim 10^{19.5}$W\,Hz$^{-1}$, or the
total space density of radio sources would exceed the space density of
galaxies brighter than $L_*$, which form the typical hosts. Cattaneo \&
Best (2009) similarly argue for a turn-over no fainter than $P_{\rm
  1.4GHz} \sim 10^{19.2}$W\,Hz$^{-1}$ by comparison with the local space
density of supermassive black holes above $10^6$ M$_{\odot}$. Thus, the
observed radio luminosity functions now broadly constrain the entire
distribution of AGN radio luminosities.

\begin{figure}[!t]
\begin{center} 
\psfig{file=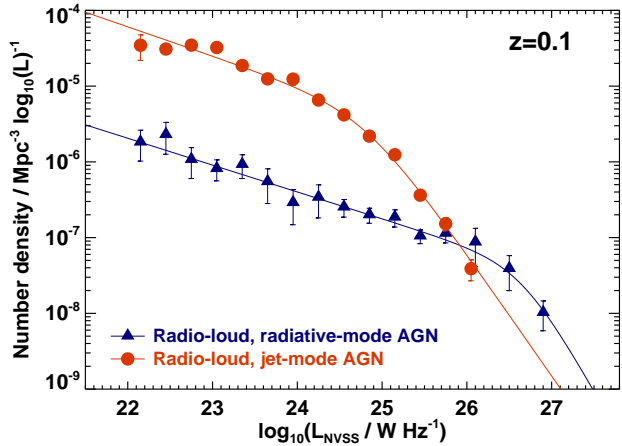,width=9cm,clip=} 
\end{center} 
\caption{\label{localradfunc} The local radio luminosity function of
  radio-loud AGN, split into radiative-mode and jet-mode sources. The data
  are largely drawn from the results of Best \& Heckman (2012) who split
  these two populations and derived luminosity functions using
  radio-selected AGN in the SDSS main galaxy sample. However, as shown by
  Gendre et~al.\ (2013), these results under-estimate the luminosity
  function of the radiative-mode AGN above $10^{26}$ W\,Hz$^{-1}$, where
  Type-1 AGN dominate. Therefore the results of Gendre et~al.\ are adopted
  for the three highest luminosity points of the radiative-mode
  population. The local radio luminosity functions are well-fitted by
  broken power law models.}
\end{figure}

The AGN population as revealed by radio surveys is composed of two
distinct categories of sources: a population with strong QSO/Seyfert-like
emission-lines (historically refered to as high-excitation sources) and a
population with weak LINER-like emission-lines (referred to as
low-excitation; e.g.\ Hine \& Longair 1979, Laing 1994). In the language
of the present review these are radiative-mode and jet-mode AGN, where the
high excitation sources form a small sub-population of radiative-mode AGN
that produce powerful jets.  Best \& Heckman (2012) separated the radio
luminosity function into these two radio source populations and showed
that both are found over the full range of radio powers they observed (see
Fig.~\ref{localradfunc}).  The jet-mode sources constitute about 95\% of
the radio-AGN population at all radio luminosities below $P_{\rm 1.4GHz} =
10^{25}$W\,Hz$^{-1}$, but the radiative-mode population becomes dominant
above $10^{26}$W\,Hz$^{-1}$. In the radio samples derived from large
spectroscopic surveys of the contemporary universe, therefore, the
radio-AGN population is predominantly composed of the jet-mode
(low-excitation) sources.

The local radio-AGN population has historically been split by Fanaroff \&
Riley (FR; 1974) radio morphology, where FR2 sources are `edge-brightened'
with jets that remain collimated until they end in bright hotspots, and
FR1 sources are `edge-darkened' with jets that flare and gradually fade.
The FR1 and FR2 populations are also found over a wide range of
overlapping radio powers, with the switch in the dominant local population
occuring at around $P_{\rm 1.4 GHz} \sim 10^{25}$ W Hz$^{-1}$ (cf.\ Best
2009, Gendre et~al.\ 2010). There is considerable overlap between the
jet-mode AGN population and those radio sources morphologically classified
as FR1s, and also between the radiative-mode AGN with strong radio
emission and the FR2s.  However, the radio morphological split and the
division in the nature of the accretion flow are not the same thing. The
large overlap arises because the population switch occurs at similar radio
powers, but there exists a large cross-population of jet-mode FR2s
(e.g.\ Laing 1994) and a smaller population of radiative-mode FR1s
(e.g.\ Heywood et~al.\ 2007), as quantified by Gendre et~al.\ (2013).

Best et~al.\ (2005b) considered the distribution of radio emission
associated with AGN activity across galaxies of different black hole
masses. They showed that the prevalence of radio-AGN activity is a very
strong function of black hole mass, rising as $f_{\rm rad} \propto M_{\rm
  BH}^{1.6}$. This offered a proper quantification of earlier widespread
statements in the literature that radio-loud AGN are found only in
galaxies at the top end of the black hole mass function. Best
et~al.\ (2005b) also considered the bivariate distribution of radio
luminosity and black hole mass.  They found that the shape of the radio
luminosity function showed only weak dependence on black hole mass.

\begin{figure}[!t]
\begin{tabular}{cc}
\psfig{file=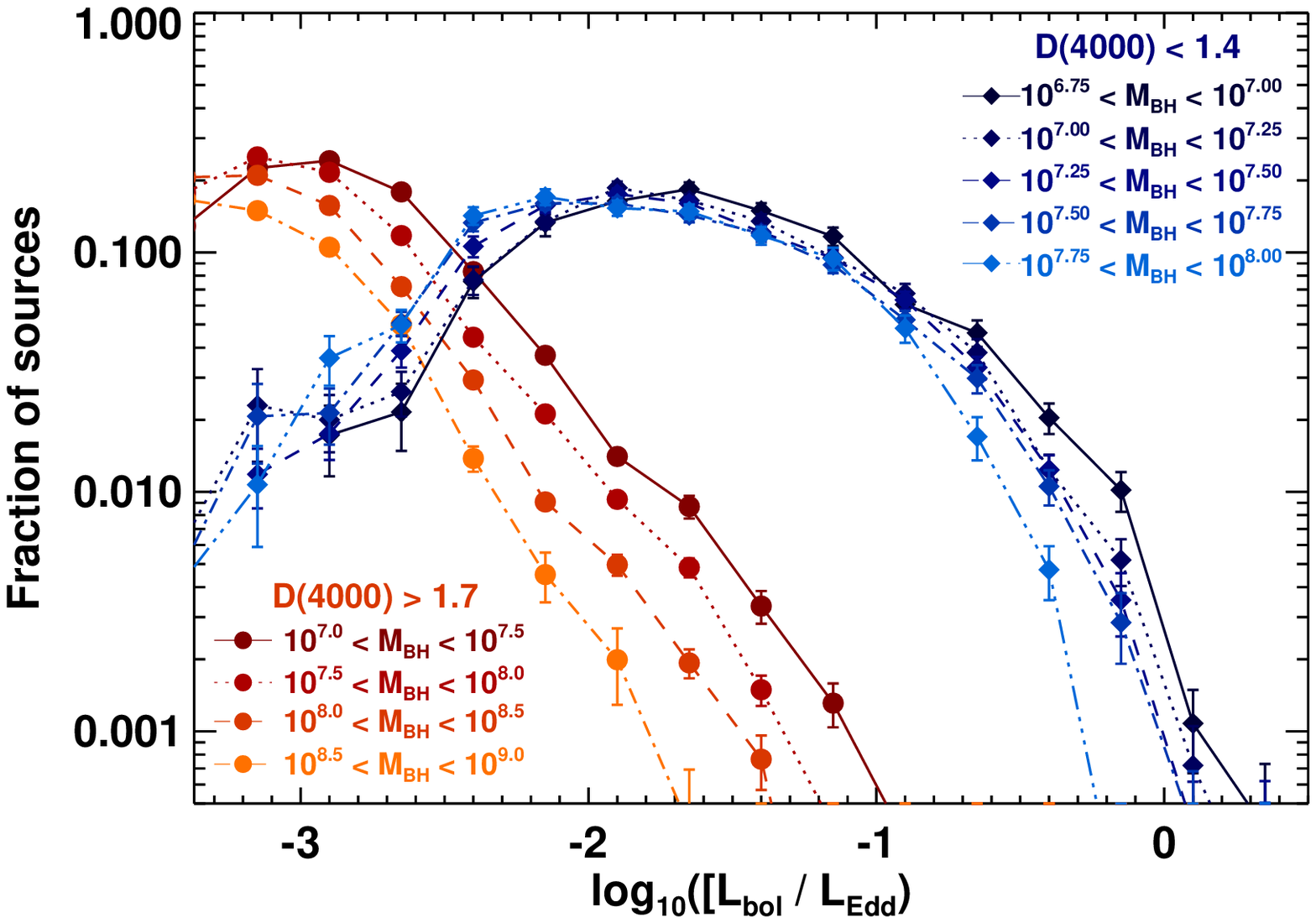,width=7.8cm,clip=} 
&
\psfig{file=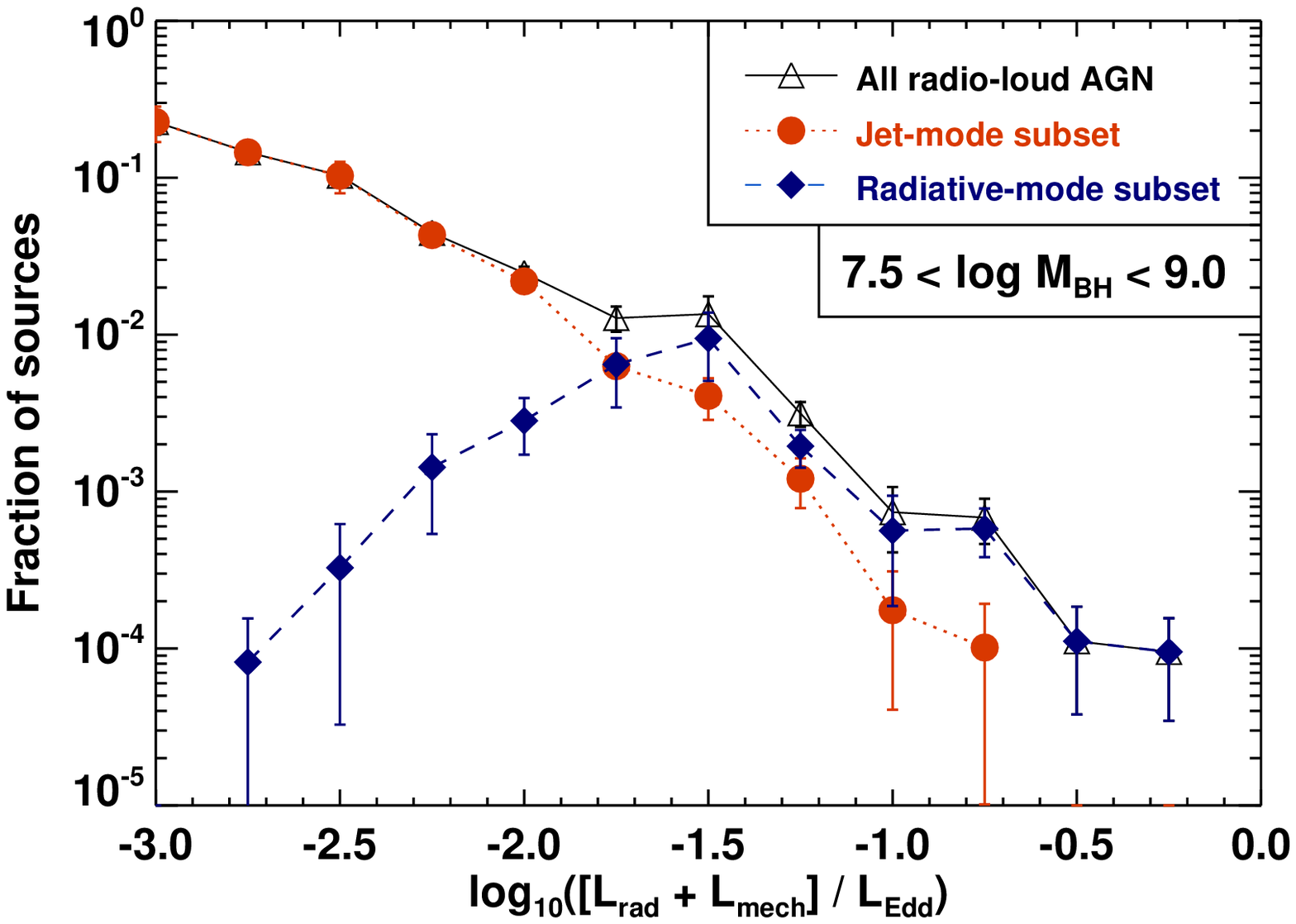,width=7.8cm,clip=} 
\end{tabular} 
\caption{\label{eddfracfuncs} Left: the distribution of Eddington-scaled
  accretion ratios for emission-line selected AGN from the SDSS main
  galaxy sample, split by star-formation activity.  Following Kauffmann \&
  Heckman (2009), the distributions are shown separately for actively
  star-forming host galaxies (D(4000)$<$1.4) and for passive host galaxies
  (D(4000)$>1.7$), each for a range of black hole masses. Fractions
  plotted are within each 0.25-dex wide bin in Eddington ratio. As found
  by Kauffmann \& Heckman, the passive galaxies show a power-law
  distribution of Eddington fractions, the amplitude of which decreases
  with increasing black hole mass. In contrast, the Eddington ratios of
  the star-forming hosts show a log-normal distribution peaking at a few
  percent of Eddington. This distribution is largely independent of black
  hole mass (though this depends slightly upon the version of the
  M-$\sigma$ relation adopted). Right: the Eddington-scaled accretion
  ratios of radio-loud AGN from SDSS, split into jet-mode and
  radiative-mode sources; again, fractions plotted are within each
  0.25-dex wide bin. Here the bolometric luminosity is calculated by
  summing the radiative and mechanical contributions.  The two classes
  show distributions remarkably similar to the populations seen for
  emission-line AGN in the left panel.}
\end{figure}

The Eddington ratio distribution function of radio-loud AGN can be
calculated by summing the jet mechanical luminosity and the bolometic
radiative luminosity (cf.\ Section 2.3) to provide the total energetic
output of the AGN. Best \& Heckman (2012) showed that the radiative-mode
and jet-mode radio sources show distinct distributions of Eddington
ratios, with the latter typically accreting below 1\% of their Eddington
rate, while the former have typical accretion rates between 1 and 10\% of
Eddington. Considering sources with black hole masses between $10^{7.5}$
and $10^9$ M$_{\odot}$, the right panel of Fig.~\ref{eddfracfuncs} shows
the distribution of Eddington ratios of radio-loud AGN in the SDSS main
galaxy sample, above $10^{-3}$ of the Eddington limit (incompleteness
prohibits analysis below this level). The jet-mode sources dominate at low
Eddington ratios and show a roughly power-law distribution with a slope of
about -1.5. In contrast, the radiative-mode population shows a
distribution peaked at a few percent of the Eddington limit.

\subsection{The Black Hole Mass Function}

The super-massive black hole mass function (BHMF) is defined as the
co-moving number density of black holes per bin in log mass at a given
redshift.  The techniques used to determine the BHMF have been reviewed
recently by Kelly \& Merloni (2012). Here we briefly summarize the
properties and implications of the BHMF.

\begin{figure}[!t]
\begin{center} 
\psfig{file=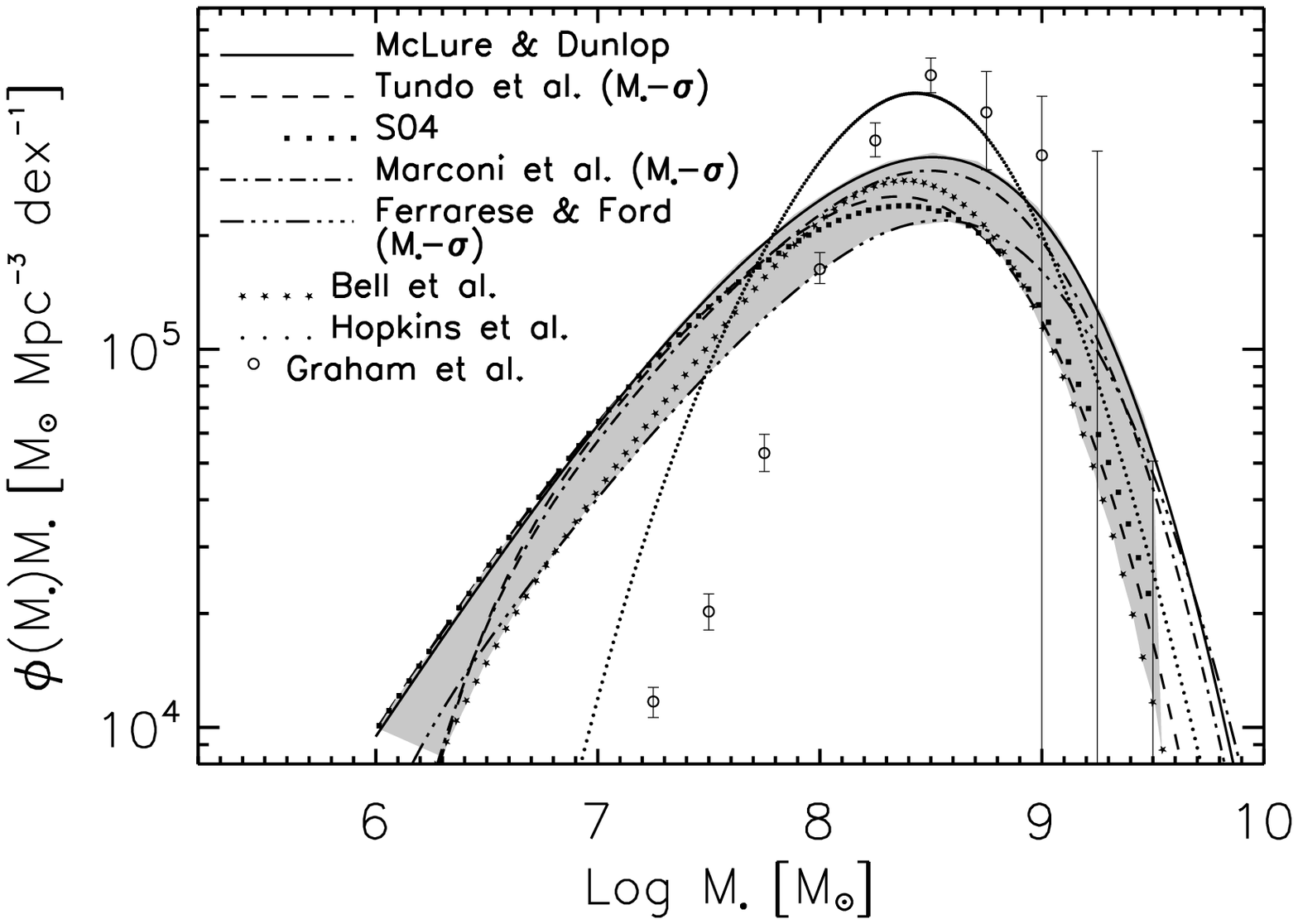,width=9cm,clip=} 
\end{center} 
\caption{\label{shankbhmf} A collection of estimates of the distribution
  of mass across the local black hole population (ie., the local black
  hole mass function scaled by black hole mass) taken from Shankar
  et~al.\ (2009). The shaded grey area represents the range of estimates
  based on local scaling relations of M$_{\rm BH}$ with $\sigma$, M$_{\rm
    Bulge}$, and L$_{\rm Bulge}$. See Shankar et~al.\ (2009) for details.}
\end{figure}

In the contemporary universe the BHMF can be measured in an empirical
model-independent fashion by adopting the measured scaling relation
between black hole mass and the bulge mass or velocity dispersion
(described in Section 2.3.3). Combined with the measured volume density of
bulges as a function of velocity dispersion or stellar mass from the SDSS,
the BHMF can then be computed. Shankar et~al.\ (2009) have compared the
BHMF derived this way by different groups (Fig.~\ref{shankbhmf}).

The two main conclusions to be drawn from this figure are that the
distribution of mass in black holes today peaks in black holes with masses
of around $10^{8.5} M_{\odot}$ and that the mass distribution is actually
quite narrow (half the total mass in black holes resides in the mass range
from $\sim 10^{7.8}$ to 10$^{8.8}$ M$_{\odot}$).  The integral of this
distribution represents the total mass per unit volume in the contemporary
universe ($\rho \sim 10^{5.7}$ M$_{\odot}$ Mpc$^{-3}$).  This mass density
is directly connected to the total energetic output of AGN over the
history of the universe (Soltan 1982).  Consistency with the cosmic
evolution of the AGN bolometric luminosity function (summarized above)
implies a mean accretion efficiency ($\epsilon = L_{\rm bol}/\dot{M}c^2$)
of-order 10\% (Yu \& Tremaine 2002, Marconi et~al.\ 2004, Hopkins
et~al.\ 2007, Shankar et~al.\ 2009).

At higher redshifts the BHMF has usually been computed in a
model-dependent fashion. These computations take the present-day BHMF as a
boundary condition and use a continuity equation to relate the evolution
of the BHMF to the accretion history as inferred from the evolution of the
AGN luminosity function. To compute BHMF(z) an assumption must be made
about the mean AGN light curve ($L(M_{\rm BH},t)$) and the accretion
efficiency ($\epsilon$). Examples of this approach are Marconi
et~al.\ (2004), Shankar et~al.\ (2009, 2013) and Cao (2010). An alternative
technique for measuring BHMF(z) uses the mass function of actively
accreting black holes and the so-called Black Hole Fundamental Plane
(Merloni \& Heinz 2008).

All these analyses reach qualitatively similar conclusions: the BHMF grows
anti-hierarchically with the population of more massive BHs being produced
at higher redshifts. The population of black holes with masses $> 10^9$
M$_{\odot}$ was largely in place by $z \sim$ 2.  Since $z \sim 1$ only the
population of black holes with masses below about $10^8$ M$_{\odot}$ has
been growing significantly. As we have summarized above, this inference is
directly confirmed by measurements of the mass-dependent growth rates of
supermassive black holes in the contemporary universe (Heckman
et~al.\ 2004, Kauffmann \& Heckman 2009).

\section{ HOST GALAXY PROPERTIES}

\subsection{Overview}

In the introduction to this review we summarized the overall landscape
defined by the population of galaxies in the contemporary universe and
described the bimodal galaxy population (lower mass, lower density,
star-forming (blue) {\it vs.}\ higher mass, higher density quiescent (red)
galaxies. Here we want to see where the population of growing black holes
today lives within this landscape.  As we will see, both the structural
properties of the galaxies and their stellar populations turn out to be
crucial in this regard.  We begin with an overview of both the
radiative-mode and jet-mode AGN in this section, and will then follow with
detailed discussions of the two populations in the two sections to follow.

\begin{figure}[!t]
\begin{center} 
\psfig{file=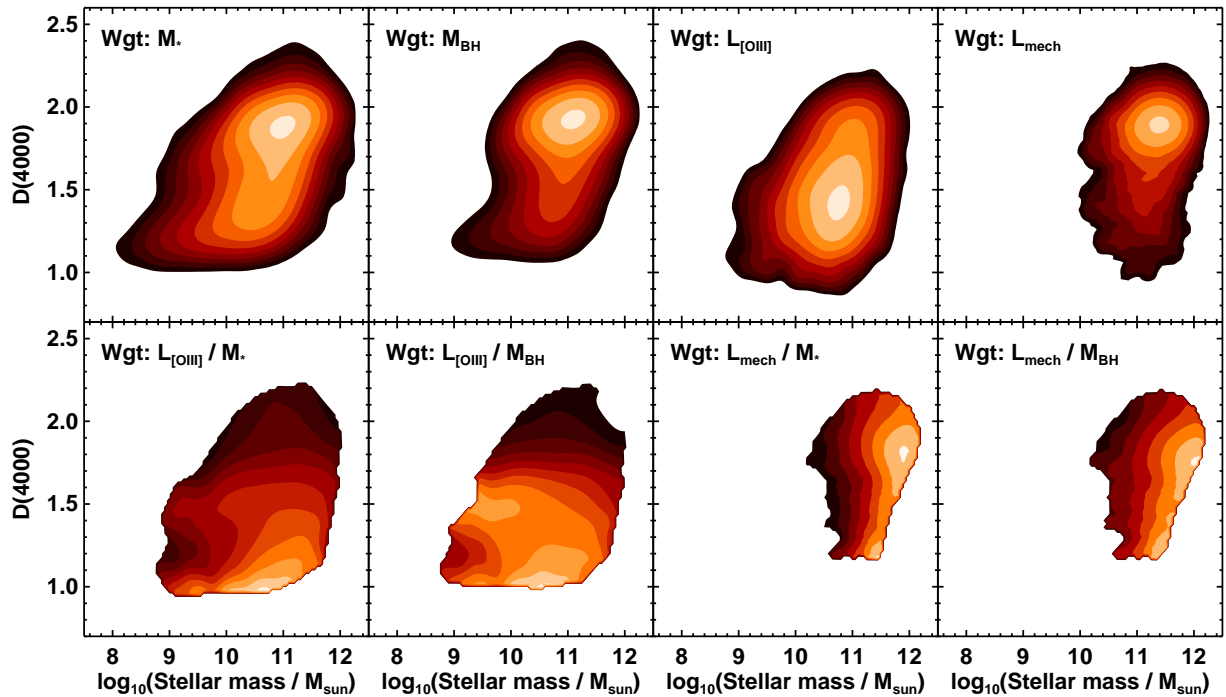,angle=90,width=\textwidth,clip=} 
\end{center} 
\caption{\label{agnfracs} Top panels: the distribution of galaxies from
  the SDSS main galaxy sample in the plane of stellar mass {\it vs.}
  4000\AA\ break strength, weighted according to (from left to right):
  stellar mass; black hole mass; reddening-corrected [OIII] line
  luminosity (ie., bolometric radiative luminosity); jet mechanical
  luminosity. Each color band corresponds to a factor of two difference,
  increasing from dark to light. Bottom panels: ratios of upper panels
  (same color coding) to illustrate the distributions of (from left to
  right): bolometric radiative luminosity per stellar mass; bolometric
  radiative luminosity per black hole mass; jet mechanical luminosity per
  stellar mass; jet mechanical luminosity per black hole mass.}
\end{figure}

\begin{figure}[!t]
\begin{center} 
\psfig{file=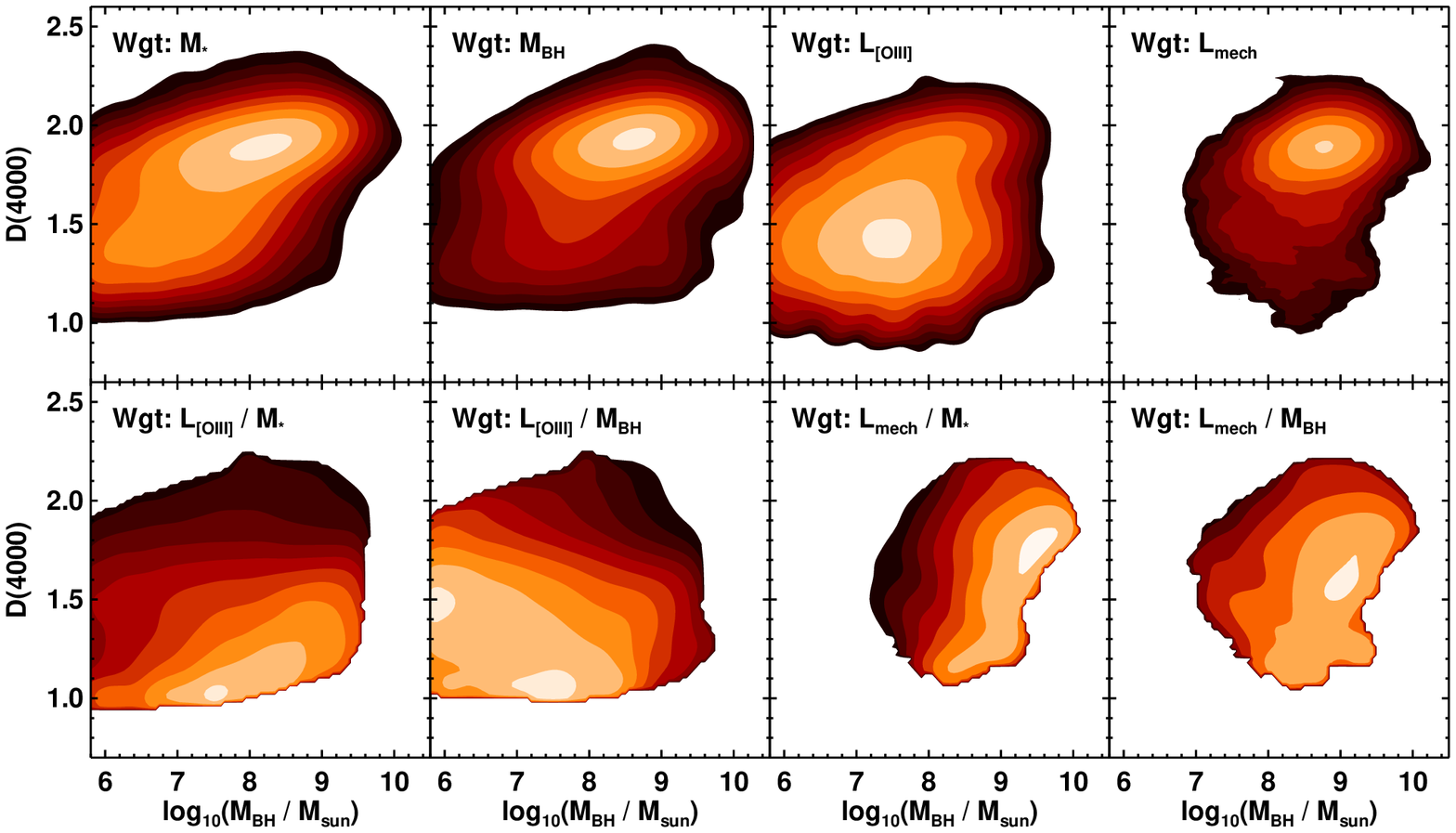,angle=90,width=\textwidth,clip=} 
\end{center} 
\caption{\label{agnfracs2} As Fig.~\ref{agnfracs}, but now showing the
  distributions in the black hole mass {\it vs.}\ 4000\AA\ break strength
  plane.}
\end{figure}

To fully appreciate the interplay between the structural properties and
the stellar population in AGN host galaxies it is worth taking a guided
tour of the pair of multi-panel plots shown in Figs.~\ref{agnfracs}
and~\ref{agnfracs2}, derived from the SDSS.  Fig.~\ref{agnfracs} plots the
volume-weighted values of various key parameters in the 2-dimensional
space of stellar mass (M$_*$) and D(4000).  The top four panels show the
distributions of M$_*$, black hole mass (M$_{\rm BH}$),
extinction-corrected [OIII] luminosity from AGN (L$_{\rm [OIII]}$; tracing
bolometric radiative output), and the production of mechanical energy in
radio jets (L$_{\rm mech}$).  The bottom four panels show the ratios of
L$_{\rm [OIII]}/$M$_*$, L$_{\rm [OIII]}/$M$_{\rm BH}$, L$_{\rm
  mech}$/M$_*$, and L$_{\rm mech}$/M$_{\rm BH}$.

The plot weighted by M$_*$ shows the basic layout of the bimodal galaxy
population.  The plot weighted by M$_{\rm BH}$ shows a shift towards
higher values of M$_*$ (consistent with the steep dependence of M$_{\rm
  BH}$ on M$_*$ shown in Fig.~\ref{mass_bhmass}). The plot weighted by
L$_{\rm [OIII]}$ traces the growth of black holes through accretion (this
is essentially tracing the radiative-mode population of AGN). It is
immediately clear that black hole growth is occurring primarily in
moderately massive galaxies (M$_* \sim 10^{10}$ to a few times $10^{11}$
M$_{\odot}$) with a young central stellar population (see Kauffmann
et~al.\ 2003a; also see Fig.~\ref{mass_ssfr}).  This is further quantified
by the panels below.  The plot of L$_{\rm [OIII]}$/M$_*$ shows a strong
dependence on the age of the stellar population and only a weak dependence
on M$_*$ over the range from M$_* \sim 10^{10}$ to $10^{11.5}$
M$_{\odot}$.  The plot of L$_{\rm [OIII]}$/M$_{\rm BH}$ (a proxy for the
Eddington ratio) tells a similar story, but here the dependence is almost
entirely in stellar age with essentially no dependence on stellar mass for
a given value of D(4000).

A complementary set of plots is shown in Fig.~\ref{agnfracs2} which shows
the same set of distributions in the 2 dimensional space of M$_{\rm BH}$
{\it vs.}\ D(4000). These plots reinforce the conclusions above.  The
growth of black holes as traced by [OIII] emission is primarily occurring
in the lower mass black holes ($\sim 10^{6.5}$ to $10^{8}$ M$_{\odot}$;
Heckman et al.\ 2004), more specifically in the subset of these black
holes that live in galaxies undergoing significant central
star-formation. The plots of L$_{\rm [OIII]}$/M$_*$ and L$_{\rm
  [OIII]}$/M$_{\rm BH}$ both illustrate the fact that the strong
dependences of these ratios are primarily on the age of the central
stellar population.

Turning our attention now to the panels involving L$_{\rm mech}$, we see
that the jet-mode population of AGN is almost disjoint from the
radiative-mode AGN in the properties of the host galaxies and their black
holes (see also Fig.~\ref{mass_ssfr}).  The bulk of the jet mechanical
energy is produced by the most massive black holes (M$_{\rm BH} \sim 10^8$ to
$10^{9.5}$ M$_{\odot}$) living in the most massive galaxies (M$_* \sim
10^{11}$ to $10^{12}$ M$_{\odot}$), and generally having old stellar
populations (although there is a minority population of massive galaxies
with younger stellar populations). In direct contrast to the
radiative-mode AGN, the plots of L$_{\rm mech}$/M$_*$ and L$_{\rm
  mech}$/M$_{\rm BH}$ show strong positive correlations with both M$_*$
and M$_{\rm BH}$ (see Best et al., 2005b) but only weak dependences on the
age of the stellar population.

\begin{figure}[!t]
\begin{center} 
\psfig{file=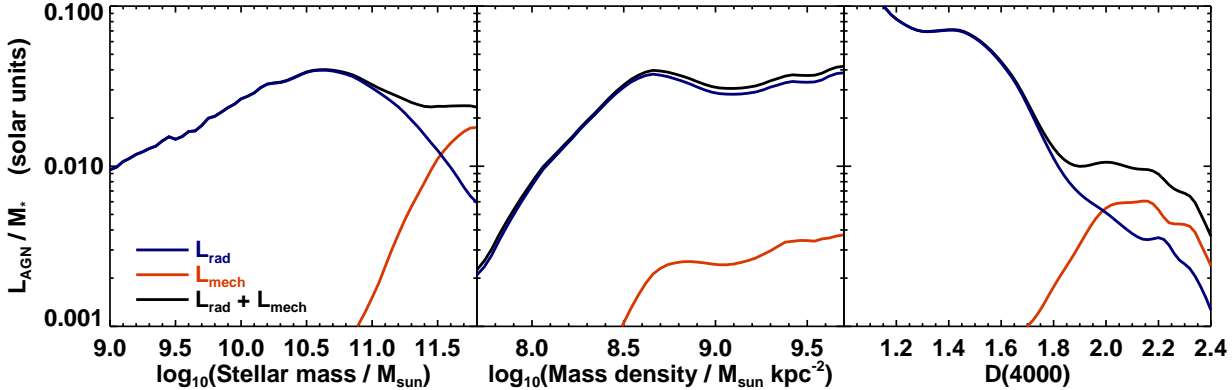,angle=90,width=\textwidth,clip=} 
\end{center} 
\caption{\label{lagnmass} The distribution of AGN luminosity per unit
  stellar mass amongst galaxies from the SDSS main galaxy sample, as a
  function of galaxy stellar mass (left), stellar surface mass density (middle)
  and 4000\AA\ break strength (right). In each plot the blue line
  indicates the radiative luminosity output (estimated from the [OIII]
  emission line), the red line shows the mechanical luminosity output in
  the radio jet, and the black line shows the sum of the two.}
\end{figure}

The main dependences revealed in these 2-D plots are summarized in a
simpler form in Fig.~\ref{lagnmass} in which the mass-normalized AGN
outputs are plotted as 1-D functions of M$_*$, $\mu_*$, and D(4000). Three
key conclusions are evident. 

First, there is only a slow rise of about 0.5 dex in L$_{\rm rad}$/M$_*$
on M$_*$ over the range from $10^9$ to $10^{10.7}$ M$_{\odot}$, but then a
steeper drop at higher masses. This can be contrasted with the very steep
increase in L$_{\rm mech}$/M$_*$ with M$_*$. This shows that the
predominant form of energy produced by black holes switches from radiation
to jet energy at the highest stellar masses ($> 10^{11.5}$
M$_{\odot}$). The cross-over corresponds to M$_{\rm BH} > 10^9$
M$_{\odot}$. It is intriguing that the sum of the mass-normalized L$_{\rm
  rad}$ and L$_{\rm mech}$ remains relatively constant over the full
stellar mass range plotted. Since the ratio of black hole to galaxy
stellar mass rises strongly with increasing mass (Fig.~\ref{mass_bhmass})
this near-constancy would not apply to AGN output normalized to black hole
mass (e.g.\ Fig.~\ref{heck04a}).

Second, both modes of black hole output plummet precipitously below the
characteristic value $\mu_* \sim 10^{8.5}$ M$_{\odot}$ kpc$^{-2}$ that
divides the bimodal galaxy population described in Section 1.3. The
effective growth of black holes evidently requires conditions of high
central density. 

Finally, the strong overall increase in L$_{\rm rad}$/M$_*$ moving from old
(D(4000) $>$ 2) to intermediate-age (D(4000) $\sim$ 1.5) stellar
populations, followed by saturation at still younger ages is very
clear. This was shown most explicitly by Kauffmann \& Heckman (2009), and
is consistent with the left-hand panel of Fig.~\ref{eddfracfuncs}.  In
contrast, the jet output is only significant compared to the radiative
output in the galaxies with the oldest stellar populations (see Best et
al.\ 2005b).
   
Taken together these figures clearly support the division of the AGN
population into two separate modes, as we have stressed throughout our
review. The broad conclusions on host galaxy properties of the different
AGN populations are summarised in Fig.~\ref{props_tab}.

\subsection{The Hosts of Radiative-Mode AGN}

\begin{figure}[!t]
\begin{center} 
\psfig{file=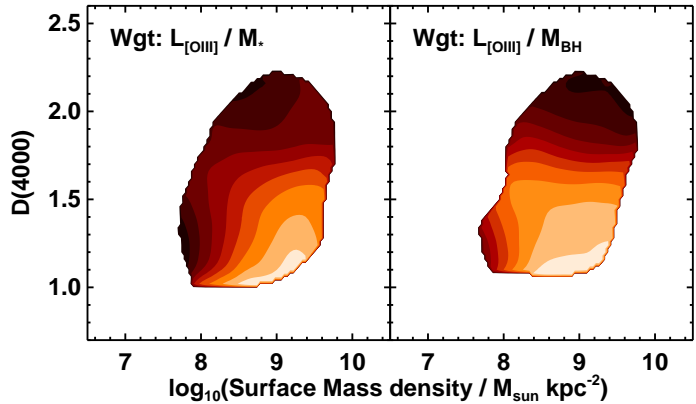,angle=90,width=10cm,clip=} 
\end{center} 
\caption{\label{mu_d4000} The distributions of AGN bolometric radiative
  luminosity per unit galaxy stellar mass (left) and bolometric radiative
  luminosity per black hole mass (right) in the stellar surface mass
  density {\it vs.}\ 4000\AA\ break strength plane. [Color-banding as in
    Fig.~\ref{agnfracs}]. Radiative-mode AGN activity clearly requires
  both an abundant supply of cold gas (indicated by recent or on-going
  star formation) and a high stellar surface mass density.}
\end{figure}

The above plots for the radiative-mode AGN lead to a simple and physically
satisfying conclusion: the accretion-driven growth of black holes in the
contemporary universe is primarily occurring in centers of galaxies where
two criteria are met: 1) there is an abundant supply of cold gas (whose
presence is implied by recent or on-going star-formation) and 2) the
stellar surface mass density is high. Putting these two criteria together
it is natural to conclude that there is really one over-arching
requirement, namely an abundant central supply of cold and dense gas. This
is shown in an explicit and compact way in Fig.~\ref{mu_d4000}. In this
section we will expand on this by considering different aspects of the
relationship between black hole growth and star formation in some detail.

\subsubsection{The Relationship to the Star-Formation Rate}

So far we have discussed the relationship between AGN and star-formation
only in a qualitative sense. Here we summarize what has been learned about
the quantitative relationship.  Let us consider the correlation between
star-formation rate and AGN luminosity. There are two complementary
approaches to this. One is to select a sample of AGN spanning a range of
luminosity and then measure the star-formation rates for each. This is
conceptually simple, but ignores the fact that many galaxies do not host
detectable AGN, and these galaxies are ignored in the above
approach. Thus, the second approach is to examine the distribution of AGN
luminosity as a function of star-formation rate for a complete sample of
galaxies selected without regard to the presence of an AGN.

As an example of the first approach, Kauffmann et~al.\ (2003a) used SDSS
to measure the distribution of the strength of the 4000\AA\ break for AGN
as a function of raw [OIII] luminosity. Based on subsequent calibration
(by Brinchmann et~al.\ 2004) of the 4000\AA\ break as a function of
specific star-formation rate (sSFR $=$ SFR/M$_*$), there is a clear but
shallow dependence: as the [OIII] luminosity increases from $10^6$ to
$10^9$ L$_{\odot}$, the median sSFR only increases from $10^{-11.3}$ to
10$^{-10}$ year$^{-1}$.  This is consistent with the SDSS-based results of
LaMassa et~al.\ (2013) who find that the mean specific star-formation rate
in Type 2 Seyferts (within a radius of about 2 kpc) only increases from
$\sim 10^{-11}$ to 10$^{-10}$ year$^{-1}$ as the ratio of $L/L_{\rm Edd}$
increases from $\sim 10^{-3}$ to $\sim$ 1. Similarly, they found that the
star-formation rate in this central region only increases from 0.1 to 1
M$_{\odot}$ year$^{-1}$ as the extinction-corrected [OIII] luminosity
increases from $10^{6.5}$ to $10^9$ L$_{\odot}$ (ie., SFR $\propto L_{\rm
  AGN}^{0.4}$).  Diamond-Stanic \& Rieke (2012) and Netzer (2009) find
steeper but still sub-linear relations between star-formation rate and AGN
luminosity (SFR $\propto L_{\rm AGN}^{0.6-0.8}$). Rosario et~al.\ (2012)
find no correlation (at z $\sim 0$) between the global (galaxy-wide) SFR
and AGN luminosity for AGN bolometric luminosities below about $10^{44}$
erg/s, and SFR $\propto$ L$_{\rm AGN}^{0.5}$ at higher luminosity.

The second approach was taken by Kauffmann \& Heckman (2009). They looked
at the distribution of the Eddington ratio ($L_{\rm bol}/L_{\rm Edd}$) as
a function of the amplitude of the 4000\AA\ break for a complete
SDSS-based sample of galaxies (see Fig.~\ref{eddfracfuncs}, left panel).
They found that for galaxies with predominantly old central stellar
populations, the probability of finding an AGN decreases with increasing
Eddington ratio as a power-law.  Using the calibration of Brinchmann
et~al.\ (2004), the results in Kauffmann \& Heckman (2009) imply that the
normalization of this power-law increases roughly linearly with the
central sSFR over the range in the latter from $\sim10^{-12}$ to
$10^{-10.5}$ year$^{-1}$.  However, at higher values of central sSFR, the
Eddington ratio distribution transitions to a log-normal that peaks at an
Eddington ratio of about 3\%. In this regime, the Eddington ratio
distribution is independent of the central sSFR. It is also largely
independent of the mass of the black hole (though the precise details
depend upon the version of the M-$\sigma$ relation adopted).  Thus,
Kauffmann \& Heckman found no simple monotonic relation between central
star-formation and black hole growth rates.  Once the central region of
the galaxy has a mass-doubling time (M$_*$/SFR) that is comparable to the
current Hubble time, the SMBH growth rate saturates and is not boosted by
additional star-formation.  We will discuss some possible interpretations
of this in Sections 5 and 6 below.

The combination of power-law plus narrow peaked distribution for the
Eddington ratio distribution found for emission-line AGN bears close
resemblance to the results of the radiative-mode {\it vs.}\ jet-mode
decomposition of Eddington ratio distributions for radio AGN (Section~3.2;
Fig.~\ref{eddfracfuncs} right panel). This fits with the picture discussed
above, that radiative-mode AGN are generally found in star-forming host
galaxies while jet-mode AGN are mostly in passive galaxies. It also
indicates that the main properties of the Eddington ratio distribution are
largely independent of whether or not an AGN is classified as
radio-loud. This further justifies our approach throughout this review of
considering just two fundamental AGN modes (cf.\ Fig.~\ref{props_tab}).

\subsubsection{The Radial Distribution of Star-Formation and Cold Gas}

The physical connection between the fueling of the SMBH and star-formation
should be clearer on smaller scales near the SMBH. Kauffmann
et~al.\ (2007) compared this relationship determined on two scales. They
found that the luminosity of the AGN (relative to the Eddington limit)
correlates much more tightly with the youth of the central stellar
population of the galaxy as probed with the SDSS spectra than with its
global stellar population as probed with SDSS and GALEX optical/UV
imaging. They concluded that on-going star-formation in the global galaxy
disk is a necessary but not sufficient condition for on-going star
formation in the center and the concurrent growth of the SMBH: young
bulges and their growing SMBH are only found in galaxies with star-forming
disks, but many galaxies with star-forming disks have dead bulges and
quiescent SMBHs.

A similar result was found by Diamond-Stanic \& Rieke (2012) based on an
analysis of Spitzer mid-IR imaging and spectroscopy of a sample of very
nearby Seyfert galaxies. They found that the correlation between the
star-formation rate and the rate of SMBH growth (measured using the
luminosity of the [OIV]25.9 $\mu$m line) was significantly stronger for
the star-formation rate measured in the inner-most kpc than for the entire
galaxy. The radial distribution of the star-formation as a function of the
AGN luminosity was mapped in a statistical sense by LaMassa et~al.\ (2013)
for a sample of 28,000 Type 2 Seyfert galaxies drawn from the SDSS.  They
examined the star-formation rate within the SDSS fiber for sub-samples
matched in both host galaxy properties and AGN luminosity but spanning a
range in mean redshift from $z \sim$ 0.05 to 0.15 (so that the fiber
mapped the star-formation over different radial scales in each
sub-sample). They found that as the AGN luminosity increased, the total
star-formation increased and became more and more centrally concentrated
(located within about 2 kpc of the nucleus).

On smaller radial scales, the stellar populations in nearby Type 2 Seyfert
nuclei have been investigated by Cid Fernandes et~al.\ (2001) using optical
spectra that probe typical radial scales of only a few hundred pc.  By
computing $L_{\rm bol}/L_{\rm Edd}$ for this sample, we find that a young
stellar population can be directly detected surrounding about 75\% of the
Seyfert nuclei with luminosities exceeding 10\% of the Eddington limit,
but only in about 25\% of those with lower luminosities.  This
investigation was extended to lower AGN luminosities and still smaller
scales using ground-based and HST spectra by Cid Fernandes et~al.\ (2004)
and Gonzalez-Delgado et~al.\ (2004) respectively. They found that the
nuclear stellar populations were usually old, except in cases whose
optical emission-line properties were consistent with an ionizing
contribution from both an AGN and young stars.

A complementary picture emerges from consideration of the radial
distribution of the cold gas associated with star-formation.  By comparing
the hosts of radio-quiet AGN with control samples of non-AGN matched in
stellar mass, structural properties and mean stellar age, Fabello
et~al.\ (2011) found that the global atomic gas (HI) content of a galaxy
had no effect on the accretion rate on the central black hole. This agrees
with earlier studies of the HI content of smaller samples of Seyfert
galaxies (Heckman et~al.\ 1978, Mirabel \& Wilson 1984, Ho et~al.\ 2008)
which did not show systematic differences in HI content in Seyferts
compared to normal galaxies.  Atomic gas in a galaxy is generally spread
out in a large-scale disk and the surface density of the HI exhibits a dip
or even a hole in the center of the galaxy.  If gas is present in the
vicinity of the black hole, it is likely to be at high enough densities
that it would exist primarily in molecular form.

Surveys of the global molecular content of AGN hosts using the mm-wave
J=1-0 CO emission-line show only small (Heckman et~al.\ 1989) or no
(Maiolino et~al.\ 1997, Saintonge et~al.\ 2012) differences in the
molecular content of AGN hosts compared to normal galaxies (in terms of CO
luminosity normalized by galaxy optical luminosity, HI mass, or
star-formation rate).  On smaller radial scales, thick, clumpy disks of
molecular gas with radii $\sim 30$ pc have been found at the centers of
nearby Seyfert galaxies (e.g.\ Hicks et~al.\ 2009). More recently, Hicks
et~al.\ (2013) examined molecular gas on 50 pc to 1 kpc scales in a
matched sample of Seyfert and quiescent galaxies and showed that the
Seyferts had more centrally concentrated H$_2$ surface brightness
profiles.
 
The presence of dust can serve as a proxy for cold gas. Kauffmann
et~al.\ (2007) showed that the central regions of the host galaxies of
more rapidly growing black holes were more heavily dust-reddened than were
the same regions in the hosts of low-power AGN.  Simoes Lopes
et~al.\ (2007) used HST imaging to show that all of the 65 AGN host
galaxies they studied have circum-nuclear dust structures on radial scales
of $\sim$ 0.1 to 1 kpc (see also Gonzalez Delgado et~al.\ 2008).  This
includes 34 early-type AGN hosts. In a carefully constructed control
sample of normal galaxies only a minority (26\%) of the early-type
galaxies had circum-nuclear dust (though all the late-type control
galaxies did). Subsequent Spitzer observations (Martini et~al.\ 2013)
imply dust masses of-order $10^6$ M$_{\odot}$, or associated gas masses
of-order $10^8$ M$_{\odot}$. In a complementary study, Lauer et~al. (2005)
used HST images to discover a strong connection between the AGN
emission-line activity and the presence, quantity and structure of central
dust in a sample of 77 early-type galaxies. All-in-all, it seems
reasonable to conclude that radiative-mode AGN reside in galaxies where
the central (kpc-scale) supply of cold dense gas is substantial.
 
\subsubsection{The Starburst-AGN Connection}

We have seen above that the connection between SMBH growth and
star-formation is clearer when the central kpc-scale region of the galaxy
is probed.  Such central star-formation can sometimes occur in a strong
episodic burst, presumably related to the short dynamical times near the
nucleus. The existence of a direct connection between the growth of a SMBH
and intense star-formation in the centers of galaxies has long been a
matter of speculation. In its most extreme form a direct starburst-AGN
connection was posited: AGN were simply the evolved descendants of
starbursts powered by a collection of supernovae and supernova remnants
rather than by a SMBH (Terlevich \& Melnick 1985, Terlevich et~al.\ 1992).
Many other authors have proposed less iconoclastic models in which the
delivery of a fresh supply of cold gas to the center of a galaxy leads to
both a strong starburst and the efficient fueling of the SMBH
(e.g.\ Sanders et~al.\ 1988a,b, Hopkins et~al.\ 2006, 2008, Di Matteo
et~al.\ 2008).

Investigations of the starburst-AGN connection using SDSS spectra have
shown that there is indeed a connection, but one that is more subtle than
might have been expected.  We have described above how SDSS spectra can be
used to identify starbursts and post-starbursts using the combination of
the 4000\AA\ break and the strength of the high-order Balmer
absorption-lines.  Based on this methodology, Kauffmann et~al.\ (2003a)
showed that high-luminosity Type 2 AGN in SDSS were more likely to have
undergone a starburst within the last $10^9$ years compared to
low-luminosity AGN or normal galaxies of similar stellar mass (see also
Goto 2006 and Yan et~al.\ 2006). This issue was examined in more detail by
Wild et~al.\ (2007). Based on a sample 34,000 early-type galaxies in SDSS,
they found that the post-starburst systems had the highest mean AGN
luminosity. However, because these are relatively rare, they represent
only about 15\% of the total amount of SMBH growth in the contemporary universe.
They found that galaxies with on-going (but less episodic) central
star-formation are the sites of the majority of SMBH growth.

\begin{figure}[!t]
\begin{center} 
\psfig{file=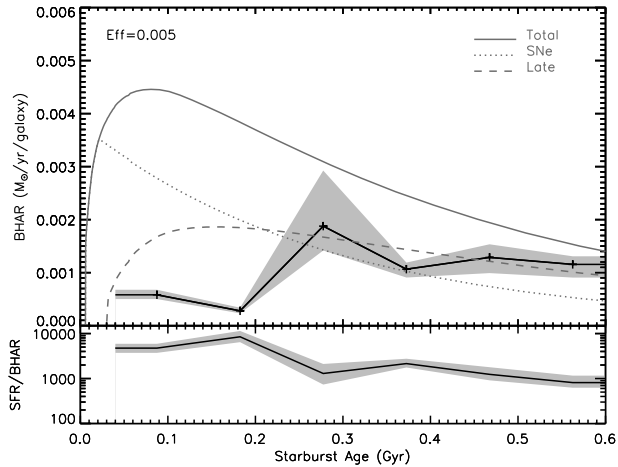,width=8.5cm,clip=} 
\end{center} 
\caption{\label{wildburst} Top Panel: the mean black hole accretion rate
  (BHAR) averaged over all galaxies in a sample of starburst and
  post-starburst galaxies (black line). The grey-shaded area shows the
  typical 10th and 90th percentile range on the total BHAR estimated from
  bootstrap resampling of the data.  The model prediction for the average
  gas mass-loss rate from the stars formed during the starburst (solid
  grey line) assumes an accretion efficiency onto the black hole of 0.5
  per cent of the ejecta from intermediate and low-mass stars between the
  age of 250 and 600 Myr. Stars return mass to the ISM through supernova
  explosions and fast winds from massive stars (dotted line) and planetary
  nebula ejections and stellar winds from lower mass stars (dashed
  line). Bottom panel: the ratio of SFR (from H$\alpha$) to BHAR, with
  mean value and errors calculated as for the top panel. The BHAR lags the
  SFR and becomes significant only after several hundred Myr. Figure from
  Wild et~al.\ (2010).}
\end{figure} 

In a subsequent paper, Wild et~al.\ (2010) examined the time history of
SMBH growth during and after an intense burst of central star formation
(involving roughly 10\% of the galaxy stellar mass). They found that the
growth of the SMBH was delayed by about 200 Myr relative to the start of
the starburst (Fig.~\ref{wildburst}). These results pertain to the region
within a radius of $\sim$ 1 to 2 kpc.  Qualitatively similar results from
SDSS were found by Schawinski et~al.\ (2007) for galaxies selected
morphologically to be ellipticals.  On much smaller scales (ten to hundred
pc) and for a much smaller sample, Davies et~al.\ (2007) also find
evidence for a delay of-order 100 Myr between a starburst and powerful AGN
activity. We will return to a discussion of the implications of these
results below.

\subsection{The Hosts of Jet-Mode AGN}
\label{jethosts}

The results in Section 4.1 above show that the host galaxies of the black
holes that dominate the production of jet mechanical energy are nearly
disjoint from the host galaxies where black holes are actively
growing. These jets are traced by their radio emission. Detailed
investigations of the structure and stellar content of the radio-selected
AGN population in SDSS were first undertaken by Best et~al.\ (2005b). They
confirmed the widely-held view that nearly all such AGN are hosted by
massive elliptical galaxies with old stellar populations: the radio source
hosts are highly bulge-dominated (their black hole mass {\it vs.}
stellar-mass relation lies on the established black hole mass {\it vs.}
bulge mass relation), with 4000\AA\ break strengths clustered around 2.0,
concentration indices typically between 2.9 and 3.4, and effective stellar
surface mass densities around $10^9$\,M$_{\odot}$\,kpc$^{-2}$. This is
entirely consistent with the results summarized in Section 4.1.

More interestingly, Best et~al.\ (2005b) were able to quantify the
prevalence of radio-AGN activity as a function of stellar mass and black
hole mass with much greater statistical robustness than previous
studies. They found that the fraction of galaxies selected via radio
emission scaled as M$_*^{2.5}$ and as M$_{\rm BH}^{1.6}$, with these
results being broadly independent of the radio luminosity limit. This is
strikingly different from the mass-dependence of the prevalence of
emission-line AGN activity (see Fig.~\ref{agnfracs}).  Best
et~al.\ (2005b) did not separate their sample into radiative-mode and
jet-mode sources, but their sample was dominated by the latter, and it is
these that drive the strong mass dependence. Janssen et~al.\ (2012)
considered the mass dependences of the two populations separately. They
showed that the prevalence of radiative-mode radio sources has a much
shallower dependence on mass, scaling as M$_*^{1.5}$, albeit that this
remains steeper than that found for radio-quiet radiative-mode AGN
selected by their optical emission-lines.

For jet-mode sources, Best et~al.\ found no evidence for any strong
relationship between the [OIII] and radio luminosities; indeed the radio
emitting fraction was found to be largely independent of whether or not
the optical galaxy is classified as an emission-line AGN.  This result
breaks down at higher radio luminosities, when the radiative-mode radio
sources begin to dominate. These have been long known to show a strong
correlation between radio and emission line luminosities (e.g.\ Rawlings
\& Saunders 1991, Xu et~al.\ 1999).  Kauffmann et~al.\ (2008) investigated
the subset of radio sources within the SDSS that display emission lines,
and confirmed this correlation, finding also that it became tighter if
Eddington-scaled quantities (ie., $L_{\rm 1.4GHz}/M_{\rm BH}$ and $L_{\rm
  [OIII]}/M_{\rm BH}$) were considered.

Lin et~al.\ (2010) compared the host galaxies of radio sources separated
into different radio morphological types.  As previously discussed, these
two radio morphological classes show a broad overlap with the jet-mode and
radiative-mode classes, with most FR1s being jet-mode and most FR2s being
radiative-mode; however, there are substantial differences between the
segregation methods, and a significant population of jet-mode
(low-excitation) FR2 sources exists. Lin et~al.\ found that the most
significant differences between FR1s and FR2s occurred for the subset of
the FR2 population which displayed both strong high-excitation optical
emission lines and hotspots close to the end of the radio lobes. They
found this FR2 subset to be hosted by lower mass galaxies, with bluer
colors, living in poorer environments and with higher accretion rates than
the rest of the radio source population. Similarly Kauffmann
et~al.\ (2008) found that their radio-AGN with emission lines showed lower
stellar masses, lower velocity dispersions, lower 4000\AA\ break
strengths, and stronger Balmer absorption features than those radio-AGN
without emission lines.

Best \& Heckman (2012) did an explicit separation of jet-mode and
radiative-mode radio sources and confirmed these results. They also
constructed matched samples of the two radio source classes, to
investigate whether observed differences in host galaxy colors and
morphologies were just secondary effects driven by, for example, the
different stellar masses of the two samples. They found that the hosts of
radiative-mode radio-AGN have bluer colors, smaller half-light radii,
lower concentration indices, and lower 4000\AA\ break strengths than the
hosts of jet-mode radio-AGN of the same stellar mass, black hole mass and
radio luminosity. These results imply that the radiative-mode radio
sources are associated with on-going star-formation activity, as is also
seen in the radio-quiet AGN samples. Further evidence in support of this
comes from the work of Sadler et~al.\ (2013), who investigated differences
in the mid-IR colors of the two radio source classes using WISE and
concluded that star-formation activity was present in the radiative-mode
AGN, but not the jet-mode AGN (see also G\"urkan et~al.\ 2013).

The larger scale lengths and higher concentration indices of the hosts of
the jet-mode AGN can be explained if these sources are preferentially
located toward the centers of groups and clusters (as suggested by their
clustering and environmental properties; see Section 4.4), where host
galaxies are typically larger and may have more extended (cD-type) light
profiles. Mannering et~al.\ (2011) compared the scale lengths of radio-AGN
at both red and blue wavelengths and found that the ratio of the two was
higher than for control galaxy samples, suggesting the possible presence
of extended red envelopes (although their control samples were matched
only in redshift and magnitude, so this result might be driven by
morphology differences between the radio-AGN and control samples).

Janssen et~al.\ (2012) considered the relationship between AGN
classification, radio luminosity, and galaxy color. Red (passive) galaxies
show a higher probability of hosting jet-mode radio-AGN than blue
(star-forming) galaxies of the same stellar mass, by a factor of a few.
Blue galaxies show a higher probability of hosting radiative-mode
radio-AGN at all radio luminosities. For blue galaxies, the likelihood of
hosting a radio-AGN of either class is a strong positive function of the
star-formation rate. This suggests that the presence of cold,
star-forming gas in a galaxy enhances the probability of its central black
hole becoming a radio-loud AGN. This means that jet-mode AGN activity,
especially at high radio luminosities, is not solely related to hot halo
gas accretion (see Section 5 below).

\subsection{Clustering, Environments, and Halo Masses of AGN}

Since early in the study of Seyfert galaxies, QSOs, and powerful radio
sources, attempts have been made to categorize the typical environments in
which these sources are found (e.g.\ Gisler 1978, Longair \& Seldner 1979,
Yee \& Green 1984).  Analyses mostly relied on relatively small fields
around the AGN, and either considered simple galaxy counts out to some
projected radius, or divided those counts into annuli to enable a
cross-correlation analysis. The majority of analyses indicated that AGN
are found in a wide range of environments, with powerful QSOs and radio
sources being often located in group or poor cluster environments, but
Seyfert galaxies generally avoiding the densest environments. However, AGN
samples were relatively small and contradictory results were obtained in
different studies. Furthermore, AGN activity depends strongly on host
galaxy properties such as stellar mass (especially for radio-selected AGN;
see previous sections); therefore, in order to reliably isolate any trends
of AGN activity that are explicitly caused by environment, large and
homogeneous datasets like that offered by SDSS are required so that the
AGN samples can be compared against control samples carefully matched in
other galaxy properties. Even so, there has remained much debate as to the
nature of these trends, largely because of the different definitions
adopted both for AGN samples (all emission line AGN; just Seyfert sources;
radio-AGN; etc) and for environment (local galaxy density; membership of,
or distance from center of, a group or cluster).

Once stellar mass effects are properly accounted for, the incidence of
optical emission-line AGN is generally found to be lower in denser
environments and in groups and clusters, than in the field (Miller
et~al.\ 2003, Kauffmann et~al.\ 2004, von der Linden et~al.\ 2007): the
fraction of AGN among galaxies of given stellar mass decreases by a factor
of two from the field towards the cluster/group center. This is matched by
a decrease in star-formation activity of these galaxies. Hwang
et~al.\ (2012) investigate these trends for early and late-type galaxies
separately, and find that the prevalence of AGN activity in early-type
galaxies already begins to decrease at the virial radius of the cluster,
whereas in late-type galaxies it only decreases as the cluster center is
approached. Chandra has revealed significant samples of X-ray-detected AGN
in nearby galaxy clusters (e.g.\ Martini et~al.\ 2006), the majority of
which do not show emission lines and so are not identified as AGN in
optical observations. These may be obscured AGN, or more likely may be
jet-mode AGN. Haines et~al.\ (2012) argues that luminous X-ray AGN in
massive clusters are mostly an infalling population. These results are all
generally consistent with a picture in which there is a gradual decline in
emission-line AGN activity as the denser environment decreases the cold
gas supply for an AGN.

In contrast to emission-line AGN, radio-selected AGN are found to be
preferentially located in group and cluster environments (e.g.\ Best
et~al.\ 2005b, Sabater et~al.\ 2013), confirming the results of many
previous studies of small radio-loud AGN samples. This is particularly
true of the jet-mode subset of the radio-AGN population (Best
et~al.\ 2004, Reviglio \& Hefland 2006, Sabater et~al.\ 2013), again
suggesting a different mechanism for the triggering of AGN activity in
these sources than for radiative-mode AGN. The prevalence of radio-AGN
activity amongst brightest cluster galaxies (BCGs) is especially high (Best
et~al.\ 2007, Croft et~al.\ 2007; cf.\ Burns 1990), while that of optical
AGN is lower than in non-BCG galaxies of the same mass (von der Linden
et~al.\ 2007).

An alternative view of the environments of AGN can arise from looking at
the two-point correlation function. This describes the excess probability
of finding a galaxy at a given distance R from another galaxy, due to
clustering. In the widely adopted halo-model description of large-scale
structure, on small scales ($<$ 1Mpc) the signal in the correlation
function is dominated by pairs of galaxies within the same dark matter
halo (known as the one-halo term; e.g.\ Seljak 2000). This therefore
measures the prevalence of a particular galaxy type amongst satellite
galaxies in halos. Li et~al.\ (2006) computed the cross-correlation
function between Seyfert galaxies and the SDSS spectroscopic main galaxy
sample, and compared this against the results for a non-AGN control sample
that was matched in redshift, stellar mass, black hole mass (from velocity
dispersion), concentration index and 4000\AA\ break strength.  They found
that on scales between 100 kpc and 1 Mpc the AGN showed weaker clustering
than the control sample. By comparison with simulations, they argued that
this could be understood if AGN activity was dis-favored within satellite
galaxies. This interpretation is consistent with the results arising from
the study of AGN activity as a function of cluster membership or local
galaxy density, discussed above.

On much larger scales (10-20 Mpc) the two-point correlation function
measures the correlation between different dark matter halos (the two-halo
term), which depends upon the properties of the parent halo for a given
galaxy type (see review by Cooray \& Sheth 2002). If the luminosity
function and two-point correlation function of a set of objects is known,
then (for a given cosmological model) it is possible to estimate the mean
mass of the parent dark matter halo (e.g.\ Sheth, Mo \& Tormen
2001). Mandelbaum et~al.\ (2009) used this method to estimate dark matter
halo masses for ($z<0.3$) Seyfert galaxies from the SDSS, and confirmed
the reliability of their measurements by using an alternative estimate
from weak gravitational lensing of background galaxies. They found a mean
halo mass of $\sim 10^{12}$ M$_{\odot}$.  They compared these results
against a control sample of non-AGN, and found that the halo mass versus
stellar mass relation was independent of whether or not Seyfert activity
was present.

Mandelbaum et~al.\ (2009) also analyzed radio-loud AGN in the same
redshift range. They found an average halo mass for the radio-loud AGN of
$10^{13}$ solar masses, nearly an order of magnitude higher than that of
optically-selected AGN. Much of this increase is because radio-loud AGN
are hosted by more massive galaxies, but Mandelbaum et~al.\ also found
that the dark matter haloes of radio-loud AGN are about twice as massive
as those of control galaxies of the same stellar mass. A similar
enhancement in the clustering strength of radio-loud AGN relative to a
carefully-matched control sample of massive galaxies was found by Donoso
et~al.\ (2010) at redshifts $0.4<z<0.8$ using the SDSS Luminous Red Galaxy
sample. Donoso et~al.\ also showed that the clustering strength of the
radio galaxy population is luminosity dependent. At low radio luminosities
where the jet-mode AGN dominate the radio-AGN samples (these also
dominated the Mandelbaum et~al.\ sample), the clustering strength is
enhanced relative to radio-loud QSOs.  At higher radio luminosities the
radio galaxy population is dominated by radiative-mode sources and the
clustering amplitude is more in line with that of the radio-loud QSOs, as
expected from the standard AGN Unified Model.

The 2dF and SDSS QSO redshift surveys have been used to study the dark
matter haloes of powerful AGN to even higher redshifts (Porciani
et~al.\ 2004, Croom et~al.\ 2005, Ross et~al.\ 2009). Intriguingly, the
mean mass of haloes hosting QSOs remains roughly constant (at around $2
\times 10^{12}$ M$_{\odot}$) over the entire redshift range $z < 2.5$, and
shows only weak dependence on QSO properties like luminosity, color or
black hole mass (Croom et~al.\ 2005, Shen et~al.\ 2009). Only at the
highest QSO luminosities is the clustering amplitude seen to
increase. Shen et~al.\ (2009) find that radio-loud QSOs reside in halos
almost an order of magnitude larger than those of radio-quiet QSOs (though
note that Donoso et~al.\ 2010 do not find this strong signal amongst their
$z \sim$ 0.6 sample). Shen's result is in line with the local results of
Mandelbaum et~al.: a large part of this is likely to be driven by the
higher typical stellar mass of radio-loud QSO hosts, but radio-AGN
activity also seems to be preferentially found in denser environments.

In summary, radiative-mode AGN are located in dark matter haloes with a
mean mass of $10^{12} M_{\odot}$, which is typical for their stellar
masses. Once stellar mass biases are account for, the prevalence of
radiative-mode AGN is lower in denser environments than in the field,
because they are dis-favored in satellite galaxies; this mirrors the
decrease in star formation activity in denser environments, with both
trends being driven the reduced cold gas supply. Radiative-mode AGN that
are found in denser environments are more likely to be radio-loud. This
partly reflects the trend for radio-loud radiative-mode AGN to be hosted
by more massive galaxies (see Fig.~\ref{props_tab}), but may also be
enhanced by a boosting of the radio luminosity in dense environments, where
confinement of the radio lobes reduces adiabatic losses (e.g.\ Barthel \&
Arnaud 1996).  In contrast to radiative-mode AGN, jet-mode AGN have an
increased prevalance in denser environments, and especially in the central
galaxies of groups and clusters. This indicates a different fueling
mechanism.

\section{BLACK HOLE FUELING}

\subsection{The nature of the accretion flow}

There are three main features that we need to consider when examining the
fueling of AGN: the origin of the fueling gas, the mechanism by which it
is transported to the vicinity of the black hole (e.g.\ Hopkins \& Quataert
2011), and the nature of the accretion flow onto the black hole. With
regards to the last of these, the most important determining properties
are the black hole mass, its spin, and the nature and rate of the
inflow. Note that even if the AGN is powered by tapping the black hole
spin via the Blandford \& Znajek (1977) mechanism, accretion is required
for this to occur.

Before embarking upon a discussion of the effects of different modes of
accretion, it is worth summarizing the lessons that have been learned from
studying accretion onto stellar-mass black holes in galactic X-ray binary
systems. These are systems consisting of a black hole of between roughly 4
and 20 solar masses in a close binary system with a normal star. At least
in terms of the black hole properties, these represent analogs of
accreting supermassive black holes, scaled down in both luminosity and the
accretion timescales. They are not precise equivalents of AGN because the
temperature of the accretion disk around a stellar mass black hole is
typically two orders of magnitude higher than that around an AGN which can
lead to differences in processes at atomic levels, such as
ionization-dependent accretion instabilities. Nevertheless, stellar mass
black holes offer the considerable advantage that variations in the
accretion flow can be directly tracked, with changes in the accretion
happening on timescales of less than days. The observed accretion rates
vary between the Eddington rate and values several orders of magnitude
below this.

X-ray binaries have been shown to display two fundamentally different
X-ray spectral states (see the review by Remillard \& McClintock 2006 in
this journal). In the low-hard state, the source spectrum is characterised
by low-luminosity, hard (predominantly $>$ 10 keV) X-ray emission, and
low-power radio jets are ubiquitous. The power of the radio jets is
correlated (albeit non-linearly) with the X-ray luminosity (Gallo
et~al.\ 2003). In the high-soft state, sources are dominated by a
high-luminosity thermal X-ray component, peaking at lower energies and
with the characteristics of a standard geometrically-thin, optically-thick
accretion disk. The primary difference between these two accretion modes
is observed to be the accretion rate onto the black hole relative to the
Eddington rate: the low-hard state is associated with accretion rates
below a few percent of the Eddington rate, whereas the high-soft state
arises when the accretion rate exceeds that value (Maccarone 2003). X-ray
binaries are observed to traverse from one state to another. Within the
low-hard state, there are indications that the transition radius between
an inner advection-dominated flow and an outer thin disk
(cf.\ Fig.~\ref{agnschematic}) decreases with increasing luminosity
(Narayan 2005).  As the accretion rate increases, the luminosity increases
in all bands, until they reach a transition or intermediate phase where
they undergo a short-lived but strong radio outburst in which both a
thermal accretion disk and powerful and highly relativistic radio jets are
seen (e.g.\ Fender et~al.\ 2004). The radio emission then ceases as the
source moves into the high-soft state. A further important property of
X-ray binaries is that winds and jets appear to anti-correlate (Miller
et~al.\ 2008, Neilsen \& Lee 2009, Ponti et~al.\ 2012): winds occur in the
high-soft state when jets are not observed, but are not observed in the
low-hard state where jets are ubiquitous. 

These states of X-ray binary activity each have an analog amongst the AGN
population. The low-hard state X-ray binaries are analogs of the jet-mode
radiatively-inefficient AGN. Indeed these AGN and low/hard X-ray binaries
fall on the same fundamental plane of black hole accretion: a relationship
between the radio luminosity, X-ray luminosity and black-hole mass of
black holes (Merloni et~al.\ 2003, Falcke et~al.\ 2004). The high-soft
state X-ray binaries are a direct analog of the radiative-mode AGN
(Seyferts and QSOs). The transition-phase objects have properties that
make them a broad analog of the radio-loud radiative-mode AGN, although in
AGN there is no observational evidence for these sources being a
transition phase that pre-empts radio-quiet radiative-mode activity.

The critical accretion rate for the switch between X-ray binary states (a
few percent of the Eddington rate) also appears to be broadly comparable
in AGN. As shown in Section 4.2 (left panel of Fig.~\ref{eddfracfuncs})
the emission-line AGN (predominantly Seyferts) in gas-rich (star-forming)
galaxies in SDSS show a log-normal distribution of Eddington rates peaking
at a few percent of Eddington.  This result is also confirmed at higher
redshifts with QSOs seen to have lower limits to their accretion rates at
around 1\% of Eddington (e.g.\ Kollmeier et~al.\ 2006, Trump
et~al.\ 2009).  Radiative-mode (Seyfert/QSO-like) radio-AGN show a similar
Eddington ratio distribution (Best \& Heckman 2012; see Section 3.2 and
right panel of Fig.~\ref{eddfracfuncs}). In contrast to this, the jet-mode
radio-AGN have accretion rates distributed below a few percent of
Eddington (Best \& Heckman 2012; see Section 3.2). If powered by the
Blandford \& Znajek mechanism then these accretion rates would be lower
still.  This last result is also supported at higher redshift, since BL
Lac objects (believed to be beamed versions of the jet-mode sources) are
also found to have luminosities below about 1\% of Eddington, while
Russell et~al.\ (2013a) find a transition from jet-mode to radiative-mode
AGN at a similar Eddington ratio amongst brightest cluster galaxies (see
also Churazov et~al.\ 2005, Merloni \& Heinz 2007).

As in the case of X-ray binaries, the spectra of the AGN subclass of
LINERs (jet-mode AGN) are best explained if the inner ADAF is supplemented
by a truncated thin disk at larger radii, with the transition between the
two being dependent upon the Eddington-scaled accretion rate (Yuan \&
Narayan 2004, Ho 2005). Sadler et~al.\ (2013) found a weak signature of
mid-IR AGN emission at the highest radio luminosities in jet-mode
radio-AGN, supporting this idea.

All of these results are in line with theoretical predictions of
advection-dominated flows, which predict a change in the nature of the
inner accretion flow below a critical fraction of the Eddington rate
(e.g.\ see the review by Narayan 2002, or by Yuan \& Narayan in this
volume), with the precise value of that 
fraction depending strongly upon the viscosity of the accretion disk
(e.g.\ Mahadevan 1997, Qiao \& Liu 2009). This is consistent with the
picture we have advocated in this review: the high Eddington-rate
(radiative-mode) AGN, and the low-Eddington rate (jet-mode) are two
fundamentally distinct populations. We include the small population of
radio-loud radiative-mode AGN together with the (radio-quiet) rest of the
radiative-mode population: the results in the previous sections have shown
that their host galaxy properties and Eddington-scaled accretion rates are
broadly similar, albeit that the radio-loud QSOs are predominantly located
at the upper end of the range of stellar and black hole masses seen in the
radio-quiet population (see Fig.~\ref{props_tab}).

\subsection{The Fueling of Radiative-Mode AGN}

\subsubsection{Mergers and Interactions}

The idea that AGN are largely triggered through the process of galaxy
mergers or strong tidal interactions has long been a key component of most
theoretical models (e.g.\ Kauffmann \& Haehnelt 2000, Hopkins
et~al.\ 2006, 2008). Nevertheless, it has proven difficult to establish
this idea observationally, with many claims and counter-claims in the
literature. As we will explain below, this is a clear case in which
arriving at the correct answer requires a superb control of the
systematics, and this is made possible by a very large and homogeneous
data set like that provided by the SDSS.  We consider two types of
observational tests for a link between interactions/merger and AGN
fueling. The first is to measure the incidence of close companion galaxies
and the second is to determine the morphological structure of the host
galaxy.

The amplitude of the cross-correlation function on scales less than $\sim
100$\,kpc probes interacting pairs of galaxies where tidal forces begin to
influence the internal structure of the pair. In two companion papers, Li
et~al.\ (2008a,b) used the amplitude of the cross-correlation function on
scales of tens of kpc as a probe of the effects of galaxy interactions on
star formation and AGN activity in galaxies.  They first showed that on
scales less than 100 kpc, the amplitude of the cross-correlation function
exhibits a strong dependence on the sSFR of the galaxy, reflecting the
well-known result that galaxy interactions and mergers trigger star-bursts
in galaxies (Larson \& Tinsley 1978). As the separation between the two
galaxies in the pair decreased, their sSFR also increased relative to the
mean value for field galaxies of the same mass. In a follow-up paper, Li
et~al.\ applied the same techniques to active galactic nuclei (AGN) in the
survey, showing that interactions do not lead to enhancement of nuclear
activity over and above that expected as a result of the boost in the
central star formation rates.  In other words, the relation between
central star formation and AGN activity in interacting galaxies is no
different from that observed for non-interacting galaxies.

Mergers can be identified quantitatively through measures of the global
asymmetry (lopsidedness) of the light distribution of the galaxy
(e.g.\ Conselice et~al.\ 2000).  An analysis of the SDSS population of
normal and AGN-host galaxies by Wild et~al.\ (2007) showed that the
majority of black hole growth in the contemporary universe occurs in galaxies
with relatively quiescent recent star formation histories that have not
experienced recent mergers or strong interactions. In subsequent work,
Reichard et~al.\ (2009) followed up on the Li et~al.\ analysis by showing
that AGN activity is not enhanced in lopsided galaxies, compared to
symmetric galaxies of the same central star formation rate.  They found
that even among AGN radiating near the Eddington limit, only a small
minority exhibited the highly lopsided morphologies of a major merger.

\begin{figure}[!t]
\begin{center} 
\psfig{file=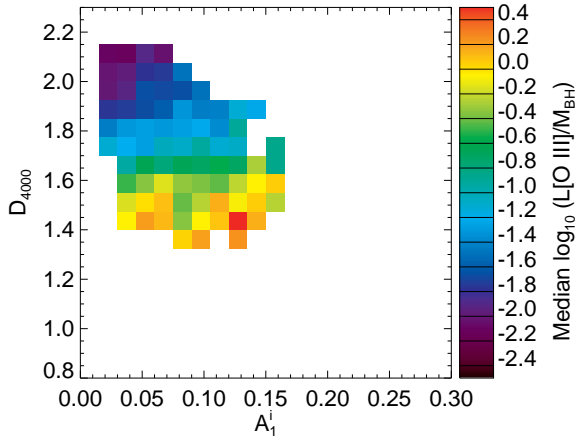,width=9cm,clip=} 
\end{center} 
\caption{\label{reichard} The relationship between $L_{\rm [OIII]}/M_{\rm
    BH}$, D4000, and galaxy lopsidedness ($A_1^i$). The color coding of
  these two-dimensional histograms indicates the median $L_{\rm
    [OIII]}/M_{\rm BH}$. The primary correlation is between stellar age
  and $L_{\rm [OIII]}/M_{\rm BH}$, as is apparent from the horizontal
  nature of the color contours. There is no independent correlation
  between $L_{\rm [OIII]}/M_{\rm BH}$ and lopsidedness in these
  data. Figure from Reichard et~al.\ (2009).}
\end{figure}

There are many papers reaching seemingly contrary conclusions, of which
some recent examples are Koss et~al.\ (2011, 2012), Alonso et~al.\ (2007),
Ellison et~al.\ (2011), Hwang et~al.\ (2012), Liu et~al.\ (2012), and
Sabater et~al.\ (2013).  We stress that these results are actually not at
odds with the results above.  As Reichard et~al.\ (2009) showed
explicitly, the hosts of strong AGN are in fact more lopsided on-average
than normal galaxies with the same stellar mass, radius, and
concentration. However they showed that this link is a secondary
unphysical one induced by much stronger correlations between lopsidedness
and central star formation and between central star formation and the
presence of a strong AGN (Fig.~\ref{reichard}).  They interpreted this as
implying that the fueling of the SMBH is enhanced by the presence of
circum-nuclear star-formation (and its associated cold gas), but this SMBH
fueling process does not depend upon the way in which this central
star-formation arises.

In summary, contrary to the assumption of many theoretical models, most of
the AGN activity in the contemporary universe appears to occur without the need
for major mergers or strong tidal interactions. All that is required is an
abundant supply of cold, dense gas in the central regions of the galaxy
(regardless of its origin).

\subsubsection{Fueling via Secular Evolution}

The radial transport of gas inward in a disk galaxy requires torque on the
gas that allows angular momentum to be transferred from the gas to the
stars or dark matter.  While this can be accomplished in a sudden and
spectacular way in a major merger or strong tidal interaction, a slower
but significant inflow can be driven by non-axisymmetric perturbations in
the underlying mass distribution that arise through internal dynamical
processes in the disk. Examples of such non-axisymmetric perturbations are
bars, oval distortions, and even spiral arms (e.g.\ Athanassoula 2008,
Kormendy \& Kennicutt 2004, Sellwood 2014).

Kormendy \& Kennicutt (2004) argued that the consequence of this secular
evolution for the inner regions of disk galaxies is the creation of a
pseudo-bulge (perhaps more descriptively called a disky bulge by
Athanassoula 2005, 2008).  These are defined observationally as central
regions of enhanced surface brightness relative to an inward extrapolation
of the radial surface brightness profile of the galaxy disk.  Unlike
classical bulges, they are dynamically-cold structures as revealed by the
presence of spiral arms, rings and bars. Unlike typical classical bulges,
they are the sites of significant on-going star-formation (e.g.\ Gadotti
2009).  They have exponentially-declining radial surface-brightness
profiles whose functional form is more similar to a classic exponential
Freeman disk than the $r^{1/4}$ de Vaucouleurs law for elliptical
galaxies.  The boundary between classical and pseudo-bulges is typically
set at a value for the Sersic index of $n \sim 2$ for the bulge component
(e.g.\ Gadotti 2009, Kormendy \& Kennicutt 2004, Athanassoula 2008, Fisher
\& Drory 2011).

If the growth of SMBH in the contemporary universe is largely fueled by
secular processes that transport gas inward, we would therefore expect
rapidly growing SMBHs to be found primarily in pseudo-bulges.  We have
summarized above the general link between the presence of a young central
stellar population (one of the defining characteristics of a pseudo-bulge)
and the rapid growth of the SMBH.  Moreover, Figure 17 in Kormendy \& Ho
(2013) shows that the transition from classical- to pseudo-bulges occurs
below a bulge velocity dispersion $\sigma \sim 150$ km s$^{-1}$, and below
a SMBH mass $M_{\rm BH} \sim$ few $\times 10^7$ M$_{\odot}$ (cf
Fig.~\ref{mass_bhmass}; and see also Fisher \& Drory 2011).  This is just the
regime where SDSS shows the bulk of SMBH growth to be occurring today
(Heckman et~al.\ 2004, Kauffmann \& Heckman 2009).

\begin{figure}[!t]
\begin{center} 
\psfig{file=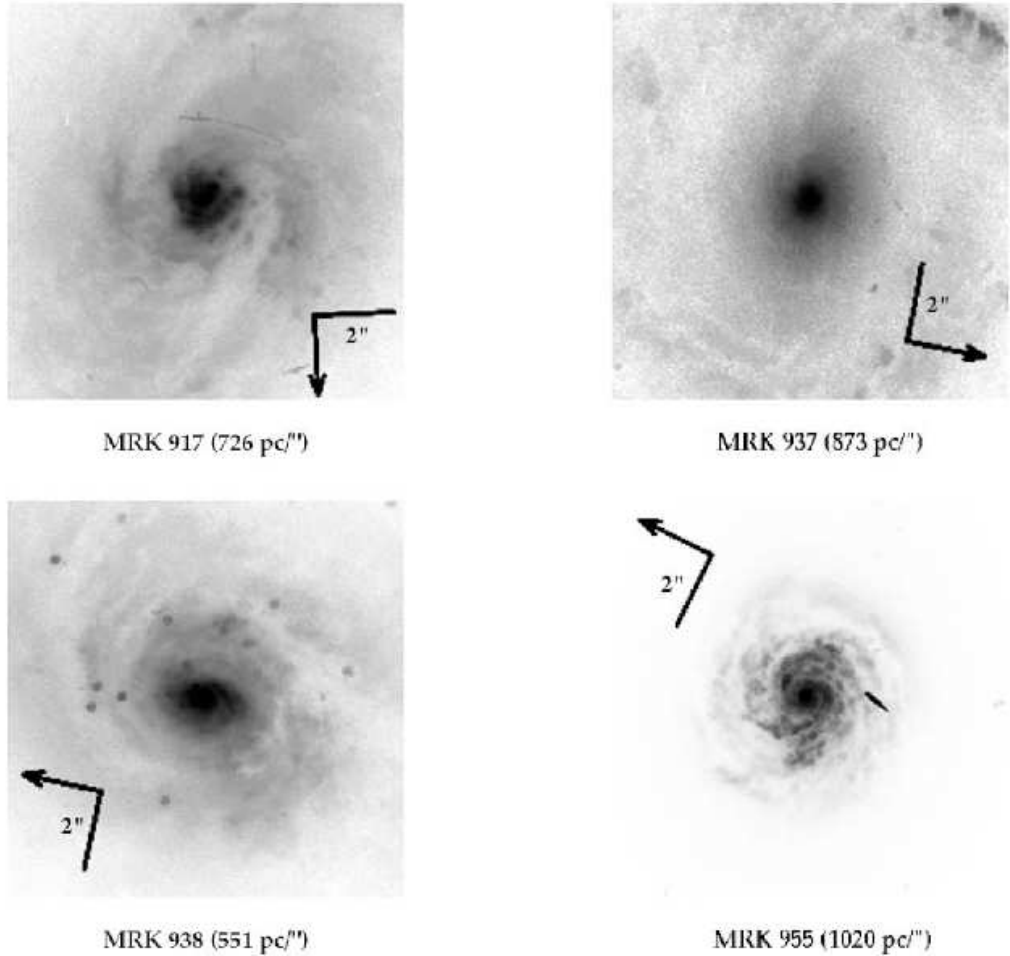,width=14cm,clip=} 
\end{center} 
\caption{\label{malkan} HST optical images of the central regions of four
  typical Type 2 Seyfert galaxies from Malkan et~al.\ (1998). The
  arrowhead points North and the bar points East, and the angular scale is
  indicated. Note that the linear scale (pc per arcsec) quoted in each
  case assumed H$_0 =$ 50 km\,s$^{-1}$Mpc$^{-1}$. In all these examples
  the central few-kpc-scale region shows the spiral arms and dust lanes
  typical of pseudo-bulges. }
\end{figure}

Is there other more direct evidence that growing SBMHs are preferentially
found in pseudo-bulges?  Intriguingly, the compilation of SMBHs in
Kormendy \& Ho (2013) shows that Seyfert nuclei are present in only 1 of
the 45 cases classified as an elliptical galaxy (the powerful radio source
Cygnus A), in only 1 of the 20 cases classified as classical bulges
(NGC4258), but in 11 of the 22 cases classified as pseudo-bulges.  The
preference of Seyfert nuclei for pseudo-bulges is confirmed by the HST
imaging survey of nearby active galaxies by Malkan et~al.\ (1998).  Simple
visual inspection of these images shows that spiral arms, rings,
dust-lanes, and bars are present in the majority of the AGN hosts on
radial scales of-order a kpc or less from the nucleus (see
Fig.~\ref{malkan}). Malkan et~al.\ assigned an inner-Hubble-type for each
galaxy based on these images, and only 20 of the 107 (19\%) of the Type 2
Seyferts had inner Hubble types of E or S0 as expected for a classical
bulge. Jiang et~al.\ (2011) reached similar conclusions for an SDSS sample
of Type 1 AGN associated with black holes with estimated masses below
$10^6$ M$_{\odot}$.

If AGN are typically fueled via secular processes, one might naturally
expect that they would be preferentially found in barred galaxies (since
bars are one of the major mechanisms for driving radial gas flows;
e.g.\ Shlosman et~al.\ 1990). As in the case of establishing a link between
interactions and AGN, there have been many studies over the years that
have reached contradictory conclusions on this matter. The most recent
studies based on SDSS do not show evidence for a link once a proper
control sample has been identified (Lee et~al.\ 2012, Cisternas
et~al.\ 2013; and references therein).  Finally we note that there are
rare cases of AGN in galaxies without any detectable bulge at all
(Filippenko et~al.\ 1993, Satyapal et~al.\ 2009, Simmons
et~al.\ 2013). The fueling mechanism for these systems is mysterious.

\subsection{Fueling of Jet-Mode AGN}

The idea that supermassive black holes can be fed through the accretion of
gas shed by evolved stars in the central region of the galaxy is an old
one (e.g.\ Bailey 1980, David et~al.\ 1987, Norman \& Scoville 1988, Ciotti
\& Ostriker 1997).

As discussed in Section 4.2, Kauffmann \& Heckman (2009) found evidence
for two distinct regimes of black hole growth in nearby galaxies. We
highlight here the regime in which the Eddington ratio distribution is a
steep power-law. This regime dominates for the more massive black holes
living in the older bulges and typically accreting at the low Eddington
rates associated with jet-mode AGN (see Fig.~\ref{eddfracfuncs}).  For
this population, Kauffmann \& Heckman found that at fixed 4000\AA\ break
strength (broadly, fixed mean stellar age), the time-averaged accretion
rate on to the black hole scales in proportion with the stellar mass of
the galaxy bulge, and the overall rate is higher in galaxies with younger
mean stellar ages.  They argued that this is consistent with the black
hole being fueled from the recycled hot gas arising from stellar mass
loss.

In fact, hot gas has been widely argued to be the fueling source of the
jet-mode radio-AGN (e.g.\ Hardcastle et~al.\ 2007). These radio-AGN are
found predominantly in massive galaxies, often at the centers of groups
and clusters, with associated hot gaseous X-ray emitting haloes.  These
hot gas halos offer not only a potential source of fueling gas for the
radio-AGN, but also a confining medium for the radio source to expand
against and a depository for the bulk kinetic energy emitted in the radio
jet: this allows for the possibility of the radio-AGN feedback cycle that
is broadly supported by both observations and galaxy formation models (as
we will discuss in Section 6; see also the reviews by Cattaneo
et~al.\ 2009 and Fabian 2012).

The rate at which hot gas accretes on a black hole was first calculated by
Bondi (1952) for a spherically symmetrical geometry with negligible
angular momentum or magnetic fields. The Bondi accretion rate from a gas
cloud of uniform density ($\rho$) and pressure is $\dot{M_{\rm B}} = 4 \pi
\lambda (G M_{\rm BH})^2 \rho / c_{s}^{3}$, where $\lambda$ is an
order-unity scaling factor and $c_s$ is the sound speed. In reality, the
gas density around a black hole is far from uniform, but the Bondi formula
remains approximately correct if $\rho$ and $c_s$ are calculated at the
Bondi radius (the radius within which the gravitational potential energy
of the black hole dominates over the thermal energy of the surrounding
gas). Allen et~al.\ (2006) showed that for a sample of 9 nearby systems, a
strong correlation exists between the calculated Bondi accretion rate and
the jet mechanical energy estimated from the cavities inflated in the
surrounding gas. Both the slope and normalization of the relation were
found to be close to unity, indicating that simple Bondi accretion of hot
gas could be powering these objects.

More recently, however, Russell et~al.\ (2013a) have found a much weaker
correlation between Bondi accretion rates and jet powers. Furthermore, for
the most luminous radio sources, Bondi accretion rate estimates fall more
than two orders of magnitude short of providing the required jet powers
(e.g.\ Cao \& Rawlings 2004, McNamara et~al.\ 2011). To produce such
powerful sources with direct Bondi accretion would require the black hole
masses to be far higher than currently estimated (requiring black hole
masses significantly above $10^{10}$ M$_{\odot}$). Such large black hole
masses would, however, help to explain how the nuclei of these powerful
objects can be radiatively inefficient (i.e. below $\sim1$\% Eddington;
see Hlavacek-Larrondo \& Fabian 2011).

A more plausible alternative to direct Bondi accretion of hot gas, is that
the gas cools out of the hot phase prior to accretion. The most powerful
jet-mode radio sources are invariably located in dense galaxy groups or
clusters, frequently with strong cooling flows. Hydrodynamic simulations
(most recently by Gaspari et~al.\ 2013) including gas cooling, turbulence
and feedback from AGN heating, reveal that when the hot gas is cooling
then accretion on to the black hole occurs predominantly in a cold and
chaotic manner. Cold clouds and filaments form out of the cooling gas in
the simulations, decouple from the hot gas, and enter a quasi free-fall
regime. Multiple cloud-cloud collisions reduce the angular momentum of the
cold gas, and lead to stochastic accretion at rates which can be two
orders of magnitude higher than the Bondi rate. This is sufficient to
produce the observed jet powers of even the most luminous jet-mode
sources.  Hillel \& Soker (2013) suggest that accretion of cooled gas may
also dominate over direct Bondi accretion in less extreme systems as well:
they argue that even the stellar winds of stars in orbit near the central
supermassive black hole would violate the zero-pressure assumption of the
Bondi formulism, and that the cooling shocked gas from such winds would
probably accrete in cold clumps.

Observationally, cool clouds and filaments are almost universally observed
in galaxies, groups and clusters with cooling times significantly below
the Hubble time, but not in systems with higher gas entropy (e.g.\ Hu
et~al.\ 1985, Cavagnolo et~al.\ 2008). As extensively reviewed by Fabian
(2012), the cooled material is predominantly cold molecular gas, which can
extend tens of kpc around the brightest cluster galaxy. This is in good
agreement with the numerical simulations. The observed masses of cooled
molecular gas are comfortably high enough to provide sufficient accretion
energy to fuel the inflated cavities (e.g.\ McNamara et~al.\ 2011); in many
objects they are two or more orders of magnitude higher, indicating that
the bulk of the cooled molecular gas does not feed the black hole, but
rather fuels star formation activity, which can also be seen in these
systems (e.g.\ Rafferty et~al.\ 2006, O'Dea et~al.\ 2008).

McNamara et~al.\ (2011) show that the mass of cooled molecular gas does
not correlate well with the radio jet power. This implies that either the
efficiency with which the cooled gas gets funneled down to the black hole
must vary greatly (which is at least plausible, given that only a small
fraction of the gas reaches the black hole), or that another factor as
well as accretion rate is important in determining the jet power. An
obvious `hidden factor' is black hole spin. Blandford \& Znajek (1977)
showed that when material accretes onto rapidly rotating black holes
threaded by high-power magnetic fields, the spin energy of the black hole
can additionally be extracted. Even if the Blandford-Znajek mechanism does
not operate, the jet power can still be a strong function of the black
hole spin, because frame-dragging from the black hole's rotation can
contribute to the twisting of the magnetic field lines, amplifying any
outflow generated by the Blandford-Payne model (Blandford \& Payne 1982;
see also Punsly \& Coroniti 1990, Meier 1999).

Black hole spin has been widely cited as a possible explanation for the
wide range of radio-loudness observed in the QSO population (the so-called
spin-paradigm of Wilson \& Colbert 1995). According to current models,
rapidly-spinning black holes are required in order to produce sufficient
jet power to account for the most radio luminous sources (e.g.\ Meier 2001,
Nemmen et~al.\ 2007). Using these models, it is possible to reproduce both
the QSO radio loudness distribution and the SDSS-derived local radio
luminosity functions of both jet-mode and radiative-mode radio sources
(Mart{\'{\i}}nez-Sansigre \& Rawlings 2011). This requires a bimodal
distribution of black hole spins, with black holes being either close to
maximally-spinning or with very low spin parameters. Nevertheless, despite
the success of these models, the role of black hole spin in powering radio
jets remains unproven. Even at the scale of X-ray binaries, there is
ongoing debate in the literature as to whether the power of the radio jet
produced depends upon the spin of the black hole (e.g.\ Russell
et~al.\ 2013b, Steiner et~al.\ 2013).

In conclusion, evidence is strong that a majority of low Eddington
fraction AGN are fueled directly or indirectly from hot gas.  The
accretion flows are radiatively inefficient, and a significant fraction of
the energetic output occurs through jet outflows, the strength of which
may be enhanced by rapidly spinning black holes. For many galaxies, the
fueling hot gas arises from recycled material from stellar mass loss. In
more massive systems (the most massive ellipticals, and galaxies at the
centers of groups and clusters) extended hot gas halos provide a more
abundant gas source, and large-scale cooling flows can greatly enhance the
accretion rate.  In such systems the extended hot gas halo also provides a
working surface for a jet, and a medium to confine the expanding radio
lobes, minimizing adiabatic losses and leading to luminous radio
emission. The hot halo also acts as a repository for the radio jet energy,
offering the ideal conditions for a radio-AGN feedback cycle (see Section
6).

\section{FEEDBACK PROCESSES}

As recently reviewed by Cattaneo et~al.\ (2009) and in this journal by
Fabian (2012), feedback from AGN is currently invoked in both
semi-analytic models and numerical simulations in order to successfully
reproduce the observed properties of massive galaxies (e.g.\ Di Matteo
et~al.\ 2005, Springel et~al.\ 2005, Bower et~al.\ 2006, Croton
et~al.\ 2006, Hopkins et~al.\ 2006, Ciotti et~al.\ 2010).  While AGN
feedback is generally assumed to be negative (inhibiting star-formation
and/or black hole growth), it could also be positive in nature
(e.g.\ Ishibashi \& Fabian 2012, Silk 2013, Zinn et~al.\ 2013).  Feedback
from AGN is generally invoked in two flavors, which relate to the two
fundamental modes of AGN activity. The first is a mode of powerful
AGN-driven winds that are postulated to occur in galaxies with actively
growing black holes and is argued to be responsible for the termination of
star-formation and the migration of the galaxy from the blue star-forming
main sequence to the red sequence. It may also be responsible for setting
up the correlation between black hole mass and bulge properties (see
review by Kormendy \& Ho 2013). We discuss this in Section
6.1. Complementary to this is the requirement to keep the galaxies as red
and dead once they have arrived at the red sequence, by preventing further
gas cooling. Recurrent low-luminosity radio-AGN activity is often posited
to be responsible for this kinetic mode of feedback, which is often
referred to as maintenance-mode or radio-mode feedback. We examine this in
Section 6.2

\subsection{ Feedback and Radiative-Mode AGN}

We have reviewed above the evidence that the growth of supermassive black
holes is related to the formation of massive stars in the central
(few-kpc-scale) region of the host galaxy. Thus, we will first summarize
the evidence for feedback associated with short-lived massive stars and
then the evidence for feedback from the black hole itself.

\subsubsection{Feedback from Massive Stars} 

Could the feedback provided by massive stars formed in the central region
of the host galaxy affect the fueling of the black hole or the evolution
of the host galaxy?  To set the scale, note that for a normal
Kroupa/Chabrier stellar IMF, massive stars return about 1.4 $\times
10^{49}$ ergs in kinetic energy per solar mass of star formation (based on
Starburst 99; Leitherer et~al.\ 1999).  This amount of kinetic energy per
unit mass is equivalent to a characteristic velocity of about 1200 km/s
(far larger than the galaxy escape velocity).  Provided that this energy
can be efficiently coupled to the surrounding gas, it could indeed result
in significant feedback.

\begin{figure}[!t]
\begin{tabular}{cc}
\raisebox{0.8cm}{\psfig{file=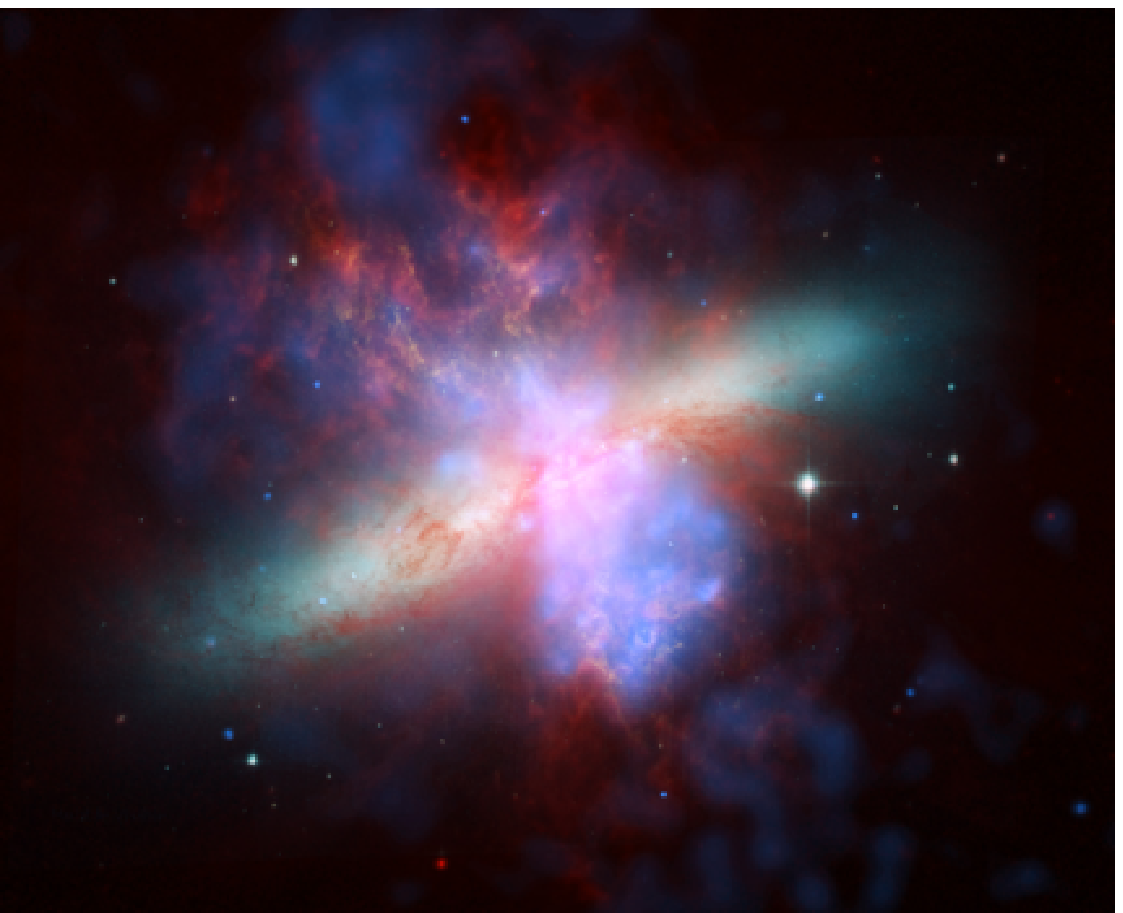,width=8.1cm,clip=}} 
\psfig{file=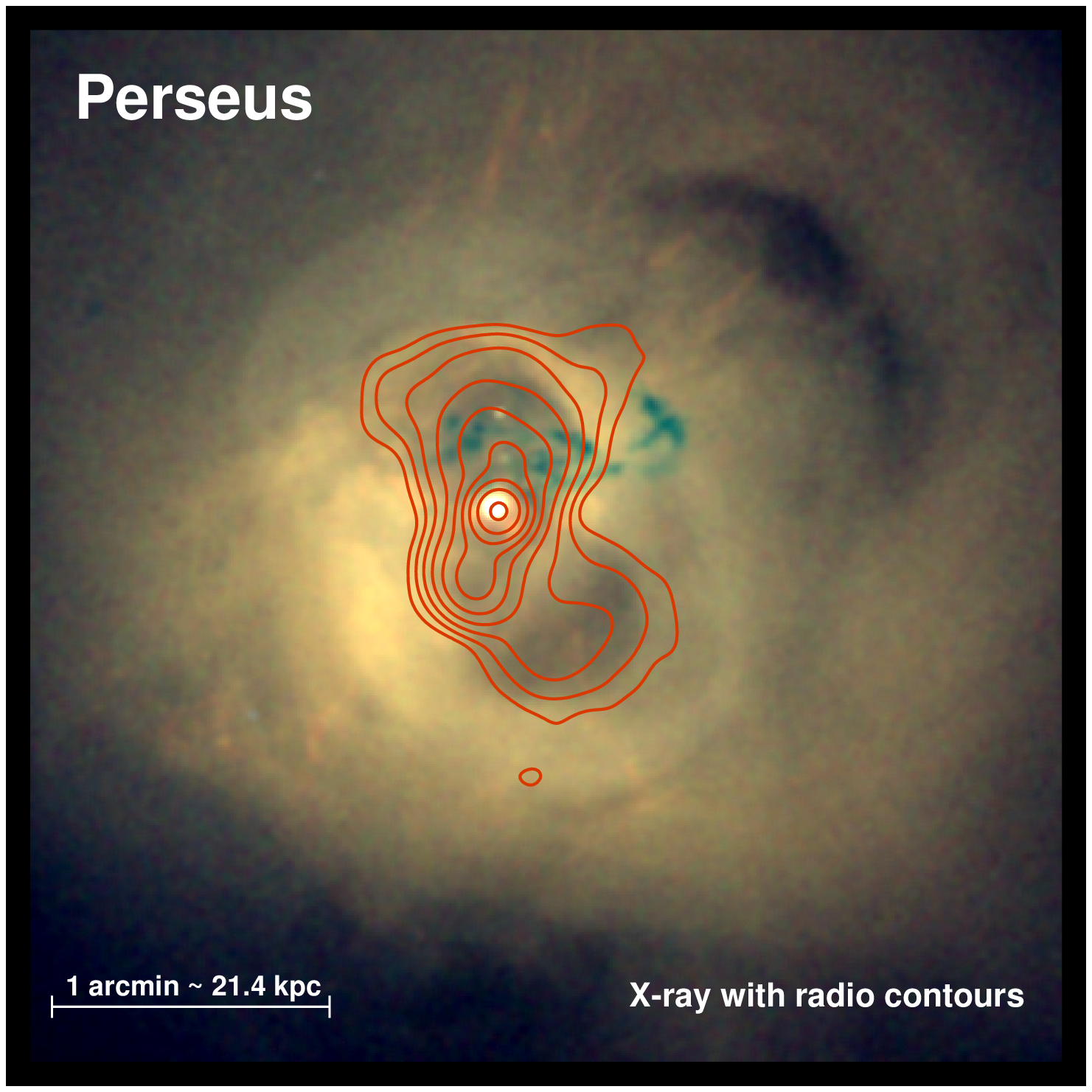,width=8.2cm,clip=} 
\end{tabular}
\begin{tabular}{ccc}
\psfig{file=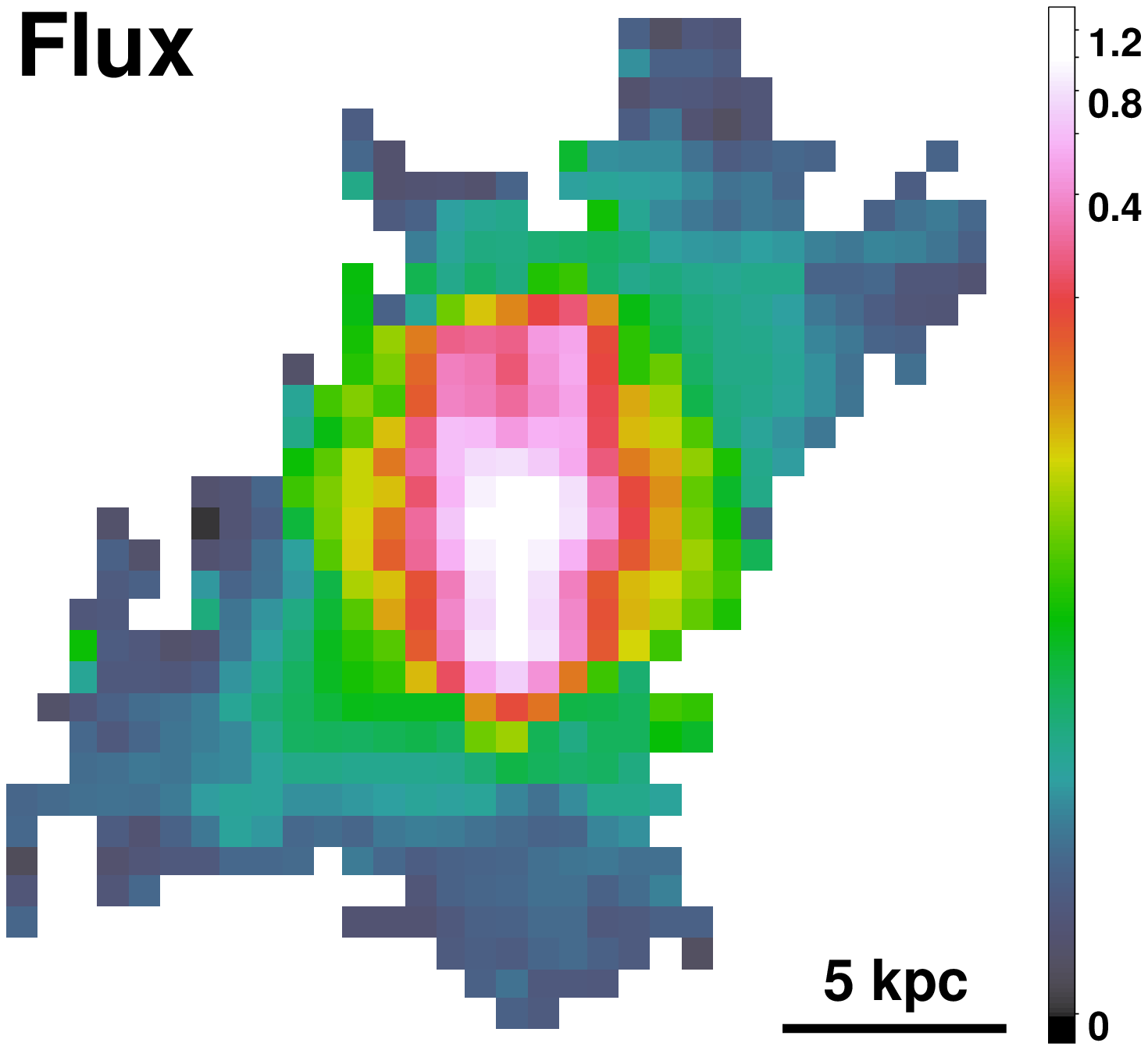,width=5.5cm,clip=} 
\psfig{file=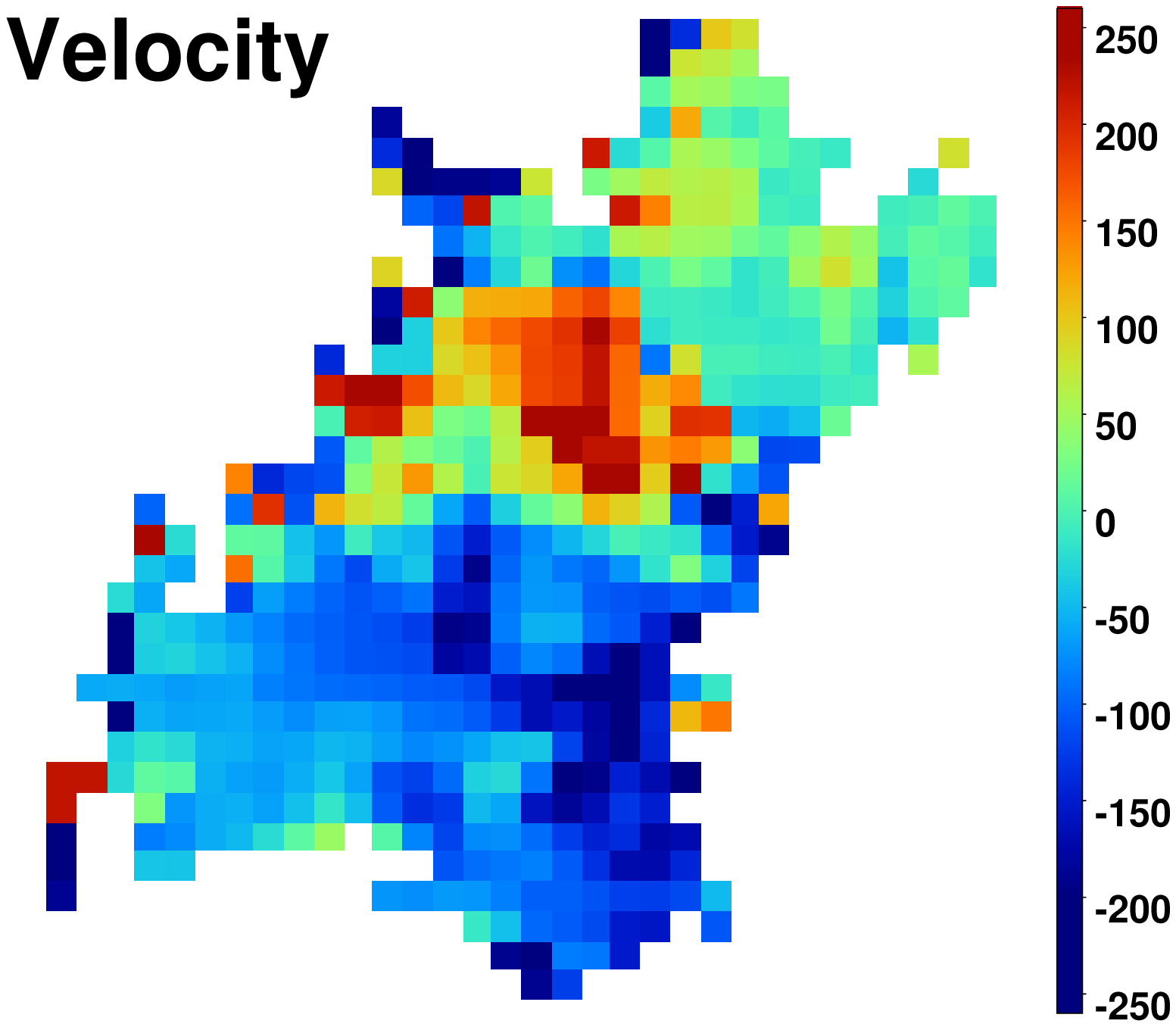,width=5.5cm,clip=} 
\psfig{file=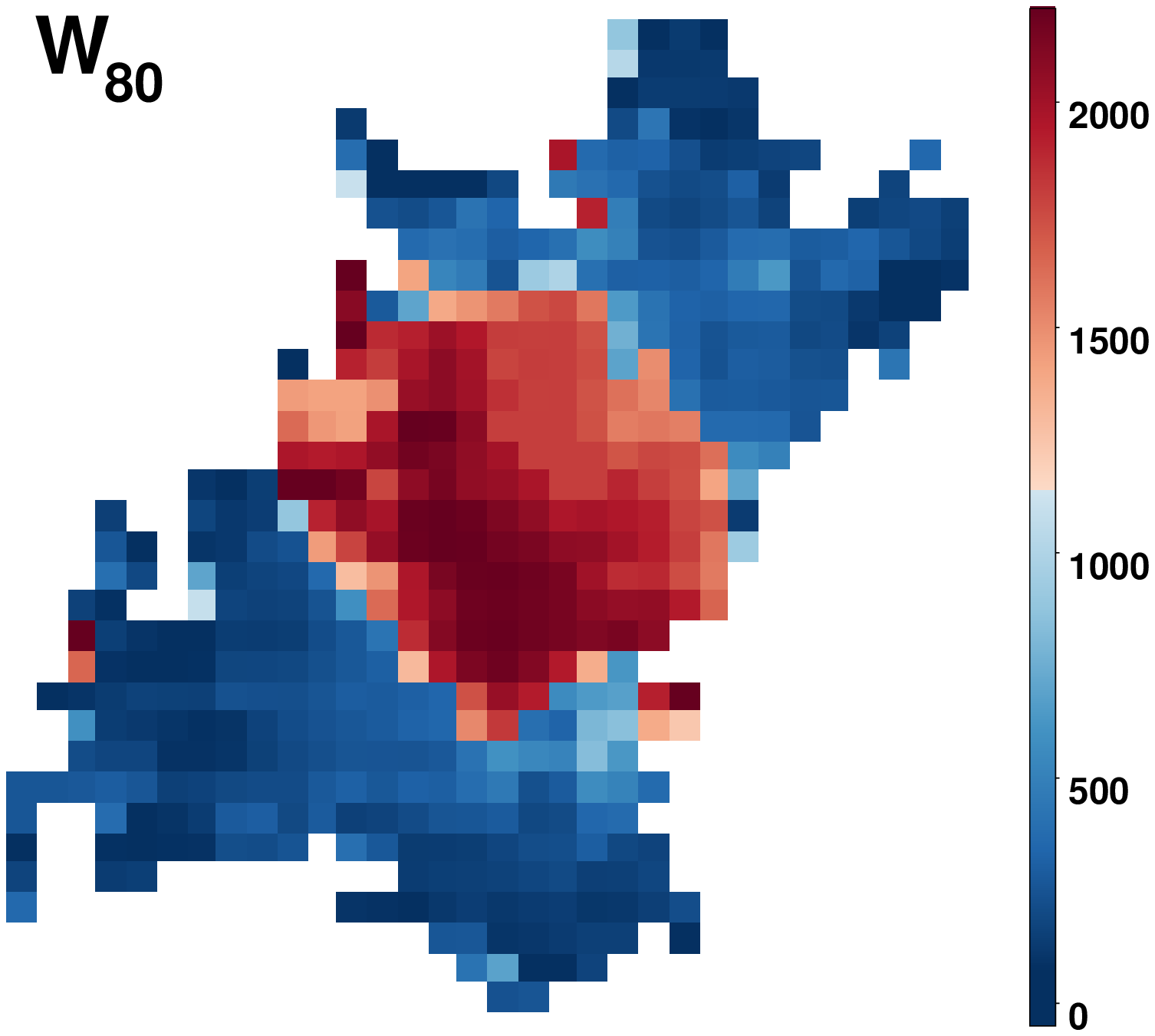,width=5.5cm,clip=} 
\end{tabular}
\caption{\label{outflow_compos} Illustrative examples of feedback effects
  at work. The upper-left panel shows the `Great Observatories' composite
  image of the starburst-driven galactic wind in M82: visible light is
  shown in yellow-green, infrared emission in red, H$\alpha$ emission in
  orange and X-ray emission in blue (Credit:
  NASA/JPL-Caltech/STScI/CXC/UofA/). The upper-right panel shows jet-mode
  feedback in the Perseus cluster (data from Fabian et~al.\ 2006). The
  color-scaling shows the X-ray emission, colored such that hard X-rays
  are in blue and soft X-rays in red. The contours indicate the radio
  emission, which can be seen to have evacuated cavities in the X-ray
  gas. The three lower panels illustrate AGN-driven feedback from a high
  luminosity radiative-mode AGN (reproduced from Liu et~al.\ 2013b). They
  show, respectively, the surface brightness (in units of
  10$^{-14}$erg\,s$^{-1}$cm$^{-2}$arcsec$^{-2}$), relative
  velocity (in km\,s$^{-1}$), and velocity width (W80 is the velocity
  width in km\,s$^{-1}$ containing 80\% of the emitted line flux) of the
  [OIII] emission line in the Type 2 QSO SDSS J0319-0019.}
\end{figure}

There is certainly compelling evidence that this can occur. The most
dramatic effect is the production of galactic winds that can be traced in
emission and absorption in both the local (e.g.\ Heckman et~al.\ 1990) and
high-redshift (e.g.\ Steidel et~al.\ 2010, Shapley 2011) universe.  These
outflows are driven by the energy and/or momentum supplied by massive
stars through their stellar winds, their supernova ejecta, and their
radiation (e.g.\ Chevalier \& Clegg 1985, Murray et~al.\ 2011).  While
there are detailed multi-waveband studies of relatively small samples of
local galactic winds (see the review by Veilleux et~al.\ 2005), the
largest census of galactic winds in the local universe was undertaken by
Chen et~al.\ (2010) based on using SDSS galaxy spectra to measure the
interstellar absorption feature produced by neutral Na atoms that trace
the outflow. Their work provided statistically robust evidence that local
galactic winds are wide-angle bipolar flows along the galaxy minor axis
and are associated with galaxies with high star-formation rates per unit
area (cf. Fig.~\ref{outflow_compos}).

Investigations of starburst-driven outflows in the local universe have
only been able to directly detect the flows out to radii of-order ten kpc
(e.g.\ Grimes et~al.\ 2005). However, recent work using absorption-line
spectra of background QSOs shows that the wind affects the properties of
the surrounding gaseous halo of a starburst galaxy all the way out to the
galaxy's virial radius (Borthakur et~al.\ 2013). This is consistent with
similar observations of Lyman Break Galaxies at z $\sim$ 3 (Steidel
et~al.\ 2010).  While global feedback from massive stars clearly exists,
the evidence that it limits the growth of black holes is only indirect so
far. We have described above the result from Wild et~al.\ (2010) that
showed that in a large SDSS-derived sample of starburst and post-starburst
galaxies, the growth of the supermassive black hole was delayed by about
200 Myr relative to the onset of the starburst.  Wild et~al.\ showed that
the turn-on in black hole fueling in these galaxies occurred at the time
at which the primary source of mass-loss from stars in the galaxy's central
region switched from supernovae (with ejecta velocities far above the
local escape velocity) to intermediate-mass AGB stars (with ejecta
velocities far below the local escape velocity). On this basis they
speculated that supernova feedback was preventing or inhibiting the
accretion of gas by the black hole (Fig.~\ref{wildburst}). This idea had
been previously suggested both theoretically (Norman \& Scoville 1988) and
based on detailed studies of small samples of local AGN (Davies
et~al.\ 2007).

It is also possible that feedback from massive stars is responsible for
the saturation effect found by Kauffmann \& Heckman (2009): once the
specific star-formation rate in the bulge exceeds a value of
$\sim10^{-10}$ yr$^{-1}$ the distribution of the Eddington ratio for the
black hole assumes a universal log-normal form that does not change at
still higher star formation rates. While higher star-formation rates are
associated with a larger gas supply, they are also associated with higher
levels of feedback.  Kauffmann \& Heckman (2009) presented an alternative
interpretation: the saturation effect could be understood if the black
hole regulated its own growth at a fixed average rate of a few percent of
the Eddington limit when its fuel supply was plentiful. In a subsequent
paper, Hopkins \& Hernquist (2009) compared their observational results
with their models for self-regulated BH growth, in which feedback produces
a self-regulating phase after the AGN reaches some peak luminosity/BH mass
and begins to expel gas and shut down accretion, and found good agreement
at the bright end.  On that note, we now describe the direct observational
evidence for black-hole-driven feedback in radio-quiet local AGN.

\subsubsection{Feedback from the Black Holes in Radiative-Mode AGN}

As reviewed by Fabian (2012) there is ample direct evidence that
radio-quiet AGN drive outflows. This is primarily based on the frequent
presence of blue-shifted absorption-lines in the optical, ultraviolet, and
soft X-ray spectra of Type 1 AGN (see Crenshaw et~al.\ 2003 and references
therein). These trace highly ionized gas with outflow velocities ranging
from several hundred to several thousand km s$^{-1}$ in low-luminosity AGN
(Type 1 Seyfert nuclei) to tens-of-thousands of km/s in powerful QSOs (the
Broad-Absorption-Line systems).

The uncertainty in the feedback effect of these flows arises due to the
difficulty in measuring their basic physical properties. Simple arguments
imply that the mass (energy) outflow rate will be proportional to the
product of the outflow's total column density times its size times its
solid angle times its velocity (cubed).  The strong absorption-lines
measured in the AGN-driven outflows are almost certainly saturated, making
it difficult to determine even the ionic column density (let alone a total
gas column density). Moreover, the sizes of these outflows have been
largely unconstrained. The lack of the detection of such high-velocity
outflows in the spectra of typical Type 2 Seyferts (e.g.\ Krug
et~al.\ 2010) is evidence that either the flows in Seyferts do not extend
beyond the central few tens of parsec or that their column densities drop
below detectable values on these larger radial scales.  It is therefore
not surprising that the spectroscopic data in SDSS on a vast number of
local Type 2 AGN has not uncovered any direct evidence that they are
driving large-scale outflows that cannot be attributed to the feedback
from central star-formation described in Section 6.1.1 above.

Another line of evidence for outflows in typical radio-quiet AGN comes
from mapping the physical and dynamical state of the ionized gas seen in
the narrow emission-line region. This has been reviewed by Veilleux
et~al.\ (2005) in this journal and we refer the reader there for
details. In a number of local Seyfert galaxies the ionized gas on a scale
of $\sim$ a kpc from the AGN is clearly being impacted by the collimated
outflows (radio jets), which are commonly present even in radio-quiet
Seyferts (e.g.\ Wilson \& Heckman 1985, Rosario et~al.\ 2010a,b). It is
difficult to accurately estimate the total outflow rates in these cases,
but the kinetic energy flux appears to be a small fraction of the AGN
bolometric luminosity in typical local Seyfert galaxies (see Veilleux
et~al.\ 2005 and references therein).

While direct evidence for black-hole-driven feedback in typical local AGN
is lacking, recently there have been several exciting discoveries that
imply that the most luminous AGN can indeed drive highly energetic
outflows that could have a significant impact on both the fueling of the
black hole and the evolution of the host galaxy. These are AGN with
bolometric luminosities well above the knee in the local AGN luminosity
function ($L_* \sim 10^{45}$ erg/s $\sim 10^{11.5}$ L$_{\odot}$). As such
they are rare and only rather small samples have been studied so far.

The first such discovery is the prevalence of outflows of molecular gas in
Ultra-Luminous Infrared Galaxies (ULIRGs) with AGN bolometric luminosities
above about $3 \times 10^{45}$ erg/s. This evidence is of two types.
First, the PACS instrument on Herschel has been used to observe the far-IR
emission and absorption features due to the OH molecule. Initial results
were presented in Fischer et~al.\ (2010) and Sturm et~al.\ (2011). Most
recently, Veilleux et~al.\ (2013) and Spoon et~al.\ (2013) have summarized
observations of an expanded sample of ULIRGs and QSOs at $z <$ 0.3. For
each object, the fractions of the bolometric luminosity due to a starburst
and AGN were determined following the analysis of Veilleux et~al.\ (2009).
The OH feature is detected in absorption in most objects, with the
majority of these showing evidence for molecular outflows in the form of
blue-shifted OH absorption lines. The equivalent widths of the OH
absorption features (tracing the column density) do not depend upon the
AGN luminosity or the ratio of the AGN/starburst luminosities. The outflow
velocities do not depend upon the starburst luminosity but they do
increase with both increasing AGN luminosity and AGN/starburst luminosity
ratio. Characterizing the outflow speed by the midpoint in the line
profiles yields outflow speeds of-order 100 km/s in the
starburst-dominated systems and about 400 km/s in the AGN-dominated cases,
but the most rapidly outflowing gas can reach speeds of-order 10$^3$ km/s.
As noted above, deriving an outflow rate requires determination of a size
for the flow. Sturm et~al.\ (2011) have used a radiative-transfer code
(Gonzalez-Alfonso et~al.\ 2013) applied to the observed far-IR spectrum to
estimate the size scales, and hence the outflow rates in five ULIRGs.  For
an assumed OH/H$_2$ ratio, the implied outflow rates range from $10^2$ to
$10^3$ M$_{\odot}$ year$^{-1}$ (comparable to, or larger than, the SFR).

Complementing these probes are measurements of molecular gas in emission
in the mm-wave regime.  The molecular outflow in the proto-typical
AGN-dominated ULIRG Mrk 231 has been mapped in CO (Feruglio et~al.\ 2010,
Cicone et~al.\ 2012) and HCO$+$, HCN, and HNC (Aalto et~al.\ 2012) using
the Plateau de Bure mm-wave interferometer. The flow is spatially-resolved
with outflow speeds of up to 800 km/s detectable over a kpc-scale region
and an estimated outflow rate of about 700 $M_{\odot}$ year$^{-1}$, in
broad agreement with the estimates based on the OH absorption line seen
with Herschel. Similar results using CO emission as a tracer have now been
found for a small sample of other AGN-dominated ULIRGs (Cicone
et~al.\ 2013).  They find that the kinetic energy in the molecular outflow
is a few percent of the AGN bolometric luminosity and that the momentum
flux is about 20 times larger than $L_{\rm bol}/c$.  The primary
uncertainty affecting these estimates is the conversion factor used to
convert the CO luminosity into an H$_2$ mass.

These outflow rates are impressive, but the flows can only be traced out
to modest radii (kpc-scale). Recently, the effects of AGN outflows on the
global properties of the host galaxy have been discovered in samples of
Type 2 QSOs -- obscured AGN with bolometric luminosities in excess of
$10^{46}$ erg s$^{-1}$ (Zakamska et~al.\ 2004).  Greene et~al.\ (2011)
reported on the analysis of long-slit optical spectroscopy probing the
kinetics of the ionized gas surrounding 15 Type 2 QSOs at z $\sim$ 0.1 to
0.45. They found broad emission-lines of [OIII]5007 (full width at 20\% of
the maximum line intensity of $\sim$ 500 to 1500 km s$^{-1}$) that
extended across the entire extent of the galaxy (radii of-order 10 kpc).
More recently Liu et~al.\ (2013a,b) used the Gemini integral field unit to
fully map out the 2-D distribution [OIII] and H$\beta$ emission-lines
around a sample of 11 radio-quiet Type 2 QSOs at z $\sim$ 0.3 to 0.65 with
typical bolometric luminosities of $10^{47}$ erg/s.  They obtain similar
results to Greene et~al.\ (2011) but their complete spatial coverage
enables them to better characterize the structure and kinematics
(e.g.\ Fig.~\ref{outflow_compos}). They find that a simple model of a
spherically symmetric outflow at a velocity of 500 to 1000 km s$^{-1}$
reproduces the data. The estimated outflow rates are uncertain because the
density (and hence the mass) of the emitting material is not directly
measured.  It appears that the entire interstellar medium of the galaxy
could be expelled over a timescale of only a few tens-of-Myr implying a
kinetic energy flow of-order 2\% of the bolometric luminosity (similar to
the estimates in the molecular outflows described above).

In summary, there is now some persuasive evidence that the most powerful
radiative-mode AGN in the contemporary universe can drive outflows capable of
severely impacting their host galaxies.  Evidence for global outflows in
typical low-redshift radiative-mode AGN is lacking.

\subsection{Feedback from radio sources}

As discussed in Section 5.3, radio-AGN can be fueled from their
surrounding hot gaseous haloes, either directly through the Bondi process,
or via gas which has cooled out of the hot phase in a cooling flow.  The
bulk of the energy generated from the accretion is channelled into the
powerful radio jets, which drive their way outwards. The radio jets are
confined by the hot halo gas, and produce expanding radio lobes filled
with relativistic plasma, inflating bubbles or cavities in the hot
gas. The bulk of the jet energy is deposited locally, through the
mechanical work required to inflate these cavities. Since the repository
of the AGN energy is the same gas as that which fuels the AGN, this offers
the necessary conditions for a self-regulating AGN feedback cycle.

The most extreme examples of radio-AGN feedback occur in the massive
galaxies at the centers of groups and clusters (see
Fig.~\ref{outflow_compos}). In these systems the radiative cooling time
of the surrounding gas can be substantially shorter than the Hubble
time. In the absence of a heating source, a cooling flow would be expected
to develop, whereby the temperature in the central regions of the cluster
drops and gas flows inwards at rates of hundreds or even thousands of
solar masses per year (see review by Fabian 1994). In reality, however,
observations show that the gas temperatures in the cluster cores are at
most a factor 3 below the value at large radii, and the amount of cooling
gas is only about 10\% of the cooling flow prediction (e.g.\ Peterson
et~al.\ 2001, 2003, David et~al.\ 2001, Tamura et~al.\ 2001).

A heating source must be balancing the radiative cooling losses to prevent
the gas from cooling further, and this reduces the star-formation rates in
the central galaxy by an order of magnitude (though not to zero -- these
galaxies can still have star-formation rates of tens of solar masses per
year; e.g.\ O'Dea et~al.\ 2008). The heating source is almost certainly the
jet outflows from central black hole: the prevalence of radio-AGN activity
is high ($>$70\%) in central cluster galaxies (Burns 1990), and results
from SDSS show that it is enhanced relative to other galaxies of the same
mass (Best et~al.\ 2007).  Bubbles and cavities in the X-ray gas are also
seen in a high percentage of clusters, reaching nearly 100\% in those
clusters with cooling times below 3 Gyr (Dunn \& Fabian 2006, Fabian
2012). As discussed in Section 2.3.2, the mechanical energy flow rates of
the radio jets are calculable either from the observed synchrotron
emission or from the $pV$ work required to inflate the radio
lobes. Analysis of these cooling clusters indicates that the energy input
from the radio jets is in rough balance with the energy losses from
radiative cooling within the cooling radius (see review by McNamara \&
Nulsen, 2007).

For feedback processes to be efficient, it is not only the total available
energy that is important, but also how and where that energy is
deposited. Radio jets are highly anisotropic structures, but gas cooling
and inflow needs to be switched off in all directions. It is the expanding
radio bubbles, rather than direct jet interactions, which must provide the
bulk of the energy transfer to the ambient medium. This process is
observed in the form of weak shocks or sound waves surrounding some radio
bubbles, with the most striking example being the Perseus cluster where
several concentric ripples in X-ray pressure are seen (Fabian
et~al.\ 2003, 2006). These ripples have been associated with repeated
episodes of bubble-blowing activity from the central AGN. The energy in
the sound waves or weak shocks is gradually dissipated over length scales
of $\sim$100 kpc (depending upon the gas viscosity). This gentle feedback
process is able to transfer the radio jet energy into its environment in a
largely isotropic manner, without disturbing the intracluster medium to
such an extent that the observed temperature and metal abundance gradients
would be destroyed by mixing.

The radio-AGN activity appears to be switched on nearly all of the time in
cooling flow clusters, and must be able adapt quickly to changes in its
environment in order to maintain the close heating-cooling balance. This
quick adaption is not possible for standard Bondi accretion, since the
infall time of the hot gas from kpc-scales down to the black hole is more
than a Gyr (Soker 2006). However, cold clump accretion occurs on a much
faster timescale, especially in a viscous accretion flow, and therefore
allows the black hole to react quickly to changes in the the state of its
environment (Pizzolato \& Soker 2010, Gaspari et~al.\ 2013). This allows
for efficient self-regulation of the feedback process.

Although brightest cluster galaxies offer the most striking examples of
radio-AGN feedback, which can be individually studied in detail, these
massive cooling-flow clusters are relatively rare. The data from large sky
surveys have been invaluable in extending analysis to less extreme systems
and providing a direct statistical comparison of radio-AGN feedback in
systems of different scales. Best et~al.\ (2006) investigated the role of
radio-AGN feedback in elliptical galaxies using data from SDSS.  On the
assumption that all massive galaxies would go through recurrent
short-lived radio-AGN outbursts, they considered the observed prevalence
of powerful ($P_{\rm 1.4 GHz} >10^{23}$W\,Hz$^{-1}$) radio-loud AGN as a
function of black hole mass and interpreted this as a measure of the duty
cycle of powerful radio-AGN activity for galaxies of that black hole
mass. They combined this with the black hole mass dependent radio
luminosity function to estimate, probabilistically, the fraction of its
time that a black hole of given mass would spend producing a radio source
of a given radio luminosity.  Using the radio luminosity to jet kinetic
power conversion (Section 2.3.2) this then allowed a calculation of the
average rate of jet kinetic energy production of the black hole.  Note
that most of the heating is produced by radio sources of $P_{\rm 1.4 GHz}
\sim 10^{24}$--$10^{25}$W\,Hz$^{-1}$. The heating contribution from
periods of very low radio luminosity, where in this simple picture the
radio AGN is considered to be `off', can be neglected.  Best
et~al.\ derived the heating rate for galaxies of different black hole
mass, and compared these results with the X-ray luminosities (ie., cooling
rates) of the hot haloes around the galaxies. They found that for
elliptical galaxies of all masses, the radio-AGN heating rate provided a
good balance to the gas cooling rate. AGN heating is therefore able to
counter-balance cooling and thus explain the old, red and dead nature of
the ellipticals.

The Best et~al.\ (2006) analysis actually underestimated the jet-energy
production rate, as they adopted only $pV$ instead of 4$pV$ for the cavity
energies. Also, for gas cooling rates, they scaled from the total X-ray
luminosity, rather than considering only the luminosity within the cooling
radius (which is the only region where cooling needs to be compensated in
order to avoid catastrophic cooling collapse).  Therefore, the radio-AGN
provide, in a time-averaged sense, significantly more energy than is
required to suppress cooling on galaxy scales. This was shown more clearly
by Best et~al.\ (2007) who compared the time-averaged AGN-heating rate
against the rate of radiative-cooling within the cooling radius, for
systems of different masses. They suggested that the average
heating-to-cooling ratio decreased by two orders of magnitude when going
from massive ellipticals / small groups to the most massive
clusters. Updated observational determination of the black-hole mass
dependent radio luminosity function and the radio luminosity to jet power
conversion have decreased this scale-dependence, but it remains well in
excess of an order of magnitude. Thus, if the radio-AGN heating rate is at
the level that balances cooling in the most massive clusters (or at least
the cool-core subset of these), then in smaller systems an order of
magnitude more jet mechanical energy is produced than required.

Similar indications have been found by Nulsen et~al.\ (2007; see also
Figure 5 of Fabian 2012) who found that the $pV$ energy stored in cavities
seen in lower mass systems can be an order of magnitude above that
required to balance cooling losses. This result is easily understood,
because in individual galaxies or small groups a large fraction of the jet
kinetic energy may get deposited on much larger scales than the cooling
radius: the jets themselves can extend beyond the cooling flow region, and
even if they don't then the bubbles they produce will still rise buoyantly
to beyond the cooling radius, while the weak shocks produced by the
expanding cavities dissipate their energy on similarly large (hundred-kpc)
scales (e.g.\ Fabian et~al.\ 2005).

In lower mass systems the duty cycle of radio source activity is also low,
with the AGN being turned off for most of the time. The heating-cooling
balance is a pseudo-equilibrium process, whereby in the brief periods
where the radio source does turn on, the instantaneous energetic output is
far above that required to balance cooling. This leads to over-heating of
the gas, seen for example as a temperature difference observed between
galaxy groups and low-mass clusters with and without radio-AGN (Croston
et~al.\ 2005, Maglioccetti \& Br{\"u}ggen 2007; see also Shen
et~al.\ 2008). After the radio-AGN outburst, the system will experience a
period during which accretion rates are suppressed by the turbulent
motions and heating that was induced, leading to effectively an off state
for the black hole. This behavior is seen in hydrodynamic simulations
(e.g.\ Gaspari et~al.\ 2013) and is also suggested by both the episodic
history of star-formation found by Chen et~al.\ (2013) in massive galaxies
(M$_* > 10^{11}$M${_\odot}$) and by the apparent anti-correlation they
find between a recent burst of star formation and the current presence of
a radio source. This off-period lasts longer in less massive systems since
the lower binding energy and gas sound speed lead to a longer recovery
time before gas cooling and accretion re-commence. The black hole
therefore acts as a cosmic thermostat, adjusting quickly to its
surroundings and switching on powerful radio-AGN activity whenever the gas
temperature drops sufficiently to allow significant cooling to occur. Rare
intermittent activity (duty cycle 0.001 -- 0.01) suffices to control gas
cooling in typical massive ellipticals (of stellar mass $\sim 10^{11}$
M$_{\odot}$), with the system oscillating around an equilibrium state. In
contrast, near-constant activity (duty cycle $\sim$1) is required to
control gas cooling in the most massive cooling flow clusters.

\section{THE SITUATION AT HIGH REDSHIFT} 

The focus of this review has been the contemporary universe as probed
through large surveys (most notably, the Sloan Digital Sky Survey).  It is
then natural to ask whether the results we have reviewed apply only to the
present-day, or whether a qualitatively similar picture also applies at
higher redshift.

\subsection{The Representative AGN Population}

To begin, let us use the cosmic evolution of the bolometric AGN luminosity
function to frame the issues.  While it is customary to refer to the
redshift range of 2 to 3 as the QSO era, this is actually rather
misleading. It is true that the co-moving number density of the most
powerful AGN ($L_{\rm bol} > 10^{14}$ L$_{\odot}$) peaked during this
epoch. These are associated with the formation of the most massive SMBH
($>10^{9.5}$ M$_{\odot}$), and as such comprise only a minor fraction of
the black hole mass in the local fossil record (Shankar et~al.\ 2009). A
much better way to think about the growth of the black hole population is
to use the data plotted in Fig.~\ref{shankarhist} to calculate the history
of the cumulative growth of the relic mass in supermassive black holes
over time. This leads to the striking result that the growth of black
holes is actually a protracted process: 25\%, 50\%, and 75\% of the
present-day relic mass formed by cosmic ages of about 3.3 Gyr ($z \sim$
2), 5 Gyr ($z \sim$ 1.3), and 7 Gyr ($z \sim$ 0.75).  The history of the
mass build-up in galaxies via star formation is very similar (e.g.\ Shankar
et~al.\ 2009). So in a big picture sense, the physics that connected
(connects) the growth of galaxies and black holes must have been in place
over a significant fraction of the history of the universe and the bulk of
this co-evolution took place at moderate redshifts.

To focus our discussion, we will therefore consider the population of AGN
and their host galaxies over the redshift range of 0.5 to 2.5 (during
which time about 75\% of the current black hole mass was created by
accretion).  The shape of the bolometric luminosity function at $z \sim$ 2
(Hopkins et~al.\ 2007, Shankar et~al.\ 2009) implies that most of the
black hole growth then occurred for objects near $L_*$. More
quantitatively, half the black hole growth occurred for AGN with
bolometric luminosities between $L_{\rm bol} \sim 10^{12.5}$ and
$10^{13.3}$ L$_{\odot}$ (or hard X-ray luminosities of $2 \times 10^{44}$
to $8 \times 10^{44}$ erg s$^{-1}$).  At z $\sim$ 0.5, the same
calculation implies that half the black hole growth occurred in AGN with
bolometric (hard X-ray) luminosities between $10^{11.2}$ and $10^{12.3}$
L$_{\odot}$ ($3 \times 10^{43}$ to $2 \times 10^{44}$ erg/s). We will use
these fiducial values for redshift and luminosity to define what we will
henceforth call {\it representative AGN}, by which we mean the objects
most responsible for the creation through accretion of the mass locked up
in black holes in the present fossil record.

\subsection{The Host Galaxies of Radiative-Mode AGN}  

Most of the information about AGN host galaxies at high-z comes either
from deep surveys of fields that probe co-moving volumes too small to
sample the representative AGN population or from observations of
individual AGN of exceptionally high luminosity.  It is only over the past
few years that we have learned quite a bit about the properties of the
host galaxies of the representative AGN.  For the most part these surveys
have selected the AGN and quantified their properties based on
(rest-frame) hard X-ray data. One potentially important caveat is that it
is possible that the host galaxies of X-ray-selected AGN systematically
differ from the host galaxies of AGN selected in the IR or optical (Hickox
et~al.\ 2009, Juneau et~al.\ 2013, Koss et~al.\ 2011).

The most detailed investigation of the dependence of AGN properties on the
mass of the host galaxy was undertaken by Aird et al.\ (2012) for a
population of X-ray detected objects over the range $z = 0.2$ to 1.0. They
find that the distribution of AGN X-ray luminosity normalized by the
stellar mass of the host galaxy (M$_*$) is essentially independent of
M$_*$ over mass bins with centers from M$_* = 10^{9.75}$ to $10^{11.75}$
M$_{\odot}$. Within the uncertainties, this is largely consistent with the
[OIII]-based SDSS results at $z \sim$ 0.1 we have plotted in
Fig.~\ref{lagnmass}, except that our plot shows a strong drop in
normalized luminosity at high galaxy mass. As we noted, the normalized sum
of L$_{\rm rad}$ plus L$_{\rm mech}$ is roughly independent of M$_*$ out
to the highest masses. In fact, Chen et al.\ (2013) found strong evolution
in the emission-line AGN associated with very massive galaxies (M$_* >
10^{11.4}$ M$_{\odot}$) between $z \sim$ 0.1 and 0.6. They found that the
equivalent width of the [OIII] emission-line (a rough proxy for L$_{\rm
  rad}/$M$_*$) increased by an order of magnitude, and that the
emission-lines changed from LINER-like (jet-mode) at low-z to Seyfert-like
(radiative-mode) at higher-z. This is at least qualitatively consistent
with the difference between the results of Aird et al. (2012) and what is
seen in the contemporary universe. 

Aird et al.\ (2012) also conclude that their sample of AGN shows a
universal distribution in Eddington ratio that is independent of galaxy or
black hole mass. This would be inconsistent with the picture of the
downsizing of the population of black holes since the implied timescales
for black hole mass-doubling through accretion would be independent of
black hole or galaxy mass.  However, their result is based on their
assumption of a uniform ratio of black hole to total galaxy stellar mass
(M$_{\rm BH}/$M$_*$ = 0.002) as a function of mass. As we showed
explicitly in the right-hand panel of Fig.~\ref{mass_bhmass}, in the SDSS
sample at $z \sim$ 0.1 the ratio of black hole to stellar mass increases
by a factor of about 1.5 dex (from 0.0001 to 0.003) over the range in
stellar mass bins studied by Aird et al.\ (centred from log M$_*$ = 9.75
to 11.75).  If the relationship between black hole and galaxy mass is
similar at $z \sim$ 0.1 and 0.6, the Aird et al.\ results would then be
entirely consistent with the black hole down-sizing picture. This is
supported by the conclusions of Kelly \& Shen (2013) based on their
analysis of the SDSS QSO population.
    
\begin{figure}[!t]
\begin{center} 
\psfig{file=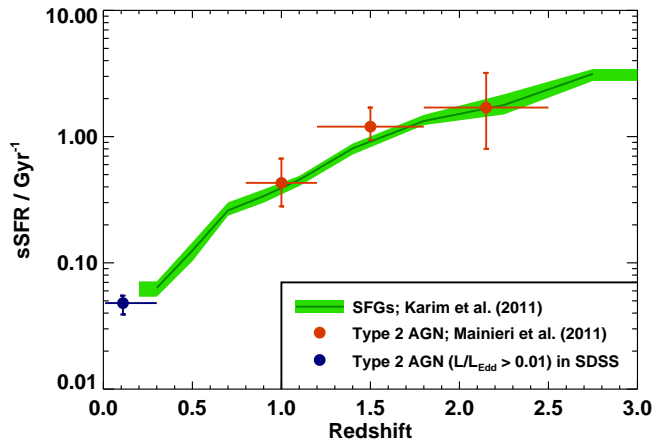,width=9cm,clip=} 
\end{center} 
\caption{\label{mainierifig} The specific star formation rate (sSFR $=$
  SFR/M$_*$) of radiative-mode AGN hosts compared to normal star-forming
  galaxies, as a function of redshift. The green shaded region shows the
  evolution of the sSFR of galaxies long the star-forming main sequence
  derived by Karim et~al.\ (2011). The red points show the sSFR of the
  host galaxies of X-ray selected Type 2 AGN from the XMM-COSMOS field
  derived by Mainieri et~al.\ (2011). The blue point indicates the average
  sSFR of emission-line selected Type-2 AGN with $L/L_{\rm Edd} > 0.01$
  (ie. broadly radiative-mode) from the SDSS. At all redshifts, the sSFR
  of radiative-mode AGN hosts are consistent with those of galaxies on the
  star-forming main sequence.}
\end{figure}

Mainieri et~al.\ (2011) studied the host galaxies of a sample of 142
X-ray-selected Type 2 AGN from the XMM-COSMOS field. The typical redshifts
are in the range 0.8 to 2 and X-ray luminosities are $10^{44}$ to
$10^{45}$ erg/s. They find few detected AGN in host galaxies with stellar
masses less than $10^{10} M_{\odot}$, and typical stellar masses of
$\sim10^{10.5}$ to $10^{11.3}$ M$_{\odot}$ (similar to local Seyfert
galaxies).  A stacked spectrum in the rest-frame optical shows values of
the 4000\AA\ break and high-order Balmer absorption-lines corresponding to
a very young star-forming population.  The masses (and the star-formation
rates derived from Herschel far-IR photometry) imply that the specific
star-formation rates are similar to those of normal galaxies on the
star-forming main sequence at the same redshifts (see
Fig.~\ref{mainierifig}).  Most host galaxies appear to be early-type
galaxies (with a significant bulge component) and only a minority show
either a prominent disk or a disturbed morphology indicative of a major
merger.

Schawinski et~al.\ (2012) investigated the host galaxy properties of a
population of IR-selected heavily obscured AGN at $z \sim2$ with typical
bolometric luminosities of $\sim 10^{12.5}$ L$_{\odot}$.  Their HST
imaging reveals that most have a significant disk component and only a
small fraction appear to be major mergers.  At lower redshifts ($z \sim$
0.7), Cisternas et~al.\ (2011) analyzed HST images of the hosts of AGN with
X-ray luminosities of $10^{43}$ to 10$^{44}$ erg/s. They found that at
least 85\% of the host galaxies have normal undisturbed morphologies
(similar to their control sample of non-AGN galaxies).

Santini et~al.\ (2012) investigated samples of X-ray selected AGN from the
GOODS N and S field and the COSMOS field over the range $z \sim$ 0.5 to
2.5. Only the COSMOS field has sufficient volume to probe the population
of representative AGN.  Based on Herschel far-IR photometry, the
star-formation rates and specific star-formation rates of the AGN hosts in
this field are enhanced by a factor of about 3 relative to the full
population of normal galaxies at the same redshift.  However, the SFRs in
the AGN hosts are consistent with those of the star-forming population of
normal galaxies (i.e.\ the AGN avoid the quiescent galaxy population).  At
$z \sim$ 0.7, Bournaud et~al.\ (2012) found a high incidence rate of
Seyfert-like AGN in a sample of star-forming galaxies with clumpy disks. They argue that the fueling of
AGN at high-z is primarily through inflows triggered by violent disk
instabilities rather than major mergers (Bournaud et~al.\ 2011, Elmegreen
et~al.\ 2008b).

Rosario et~al.\ (2013) have examined the relationship between
star-formation rate (from the far-IR luminosity) and AGN hard X-ray
luminosity over the range from $z \sim 0$ to 2.5. They find that at $z <
0.8$ there is a correlation between the AGN luminosity and star-formation
rate for AGNs with bolometric luminosities above $\sim10^{10.5}$
L$_{\odot}$. At higher redshifts the star-formation rate no longer
correlates with AGN luminosity, but rises with increasing redshift (from
$\sim$ 10 M$_{\odot}$ per year at $z \sim 1$ to $\sim$ 50 M$_{\odot}$ per
year at $z \sim$ 2).  Again, these star-formation rates are consistent
with the properties of typical massive star-forming galaxies at these
epochs. Harrison et~al.\ (2012a) reach similar conclusions.  These results
regarding the complex connection between AGN and star-formation may be
related to findings of Kauffmann \& Heckman (2009) at low-z.  As
summarized above, they found a threshold value for the specific
star-formation rate for effective black hole growth. Once this threshold
is exceeded, black hole growth is switched on, but the rate of growth does
not increase if the star-formation rate rises further.

To summarize: the hosts of the representative population of radiative-mode
AGN out to $z \sim 2$ are mostly structurally normal galaxies with
star-formation rates corresponding to those of the typical star-forming
population of galaxies at that epoch.  These results are consistent with
most other investigations of the less powerful AGN population over the
same range in redshift (e.g.\ Pierce et~al.\ 2007, Gabor et~al.\ 2009,
Hickox et~al.\ 2009, Silverman et~al.\ 2009, 2011, Schawinski
et~al.\ 2011, Kocevski et~al.\ 2012, Mullaney et~al.\ 2012a,b, Aird
et~al.\ 2012, 2013, Bohm et~al.\ 2013, Rosario et~al.\ 2013, Juneau
et~al.\ 2013, Chen et~al.\ 2013).  These results are also consistent with
the more detailed picture we have presented for radiative-mode AGN in the
contemporary universe.

\subsection{Evolution of the Radio Loud AGN Population}

The cosmic evolution of the radio luminosity function has been studied for
many years, since early work on understanding radio source counts
indicated that high luminosity sources undergo substantially stronger
cosmic evolution than low luminosity sources (Longair 1966). Seminal work
by Dunlop \& Peacock (1990) clearly established that the space density of
both flat-spectrum (i.e. beamed) and steep-spectrum powerful radio sources
is 2-3 orders of magnitude higher by $z \sim 2$ than in the contemporary
universe. Beyond this redshift the space density peaks, or declines
slowly. In contrast, the low luminosity radio sources show only a factor
$<2$ increase in space density out to $z \sim$ 0.5 (e.g.\ Sadler
et~al.\ 2007, Donoso et~al.\ 2009). Rigby et~al.\ (2011) characterized
this differential evolution, showing that the redshift at which the radio
source space density peaks is strongly luminosity dependent with low
luminosity sources showing a peak space density at $z<1$.

As discussed earlier, the radio-AGN population comprises both jet-mode
(radiatively inefficient) AGN, and radiative-mode (radio-loud
QSO/Seyfert-like) AGN. Best \& Heckman (2012) looked at the cosmic
evolution of these two populations separately within the narrow redshift
range probed by SDSS, and concluded that they showed different cosmic
evolution. This has been confirmed by Best et~al.\ (in prep), who show
that out to $z \sim$ 0.7 the radiative-mode population shows a factor
$\sim$8 increase in space density, at all radio luminosities. The jet-mode
sources, however, show only a small increase in space density out to
$z=0.5$, followed by a decline thereafter.  These results are easily
understood in terms of our earlier discussions on the fueling of these
AGN. The radiative-mode radio AGN evolve in broadly the same manner as the
upper-mass end of the radiative-mode radio-quiet population, increasing
with redshift as the availability of an abundant supply of cold gas
increases. If the jet-mode AGN are largely associated with cooling hot-gas
haloes, then their evolution will be closer to that of massive galaxies,
and thus slowly declining towards higher redshifts.

These results indicate that the luminosity-dependent cosmic evolution of
the radio luminosity function as a whole is largely driven by the
different cosmic evolution of these two AGN populations.  As a result of
this differential evolution, the dominant radio-AGN population also
changes with redshift. Locally, except at the highest radio luminosities,
the jet-mode population dominates the radio-AGN number counts. By $z > 1$
however, the QSO/Seyfert-like sources begin to dominate at all
(currently-observable) luminosities. It is for this reason that the
properties of radio-loud and radio-quiet AGN appear much more comparable
in the early universe than in the contemporary universe.

Early work to extend the study of the host galaxies of the jet-mode AGN to
higher redshifts finds a picture consistent with that in the contemporary
universe. Tasse et~al.\ (2008) and Donoso et~al.\ (2009) both investigated
the relationship between galaxy stellar mass and radio-AGN prevalence at
$z \sim$ 0.5. They found that, if consideration is restricted to only the
jet-mode AGN, then a steep power-law dependence is found, similar to that
for the contemporary universe (Section~\ref{jethosts}). Simpson et~al.\ (2013)
recently extended that analysis to $z \sim 1$, confirming the same
results.  Tasse et~al.\ (2008) also showed that the jet-mode population
was typically found in galaxy over-densities (groups or clusters). These
results indicate that the feedback balance between radiative cooling and
radio-AGN heating, seen in the contemporary universe, was already in place in
some environments when the universe was half of its current age.

\subsection{AGN Feedback at High-z}

Feedback from AGN in the form of ionizing radiation clearly plays a
fundamental role in ionizing and heating the intergalactic medium over
much of the history of the universe.  Here, we briefly summarize the
evidence regarding the existence and physical properties of bulk outflows
of mass and kinetic energy driven by AGN.  This evidence is based on
detailed investigations of small samples of AGN.

In series of papers, Arav and collaborators (Arav et~al.\ 2008, Korista
et~al.\ 2008, Moe et~al.\ 2009, Dunn et~al.\ 2010, Bautista et~al.\ 2010,
Edmonds et~al.\ 2011, Aoki et~al.\ 2011, Borguet et~al.\ 2013, Arav
et~al.\ 2013) have been able to determine the main physical properties of
outflows in high-z QSOs as traced by UV absorption-lines.  The QSOs have
bolometric luminosities of $\sim 10^{13.3}$ to 10$^{14.7}$ L$_{\odot}$ and
$z \sim$ 0.6 to 3 (they are considerably more luminous than the
representative AGN population).  Of the eight outflows they have analyzed,
four are extremely energetic with kinetic energy fluxes of 10$^{45}$ to
10$^{46}$ erg s$^{-1}$, representing $\sim$1 to 5\% of L$_{\rm bol}$.  The
typical outflows speeds are about 3000 to 8000 km s$^{-1}$ and the
inferred size-scales of the outflows range from several hundred pc to
several kpc.  Similar work by this group on less luminous AGN at lower
redshift has so far found only very weak outflows with kinetic energy
fluxes $\ll$ L$_{\rm bol}$ (Edmonds et~al.\ 2011, Borguet et~al.\ 2012).

Alexander et~al.\ (2010) and Harrison et~al.\ (2012b) have analyzed
spatially-resolved maps of the kinematics of the ionized gas surrounding
nine radio-quiet AGN at $z \sim$ 1.4 to 3.4. The objects were selected
from a parent sample of sub-mm galaxies on the basis of unusually strong
and broad [OIII]5007 emission-lines. Evidence for galactic-scale outflows
was found in the five AGN with bolometric luminosities of about $10^{13}$
L$_{\odot}$ (roughly $L_*$ at these redshifts), but not in the less
luminous cases.  The strong rest-frame far-IR emission in these objects
implies that they have much higher star-formation rates than typical AGN
at these redshifts ($\sim 10^3$ M$_{\odot}$/year compared to $\sim$ 30 to
100 M$_{\odot}$/year respectively).  Thus, it is not entirely clear
whether the outflows are driven by the AGN or the starburst in these
objects.

Using similar techniques, galaxy-scale outflows of ionized gas have been
mapped around a small sample of extremely radio-loud QSO-like AGN at $z
\sim 2$ (Nesvadba et~al.\ 2006, 2008). These outflows appear to be driven
by the mechanical energy carried by the powerful radio jets in these
systems.  At the highest redshifts, Maiolino et~al.\ (2012) have mapped an
outflow from an exceptionally luminous ($L_{\rm bol} \sim 10^{14}$
L$_{\odot}$) QSO at $z \sim$ 6.4 using the [CII] far-IR
emission-line. They estimate an outflow rate of several thousand solar
masses per year at a velocity of-order $10^3$ km/s.  We conclude that
direct evidence for AGN feedback at high-z is limited so far to objects of
exceptionally high luminosity or other unusual characteristics.  Little is
known about whether the representative population of high-z AGN produce
powerful global outflows. Again, this is all consistent with the results
derived in the contemporary universe.

\section{SUMMARY \& IMPLICATIONS}

Let us begin this final section by briefly summarizing what we regard as
the most robust conclusions about the demographics of the populations of
AGN and their host galaxies in the contemporary universe. We will then
consider the degree to which these results may be extended to higher
redshift, and conclude with some speculation about the implications these
conclusions may have for our understanding of the formation and evolution
of galaxies.

\subsection{Some Robust Conclusions}

In the contemporary universe there are two distinct families of AGN. We
have designated them radiative-mode and jet-mode AGN. The radiative-mode
population of AGN is characterized by the production of significant
amounts of radiation with typical bolometric luminosities of 1 to 100\% of
the Eddington limit.  Historically these have been called Seyfert galaxies
or QSOs depending upon their luminosity.  The AGN properties can be
explained within the theoretical context of a radiatively-efficient
geometrically-thin accretion disk.  In the contemporary universe these AGN
are characteristically found to have black hole masses of $10^6$ to $10^8$
M$_{\odot}$ and to reside in host galaxies with typical stellar masses of
a few $\times 10^{10}$ to a few $\times 10^{11}$ M$_{\odot}$ and high
stellar surface mass densities ($\mu_* > 10^{8.5}$ M$_{\odot}$
kpc$^{-2}$).  While these densities are characteristic of early-type disk
galaxies, the AGN hosts (unlike most such galaxies) have significant
amounts of on-going star-formation in the central few kpc.  When examined
at high spatial resolution these inner regions can usually be classified a
pseudo-bulges. Major mergers or strong tidal interactions are not the
dominant fueling mechanism for the central star formation and the AGN.
Internal secular processes for the radial transport of gas are evidently
more important.  This summary pertains to the typical AGN that are the
sites of the majority of the growth of supermassive black holes
today. Things may differ for the rare objects with the highest
luminosities.

The radiative-mode AGN are associated with populations of lower-mass black
holes and lower-mass (pseudo-)bulges that both have volume-averaged
mass-doubling times (via accretion and star-formation respectively) of
roughly the Hubble time. The volume-averaged ratio of star-formation to
black hole growth in these central regions is of-order a thousand (similar
to the ratio of stellar mass and black hole mass in current classical
bulges and elliptical galaxies). These are evidently living systems that
have not yet reached their evolutionary end-point. While, there is
indirect evidence that feedback is limiting the growth of the black hole,
there is little direct evidence for AGN-generated feedback for typical
objects.  On the other hand, there is recent evidence that this may occur
in the rare low-z AGNs having the highest luminosity ($L_{\rm bol} \sim
10^{12-14}$ L$_{\odot}$). The association of AGN with central
star-formation implies that feedback from massive stars will be present
but its effect on the fueling of the AGN has not been well-quantified.

The second, jet-mode, AGN class is characterized by the production of
collimated energetic outflows (jets) but with little detected emitted
radiation (characteristic luminosity below about 1\% of the Eddington
limit).  Historically these have been called (low-excitation) radio
galaxies.  The more luminous LINERs are also likely to be members of this
population. The AGN properties can be explained within the theoretical
context of a radiatively-inefficient accretion flow. The probability of a
galaxy or black hole producing a jet of a given radio luminosity is a
steep function of mass: typical radio galaxies are structurally normal
massive ellipticals ($M_* \sim 10^{11}$ to $10^{12}$ M$_{\odot}$) with
black hole masses greater than $10^8$ M$_{\odot}$. There is no evidence
for substantial recent star-formation in typical radio galaxies (although
there can be for radio-AGN in the BCGs of extreme cooling flow clusters).
There is mostly indirect evidence that the fueling of the AGN is by the
accretion of hot gas seen in X-ray observations.  Feedback produced as the
jets interact with this hot gas can be directly observed. While it is
energetically possible for this feedback to prevent the gas from cooling
rapidly, the details remain uncertain. These AGN are associated with the
populations of higher mass black holes and spheroids that both have
mass-doubling times orders-of-magnitude longer than the Hubble time. These
are effectively dead systems in which the bulk of the star-formation and
black hole growth occurred at high-redshift.

In examining the situation at higher redshift, we have emphasized the
importance of focusing on the representative population of AGN that is
responsible for the creation of the majority of the mass in relic
supermassive black holes today.  These are AGN in the redshift range $z
\sim$ 0.5 to 2.5 and with luminosities near the knee in the luminosity
function.  These representative AGN appear to be hosted by moderately massive
galaxies ($M_* > 10^{10} M_{\odot}$) with star-formation rates that are
typical of the star-forming galaxies at that epoch (they are located on
the star-forming main sequence).  They are typically found in structurally
normal galaxies rather than highly disturbed systems (major mergers).  The
population of radio-loud AGN remains prevalent among massive
galaxies. Evidence for feedback in the form of powerful AGN-driven
outflows is limited so far to objects of either exceptional luminosity or
other unusual properties (very high radio luminosity or star-formation
rate).

All of this sounds similar to the situation in the contemporary
universe. The main differences are quantitative rather than qualitative:
1) The characteristic luminosities of the actively growing black holes are
larger than the present day (by a factor of about three at $z \sim$ 0.5
and $\sim$30 at $z \sim$ 2). The characteristic masses of these black
holes are also correspondingly larger, by factors that are model-dependent
(e.g.\ Shankar et~al.\ 2013). 2) The specific star-formation rates of host
galaxies on the main sequence are higher than the current value by factors
of $\sim$3 at $z =$ 0.5 and $\sim$20 at $z =$ 2 (e.g.\ Elbaz
et~al.\ 2011). Thus, to zeroth-order, the situation at higher-z seems like
a scaled-up version of the contemporary universe.

\begin{figure}[!t]
\begin{center} 
\psfig{file=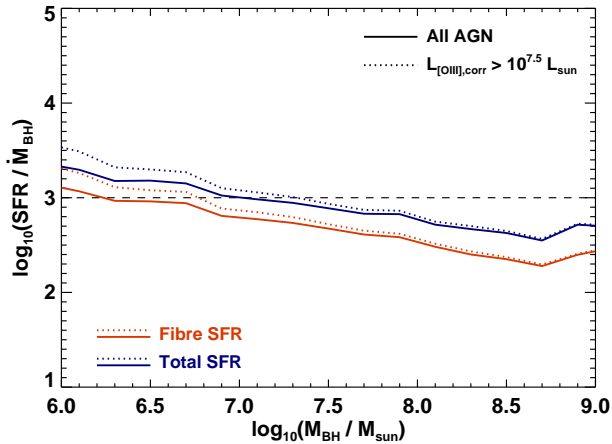,width=9cm,clip=} 
\end{center} 
\caption{\label{h04b} Ratio of the total star formation rate per unit
  volume in galaxies to the total accretion rate per unit volume onto
  black holes as traced by Type 2 AGNs, plotted as a function of the black
  hole mass. The sample studied consists of all galaxies from the SDSS
  main galaxy sample with stellar surface mass density above the
  characteristic value $\mu_* = 10^{8.5}$ M$_{\odot}$ kpc$^{-2}$. Red
  lines show the result if the SFR is calculated within the fiber aperture
  for each galaxy and the blue lines show the result using estimates of
  the total SFR. Dotted lines show results restricted to the AGNs with
  (reddening corrected and star-formation corrected) $L_{\rm [OIII]} >
  10^{7.5}$ L$_{\odot}$. The horizontal dashed line shows the fiducial
  value of stellar to black hole mass in bulges and elliptical
  galaxies. Adapted from Heckman et~al.\ (2004).}
\end{figure}

There is also (what we regard as) persuasive indirect evidence that the
physical processes that link the formation and evolution of galaxies and
black holes at high-z are still at play today.  As we noted in the
introduction, the ratio of the rate of build-up of galaxies through star
formation to the rate of the build-up of black holes via accretion has
been broadly independent of cosmic time over at least the last $\sim
10$\,Gyrs with a value of $\sim$1500 (Shankar et~al.\ 2009, 2013, Hopkins
\& Beacom 2006).  This is hard to understand unless the underlying physics
has remained basically the same over this time. See Heckman et~al.\ (2004)
and Mullaney et~al.\ (2012b) for further discussion of this
argument. Indeed, as Fig.~\ref{h04b} shows, this ratio applies not just to
the total rates, but holds over all values for the mass of the SMBH.

\subsection{ Implications and Some Speculation} 

Our current understanding of the formation and evolution of galaxies is
that the build-up of the stellar content of galaxies via star-formation is
driven primarily by the accretion of gas from the cosmic web. At least
since $z = 2$ or 3, the bulk of the star formation occurs in galaxies
lying along a strongly evolving star-forming main sequence (a tight
relationship between SFR and stellar mass). While major mergers of
gas-rich systems do occur, and lead to strong bursts of star-formation,
they are responsible for only a minority (of-order 10\%) of the total
cosmic star-formation. This picture is predicted by numerical simulations
and is consistent with observations of the evolving population of
star-forming galaxies (e.g.\ Dekel et~al.\ 2009, Bouche et~al.\ 2010,
Elbaz et~al.\ 2011, Wuyts et~al.\ 2011b, Whitaker et~al.\ 2012, Lilly
et~al.\ 2013).

Empirically, the accretion-driven mass-growth of the population of
supermassive black holes is related to the growth (via star-formation) of
the inner region of the host galaxy.  Evidently, both rates track the rate
at which gas is accreted by the galaxy (with relative efficiencies that
differ by about three orders-of-magnitude).  Mergers are not the dominant
mechanism by which gas flows to these inner regions, and hence are not the
dominant mechanism for growing the population of supermassive black holes.
The gas-flows responsible for this may instead be driven by rather slow
secular processes in the contemporary universe (e.g.\ Kormendy \&
Kennicutt 2004, Athanassoula 2008) or more violent and rapid instabilities
in clumpy gas-rich disks at high-z (e.g.\ Genzel et~al.\ 2013, Bournaud
2010, Elmegreen et~al.\ 2008a, Dekel et~al.\ 2009).  This picture would
imply that the representative population of actively growing supermassive
black hole (AGN) would live at the centers of disk-like structures (rather
than ellipticals or classical bulges).

This simple picture leaves out one critical fact: we know from
observations that the effect of this co-evolution over cosmic time must
ultimately place the relic (dead) black holes in dynamically hot
structures (ellipticals and classical bulges) that contain little cold gas
and correspondingly little star-formation.  In the classical picture for
the co-evolution of galaxies and black holes this is explained
economically by positing that the supermassive black hole grew in the
aftermath of a major merger that created an elliptical/classical-bulge and
that feedback from the rapidly growing black hole blew away the remaining
gas (e.g.\ Hopkins et~al.\ 2006, 2008, Di Matteo et~al.\ 2008). How can
this be explained in the more sedate scenario we are advocating?  Let us
try a two-part answer.

First, while low-mass bulges can be built through secular processes and
disk instabilities (e.g.\ Hopkins et~al.\ 2012, Athanassoula 2008,
Bournaud et~al.\ 2011), it appears that major mergers are the only way to
build the population of massive classical bulges and elliptical
galaxies. In this case, the sedate scenario would imply that these major
mergers only modestly augment the masses of gas accreted by the black hole
and of gas turned into stars. The $M_{\rm BH}$ vs. $\sigma$ relationship
would have to be essentially pre-built before a major merger takes place
(consistent with the recent simulations by Angl{\'e}s-Alc{\'a}zar
et~al.\ 2013). A major merger mostly rearranges the stars in the
pre-existing disks (and their black-hole-containing pseudo-bulges) into a
classical bulge or elliptical galaxy and a more massive (merged) black
hole. Subsequent dry merging between classical bulges or ellipticals can
further populate the upper end of the $M_{\rm BH}$ {\it vs.}\ $\sigma$
relation (e.g.\ Shankar et~al.\ 2010, Kormendy \& Ho 2013).

The second part of the answer requires an explanation for the dry-up of
the reservoir of cold gas and the subsequent quenching of
star-formation. In current models, this shut-down in star-formation in a
typical massive galaxy is connected to the transition from an accretion
mode in which gas reaches the galaxy in a cold form (gas temperature much
less than the virial temperature) to a mode where the accreted gas is
heated to roughly the virial temperature upon arrival and then slowly
cools (via radiation) and is gradually accreted.  This transition is
predicted to occur when the galaxy halo mass grows via mergers and exceeds
some critical value. Dekel et~al.\ (2009) argue that this critical mass
corresponds to the observed transition between high-mass mostly-quiescent
galaxy population and lower mass galaxy population that defines the
star-forming main sequence (see also Bouche et~al.\ 2010, Lilly
et~al.\ 2013, Mutch et~al.\ 2013). While current models still need to
incorporate AGN feedback to quench star formation, it is possible that
this is a result of the difficulty in incorporating the relevant physical
processes with the necessary fidelity in current state-of-the-art
numerical simulations. In this case, late feedback from AGN might only be
required to be strong enough to suppress the late cooling flows of hot gas
and keep the quiescent galaxy suitably red and dead.  This may be
accomplished by the low luminosity radio-AGN, through the self-regulating
feedback loop we have described in Section 6.2 above (the so-called
maintenance mode of AGN feedback).

As we have emphasized, this sedate model is designed to match the detailed
picture we have of the contemporary universe and the emerging picture we
have of the representative population of AGN and their host galaxies at
higher redshift.  What about the more extreme cases? More specifically,
one might ask about the physical mechanism(s) responsible for imposing a
characteristic AGN luminosity at a given epoch ($L_*(z)$) above which the
number density of AGN steeply falls. In part this must reflects the knee
in the black hole mass function. However, there is also evidence that the
most powerful AGN (at both high-z and low-z) are different from their less
luminous kin.  There are indications that the fraction of AGN host
galaxies that are major mergers increases at the highest AGN luminosities
(e.g.\ Liu et~al.\ 2009, Mainieri et~al.\ 2011, Treister
et~al.\ 2013). This is at least qualitatively consistent with the models
presented by Hopkins \& Hernquist (2009), Draper \& Ballantyne (2012),
Shankar et al.\ (2012), and Hopkins et~al.\ (2013), which partition the
amount of black hole growth associated with mergers and secular processes.
We have also seen that the most luminous AGN at low and high redshift can
drive powerful outflows that could affect the fueling of the black hole
and even the bulk properties of the gas in the host galaxy and its halo.
It is tempting to speculate that the knee of the AGN luminosity function
marks the transition from a sedate model of fueling by accretion and
subsequent internal processes with weak/no AGN feedback to a dramatic
model of fueling by mergers with strong AGN feedback.

In this review we have stressed the importance of large surveys of the
contemporary universe in revealing the way in which galaxies and their
super-massive black holes co-evolve(d). We fully expect that the next
generation of large ground- and space-based surveys will extend this type
of robust and sophisticated understanding to the era when the bulk of the
mass in galaxies and black holes was assembled.

\vspace*{0.4cm}

{\large
{\bf LITERATURE CITED}}\\[3mm]
Aalto S, Garcia-Burillo S, Muller S et~al. 2012. A\&A, 537: 44-51\\
Aird J, Coil A, Moustakas J et~al. 2012. ApJ 749: 90-112\\
Aird J, Coil A, Moustakas J et~al. 2013 ApJ 775: 41-52\\
Alexander EM, Swinbank AM, Smail I, McDermid R, Nesvadba NPH. 2010. MNRAS 402: 2211-2220\\
Allen SW, Dunn RJH, Fabian AC, Taylor GB, Reynolds CS. 2006. MNRAS 372: 21-30\\
Alonso MS, Lambas DG, Tissera, P, Coldwell G. 2007. MNRAS 375: 1017-1024\\
Anderson SF, Voges W, Margon B et~al. 2003. AJ 126: 2209-2229\\
Angl{\'e}s-Alc{\'a}zar D,{\"O}zel F, Dav{\'e} R et~al. 2013. arXiv:1309.5963\\
Antonucci, R. 1992. ARAA 31: 473-521\\
Aoki K, Oyabu S,Dunn J et~al. 2011. PASJ 63: 457-467\\
Arav N, Moe M, Constantini E et~al. 2008. ApJ 681: 954-964\\
Arav N, Borguet B, Chamberlain C, Edmonds D, Danforth C. 2013. arXiv: 1395.2181\\
Athanassoula E 2005. MNRAS 358: 1477-1488\\
Athanassoula E 2008. MNRAS 390: L69-72\\
Baade W, Minkowski R. 1954. ApJ 119: 206-214\\
Bailey ME. 1980. MNRAS 191: 195-206\\
Baldry IK, Glazebrook K, Brinkmann J et~al. 2004. ApJ 600: 681-694\\
Baldwin JA, Phillips MM, Terlevich R. 1981. PASP 93: 5-19\\
Balick B, Heckman TM. 1982. ARA\&A 20: 431-468\\
Balogh ML, Morris SL, Yee HKC, Carlberg R, Ellingson E. 1999. MNRAS 307: 463-479\\
Barthel PD, Arnaud KA. 1996. MNRAS 283: L45\\
Baumgartner WH, Tueller J, Markwardt CB et~al. 2013. ApJS 207: 19-30\\
Bautista MA, Dunn JP, Arav N et~al. 2010. ApJ 713: 25-31\\
Becker RH, White RL, Hefland DJ. 1995. Ap. J. 450: 559-577\\
Bell AR. 1978. MNRAS 182: 443-455\\
Bell EF, de Jong RS. 2001. ApJ 550: 212-229\\
Bentz MC, Denney KD, Grier CJ et~al. 2013. ApJ 767:149-175\\
Best PN. 2009. Astronomische Nachrichten 330: 184-189\\
Best PN. 2004. MNRAS 351: 70-82\\
Best PN, Kauffmann G, Heckman TM, Ivezi{\'c} \v{Z}. 2005a. MNRAS 362: 9-24\\
Best PN, Kauffmann G, Heckman TM  et~al. 2005b. MNRAS 362: 25-40\\
Best PN, Kaiser CR, Heckman TM, Kauffmann G. 2006. MNRAS 368: L67-L71\\
Best PN, von der Linden A, Kauffmann G, Heckman TM, Kaiser CR. 2007. MNRAS 379: 894-908\\
Best PN, Heckman TM. 2012. MNRAS 421: 1569-1582\\
B{\^ i}rzan L, Rafferty DA, McNamara BR, Wise MW, Nulsen PEJ. 2004. Ap.J. 607: 800-809\\
B{\^ i}rzan L, McNamara BR, Nulsen PEJ, Carilli CL, Wise MW. 2008. Ap.J. 686: 859-890\\
Blandford RD, Payne DG. 1982. MNRAS 199: 883-903\\
Blandford RD, Znajek RL. 1977. MNRAS 179: 433-456\\
Blanton MR, Hogg DW, Bahcall NA et~al. 2003. ApJ 594: 186-207\\
Blundell KM, Rawlings S. 2000. Astron. J. 119: 1111-1122\\
Bock DC-J, Large MI, Sadler EM. 1999. Astron. J. 117: 1578-1593\\
B{\"o}hringer H, Voges W, Fabian AC, Edge AC, Neumann DM. 1993. MNRAS 264: L25-28\\
Bondi H. 1952. MNRAS 112: 195-204\\
Borthakur S, Heckman T, Strickland D, Wild V, Schiminovich, D. 2013. ApJ 768: 18-37\\
Bouche N, Dekel A, Genzel R et~al. 2010. ApJ 718: 1001-1018\\
Borguet B, Edmonds D, Arav, N, Dunn, J, Kriss GA. 2012. ApJ 751: 107-121\\
Borguet B, Arav N, Edmonds D, Chamberlain, C. 2013. ApJ 762: 49-61\\
Bournaud F, Dekel A, Teyssier R et~al. 2011. ApJ 741: L33-38\\
Bournaud F, Juneau S, Le Floc?h E et~al. 2012. ApJ 757: 81-100\\
Bower RG, Benson AJ, Malbon R et~al. 2006. MNRAS 370: 645-655\\
Brinchmann J, Charlot S, White S et~al. 2004. MNRAS 351: 1151-1179\\
Brandt WN, Hasinger, G 2005. ARA\&A 43: 827-859\\
Brown TM, Ferguson HC, Stanford SA, Deharveng JM. 1998. ApJ 504: 113-138 \\
Brown TM, Smith E, Ferguson H et~al. 2008. ApJ 682: 319-335\\
Bruzual AG. 1983. ApJ 273: 105-127\\
Bruzual G, Charlot S. 2003. MNRAS 344: 1000-1028\\
Burbidge EM, Burbidge GR, Prendergast KH. 1959. ApJ 130: 26-45\\
Burns JO. 1990. Astron. J. 99:14-30\\
Cao X. 2010. ApJ 725: 388-393\\
Cao X, Rawlings S. 2004. MNRAS 349: 1419-1427\\
Capetti A, Baldi RD. 2011. A\&A 529: 126-130\\
Cattaneo A, Best PN. 2009. MNRAS 395:518-23\\
Cattaneo A, Faber SM, Binney J et~al. 2009. Nature 460:213-19\\
Cavagnolo KW, Donahue M, Voit GM, Sun M. 2008. Ap. J. Lett. 683:L107-10\\
Cavagnolo KW, McNamara BR, Nulsen PEJ et~al. Ap. J. 720:1066-72 \\
Churazov E, Br{\"u}ggen M, Kaiser CR, B{\"o}hringer H, Forman W. 2001. Ap. J. 554:261-73\\
Churazov E, Sazonov S, Sunyaev R et~al. 2005. MNRAS 363:L91-95\\
Cicone C, Feruglio C, Maiiolino R et~al. 2012. A\&A 543: 99-105\\
Cicone C, Maiolino R, Sturm E et~al. 2013. arXiv1311.2595\\
Cid Fernandes R, Heckman T, Schmitt H et~al. 2001. ApJ 558:81-108\\
Cid Fernandes R, Gonz{\'a}lez Delgado RM, Schmitt H et~al. 2004. ApJ 605:
105-126\\
Cid Fernandes R, Gu Q, Melnick J et~al. 2004. MNRAS 355: 273-296\\
Cid Fernandes R, Mateus A, Sodre L, Stasinska G, Gomes JM. 2005. MNRAS 358: 363-378\\
Cid Fernandes R, Stasi{\'n}ska G, Mateus A, Vale Asari N. 2011. MNRAS 413: 1687:1699\\
Ciotti L, Ostriker JP. 1997. ApJ 487: L105-108\\
Ciotti L, Ostriker JP, Proga D. 2010. ApJ 717: 707-723\\
Cisternas M, Jahnke K, Inskip KJ et~al. 2011. ApJ 726:57-70\\
Cisternas M Gadotti DA, Knapen JH et~al. 2013. ApJ 776:50-64\\
Chen CT, Hickox RC, Alberts S et~al. 2013. ApJ 773: 3-11\\
Chen YM, Tremonti CA, Heckman TM et~al. 2010. AJ 140: 445-461\\
Chen YM, Kauffmann G, Heckman TM et~al. 2013. MNRAS 429: 2643-2654\\
Chevalier RA, Clegg AW. 1985. Nature 317:44-45\\
Colless M, Dalton G, Maddox S et~al. 2001. MNRAS 328: 1039-1063\\
Condon JJ. 1989. Ap. J. 338: 13-23\\
Condon JJ. 1992. Annu. Rev. Astron. Astrophys. 30: 575-611\\
Condon JJ, Cotton WD, Greisen EW et~al. 1998, Astron. J. 115: 1693-1716\\
Conselice CJ, Bershady MA, Jangren A. 2000. ApJ 529:886-910\\
Cooray A, Sheth R. 2002. Phys. Reports 372: 1-129 \\
Crenshaw DN, Kraemer SB, George IM. 2003. ARA\&A 41: 117-167\\
Croft S, de Vries W, Becker RH. 2007. Ap.J. 667: L13-L167\\
Croom SM, Boyle BJ, Shanks T et~al. 2005. MNRAS 356: 415-438\\
Croom SM, Richards GT, Shanks T et~al. 2009. MNRAS 399: 1755-1772\\
Croston JH, Hardcastle MJ, Birkinshaw M. 2005. MNRAS 357:279-294\\
Croton DJ, Springel V, White S et~al. 2006. MNRAS 365: 11-28\\
Daly RA, Sprinkle TB, O'Dea CP, Kharb P, Baum SA. 2012. MNRAS 423: 2498:2502\\
David LP, Nulsen PEJ, McNamara BR et~al. 2001. Ap.J. 557: 546-559\\
David LP, Durisen RH, Cohn HN. 1987. ApJ 316: 505-516\\
Davies RI, M{\"u}ller S{\'a}nchez F, Genzel R et~al. 2007. ApJ 671:
1388-1412\\
Dekel A, Sari R, Ceverino, D. 2009. ApJ 703:785-801\\
de Vries WH, Becker RH, White RL. 2006. Astron. J. 131: 666-679\\
Diamond-Stanic AM, Rieke GH, Rigby JR. 2009. ApJ 699: 623-631\\
Diamond-Stanic AM, Rieke GH. 2012. ApJ 746: 168-181\\
Di Matteo T, Springel V, Hernquist, L. 2005. Nature 433: 604-607 \\
Di Matteo T, Colberg J, Springel V, Hernquist L, Sijacki D. 2008. ApJ 636: 33-53\\
Dong R, Greene JE, Ho LC. 2012. ApJ 761: 73-84\\
Donoso E, Best PN, Kauffmann G. 2009. MNRAS 392: 617-629\\
Donoso E, Li C, Kauffmann G, Best PN, Heckman TM. 2010. MNRAS 407: 1078-1089\\
Donoso E, Yan L, Tsai C et~al. 2012. ApJ 748: 80-90\\
Draper AR, Ballantyne DR 2012. ApJ 751: 72-84\\
Dressler A, Gunn JE. 1983. ApJ 270: 7-19\\
Dunlop JS, Peacock JA. 1990. MNRAS 247: 19-42\\
Dunn JP, Bautista M, Arav N et~al. 2010. ApJ 709: 611-631\\
Dunn RJH, Fabian AC, Taylor GB. 2005. MNRAS 364: 1343-1353\\
Dunn RJH, Fabian AC. 2006. MNRAS 373:959-71\\
Edmonds D, Borguet B, Arav N et~al. 2011. ApJ 739:7-18\\
Elbaz D, Dickinson M, Hwang HS et~al. 2011. A\&A 533: 119-144\\
Ellison SL, Patton DR, Mendel JT, Scudder JM. 2011. MNRAS 418: 2043-2053\\
Elmegreen BG, Bournaud F, Elmegreen DM. 2008a. ApJ 688: 67-77\\
Elmegreen BG, Bournaud F, Elmegreen DM. 2008b. ApJ 684: 829-834\\
Fabello S, Kauffmann G, Catinella B et~al. 2011. MNRAS 416: 1739-1744\\
Fabian AC. 1994. Annu. Rev. Astron. Astrophys. 32:277-318\\
Fabian AC. 2012. Annu. Rev. Astron. Astrophys. 50:455-89\\
Fabian AC, Sanders JS, Allen SW, Crawford CS, Iwasawa K, et~al. 2003a. MNRAS 344:L43-47\\
Fabian AC, Reynolds CS, Taylor GB, Dunn RJH. 2005. MNRAS 363:891-96\\
Fabian AC, Sanders JS, Taylor GB, Allen SW, Crawford CS, et~al. 2006. MNRAS 366:417-28\\
Falcke H, K{\"o}rding E, Markoff S. 2004. Astron. Astrophys. 414: 895-903\\
Fanaroff BL, Riley JM. 1974. MNRAS 167:P31\\
Fardal MA, Katz N, Weinberg DH, Dav{\'e} R. 2007. MNRAS. 379: 985-1002\\
Fender RP, Belloni TM, Gallo E. 2004. MNRAS 355: 1105-1118\\
Ferrarase L, Merritt D. 2000. ApJ 539: L9-12\\
Filippenko A, Ho L, \& Sargent W. 1993. ApJ 410: L75-78\\
Fiore F, Puccetti S, Grazian A et~al. 2012. A\&A 537: 16-37\\
Fischer J, Sturm E, Gonz{\'a}lez-Alfonso E et~al. 2010. A\&A 518: L41-44\\
Fisher DB, Drory N. 2011. ApJ 733: L47-51\\
Gadotti DA. 2009. MNRAS 393: 1531-1552\\
Gallo E, Fender RP, Pooley GG. 2003. MNRAS 344: 60-72\\
Gaspari M, Ruszkowski M, Oh SP. 2013. MNRAS 432: 3401-3422\\
Gelbord JM. 2003. PhD Thesis: Johns Hopkins University\\
Gendre MA, Best PN, Wall JV. 2010. MNRAS 404:1719-1732\\
Gendre MA, Best PN, Wall JV, Ker LM. 2013. MNRAS 430: 3086-3101\\
Genzel R, F{\"o}rster Schreiber, NM, Lang P et~al. 2013. arXiv: 1310.3838\\
Gebhardt K, Bender R, Bower G et~al. 2000. ApJ 539: L13-16\\
Giacconi R, Gursky H, \& Paolini F. 1962. Phys Rev L 9: 439-443\\
Gisler GR. 1978. MNRAS 183: 633-643\\
Gonz{\'a}lez-Alfonso E, Fischer J, Graci{\'a}-Carpio J et~al. 2013. arXiv: 1310.3074\\
Gonz{\'a}lez Delgado RM, Cid Fernandes R, P{\'e}rez E et~al. 2004. ApJ 606:127-143\\
Gonz{\'a}alez Delgado RM, P{\'e}rez, E, Cid Fernandes, R, Schmitt, H. 2008. AJ 135: 747-765\\
Goto T. 2006. MNRAS 369: 1765-1772\\
Goulding AD, Alexander, DM. 2009. MNRAS 398: 1165-1193\\
Graham AW, Scott N. 2013. ApJ 764: 151-169 \\
Greene JE, Ho LC. 2007a. ApJ 670: 92-104\\
Greene JE, Ho LC. 2007b. ApJ 667: 131-148\\
Greene JE, Zakamska, N, Ho, LC, Barth AJ. 2011. ApJ 732: 9-28\\
Grier CJ, Martini P, Watson LC et~al. 2013. ApJ 773: 90-103\\
Grimes JP, Heckman T, Stickland D, Ptak A. 2005. ApJ 628: 187-204\\
Gultekin K, Richstone DO, Gebhardt K et~al. 2009. ApJ 698: 198-221\\
G\"urkan G, Hardcastle MJ, Jarvis MJ. 2013. arXiv:1308.4843
Haines CP, Pereira MJ, Sanderson AJR et~al. 2012. Ap.J. 754: 97:113\\
Hainline KN, Hickox R, Greene JE, Myers AD, Zakamska NL. 2013. ApJ 774: 145-153\\
Hao L, Strauss MA, Tremonti CA et~al. 2005a. AJ 129: 1783-1794\\
Hao L, Strauss MA, Fan X et~al. 2005b. AJ 129: 1795-1808\\
Hardcastle MJ, Evans DA, Croston JH. 2007. MNRAS 376: 1849-1856\\
Haring N, Rix HW. 2004. ApJ 604: L89-92\\
Harrison CM, Alexander DM, Mullaney JR et~al. 2012a. ApJ 760: L15-19\\
Harrison CM, Alexander DM, Swinbank AM et~al. 2012b. MNRAS 426:1073-1096\\
Hasinger G, Miyaji T, Schmidt M. 2005. A\&A 441: 417-434\\
Heckman TM. 1980. A\&A 87: 152-164\\
Heckman TM, Balick B, Sullivan WTIII. 1978. ApJ 224: 745-760\\
Heckman TM, Blitz L, Wilson AS, Armus L, Miley GK. 1989. ApJ 342: 735-758\\
Heckman TM, Armus, L, Miley GK. 1990. ApJS 74: 833-868\\
Heckman TM, Kauffmann G, Brinchmann J et~al. 2004. ApJ 613: 109-118\\
Heckman TM, Ptak A, Hornschemeier A, Kauffmann G. 2005. ApJ 634:161-168\\
Heywood I, Blundell KM, Rawlings S. 2007. MNRAS 381: 1093-1102\\
Hicks EKS, Davies RI, Maciejewski W et~al. 2013. ApJ 768: 107-123\\
Hickox RC, Jones C, Forman W et~al. 2009. ApJ 696: 891-919\\
Hillel S, Soker N. 2013. MNRAS 430: 1970-1975 \\
Hine RG, Longair MS. 1979. MNRAS 188: 111-130\\
Hlavacek-Larrondo J, Fabian AC. 2011. MNRAS 413:313-21\\
Ho LC. 2005. Astrophys.Space Sci. 300: 219-225\\
Ho LC. 2008. Annu. Rev. Astron. Astrophys. 46: 475-539\\
Ho LC, Darling J, Greene JE. 2008. ApJ 681: 128-140\\
Hopkins AM, Beacom JF, 2006. ApJ 651: 142-154\\
Hopkins PF, Hernquist L, Cox TJ et~al. 2006. ApJS 163: 1-49\\
Hopkins PF, Richards GT, Hernquist L. 2007. ApJ 654: 731-753\\
Hopkins PF, Hernquist L, Cox TJ, Keres D. 2008. ApJS 175: 356-389\\
Hopkins PF, Hernquist L. 2009. ApJ 694: 599-609\\
Hopkins PF, Quataert E. 2011. MNRAS 415: 1027-1050\\
Hopkins PF, Keres, D, Murray N, Quataert E, Hernquist L. 2012. MNRAS 427: 968-978\\
Hopkins PF, Kocevski DD, Bundy K. 2013. ApJ 776: 48-65\\
Hu EM, Cowie LL, Wang Z. 1985. Ap.J.Supp. 59: 447-498\\
Hubble EP. 1925. Pop Astr 33: 252-255\\
Hwang HS, Park C, Elbaz D, Choi Y-Y. 2012. Astron. Astrophys. 538: A15-A29\\
Ibar E, Cirasuolo M, Ivison RJ et~al. 2008. MNRAS 386: 953-962\\
Ishibashi W, Fabian AC 2012. MNRAS 427: 2998-3005\\
Ivezi{\'c} \v{Z}, Menou K, Knapp GR et~al. 2002. Astron. J. 124: 2364-2400\\
Janssen RMJ, R{\"o}ttgering HJA, Best PN, Brinchmann J. 2012. Astron. Astrophys. 541: A62-A68\\
Jiang YF, Greene JE, Ho LC, Xiao T, Barth AJ. 2011. ApJ 742: 68-85\\
Juneau S, Dickinson, M, Bournaud F et~al. 2013. ApJ 764: 176-194\\
Karim A, Schinnerer E., Mart{\'{\i}}nez-Sansigre A. 2011. ApJ 730: 61-91\\ 
Kaspi S, Smith PS, Netzer H et~al. 2000. ApJ 533: 631-649\\
Kauffmann G, Haehnelt M. 2000. MNRAS 311: 576-588\\
Kauffmann G, Heckman TM, Tremonti C et~al. 2003a. MNRAS 346: 1055-1077\\
Kauffmann G, Heckman TM, White SDM et~al. 2003b. MNRAS 341: 54-69\\
Kauffmann G, White SDM, Heckman TM et~al. 2004. MNRAS 353:713\\
Kauffmann G, Heckman TM, Best PN. 2008. MNRAS 384: 953-971\\
Kauffmann G, Heckman TM. 2009. MNRAS 397:135-47\\
Kelly BC, Merloni A. 2012. AdAst 2012: 7-34\\
Kelly BC, Shen Y. 2013. ApJ 764: 45-69\\
Kerr R. 1963. Phys Rev L 11: 237-238\\
Kewley LJ, Dopita MA, Sutherland RS, Heisler CA, Trevena J. 2001. ApJ 556: 121-140\\
Kewley LJ, Groves B, Kauffmann G, Heckman T. 2006. MNRAS 372: 961-976\\
Kocevski DD, Faber SM, Mozena M et~al. 2012. ApJ 744: 148-156\\
Kollmeier JA, Onken CA, Kochanek CS et~al. 2006. Ap. J. 648:128-39\\
Korista KT, Bautista MA, Arav N et~al. 2008. ApJ 688: 108-115\\
Kormendy J, Kennicutt RC. 2004. ARA\&A 42: 603-683\\
Kormendy J, Ho LC. 2013. ARA\&A 51: 511-653\\
Koss M, Mushotzky R, Veilluex S et~al. 2011. ApJ 739: 57-76\\
Koss M, Mushotzky R, Treister E et~al. 2012. ApJ 746 L22-28\\
Krolik, J. 1999. `Active Galactic Nuclei'. Princeton NJ: Princeton University Press\\
Krug HB, Rupke DSN, Veilleux S. 2010. ApJ 708: 1145-1161\\
Lacy M, Storrie-Lombardi LJ, Sajina A et~al. 2004. ApJS 154: 166-169\\
Lackner  CN, Gunn JE. 2012. MNRAS 421: 2277-2302\\
Lagos CDP, Lacey CG, Baugh CM, Bower R, Benson A. 2011. MNRAS 416: 1566-1584\\
Laing RA, Jenkins CR, Wall JV, Unger SW. 1994. ASP Conf. Ser. 54: 201-208\\
LaMassa S, Heckman TM, Ptak A et~al. 2010. ApJ 720: 786-810\\
LaMassa S, Heckman TM, Ptak A et~al. 2012. ApJ 758: 1-27\\
LaMassa S, Heckman TM, Ptak A, Urry CM. 2013. ApJ 765: 33-38\\
Larson RB, Tinsley BM. 1978. ApJ 219: 46-59\\
Lee G-H, Woo J-H, Lee MG et~al. 2012. ApJ 750: 141-152\\
Leitherer C, Schaerer D, Goldader J et~al. 1999. ApJS 123: 3-40\\
Li C, Kauffmann G, Wang L et~al. 2006. MNRAS 373: 457-468\\
Li C, Kauffmann G, Heckman TM, White SDM, Jing YP. 2008a. MNRAS 385: 1903-1914\\
Li C, Kauffmann G, Heckman TM, White SDM, Jing YP. 2008b. MNRAS 385: 1915-1922\\
Lilly SJ, Carollo CM, Pipino A, Renzini A, Peng Y. 2013. ApJ 772: 119-137\\
Lin Y-T, Shen Y, Strauss MA, Richards GT, Lunnan R. 2010. Ap.J. 723: 1119-1138\\
Lintott C, Schawinski K, Bamford S et~al. 2011. MNRAS 410: 166-178\\
Liu X, Zakamska NL, Greene JE et~al. 2009. ApJ 702: 1098-1117\\
Liu G, Zakamska NL, Greene JE, Nesvadba NPH, Liu X. 2013a. MNRAS 430: 2327-2345\\
Liu G, Zakamska NL, Greene JE, Nesvadba NPH, Liu X. 2013b. MNRAS 436: 2576-2597\\
Liu X, Shen Y, Strauss M. 2012. ApJ 745: 94-109\\
Longair MS. 1966. MNRAS 133: 421-436\\
Longair MS, Seldner M. 1979. MNRAS 189: 433-453\\
Lupton RH, Ivezic Z, Gunn JE et~al. 2002. SPIE 4836: 350-356\\
Maccarone TJ. 2003. Astron. Astrophys. 409: 697-706\\
Machalski J, Condon JJ. 1999. Ap.J.Supp. 123: 41-78\\
Machalski J, Godlowski W. 2000. Astron. Astrophys. 360: 463-471\\
Magliocchetti M, Br{\"u}ggen M. 2007. MNRAS 379: 260-274\\
Magliocchetti M, Maddox SJ, Lahav O, Wall JV. 1998. MNRAS 300: 257-268\\
Mahadevan R. 1997. Ap.J. 477: 585-601\\
Maiolino R, Ruiz M, Rieke, GH, Papadopoulos P.1997. ApJ 485: 552-569\\
Maiolino R, Gallerani S, Neri R et~al. 2012. MNRAS 425: L66-70\\
Mainieri V, Bongiorno , Merloni A et~al. 2011. A\&A 535: 80-106\\
Malkan M, Gorjian V, Tam, R. 1998 ApJS 117: 25-88\\
Mandelbaum R, Li C, Kauffmann G, White SDM. 2009. MNRAS 393: 377-392\\
Mannering EJA, Worrall DM, Birkinshaw M. 2011. MNRAS 416: 2869-2881\\
Marconi A, Hunt LK. 2003. ApJ 589: L21-24\\
Marconi A, Risaliti G, Gilli R et~al. 2004. MNRAS 351: 169-185\\
Martin DC, Fanson, J, Schiminovich D et~al. 2005. ApJ 619: L1-6\\
Mart{\'{\i}}nez-Sansigre A, Rawlings S. 2011. MNRAS 414: 1937-1964\\
Martini P, Kelson DD, Kim E, Mulchaey JS, Athey AA. 2006. Ap.J. 644: 116-132\\
Martini P, Dicken D, Storchi-Bergmann T. 2013. ApJ 766: 121-143\\
Mauch T, Sadler EM. 2007. MNRAS 375: 931-950\\
McConnell NJ, Ma CP 2013. ApJ 764: 184-197\\
McMahon RG, White RL, Helfand DJ, Becker RH. 2002. Ap.J.Supp. 143: 1-23\\
McNamara BR, Rohanizadegan M, Nulsen PEJ. 2011. ApJ. 727: 39-46\\
McNamara BR, Nulsen PEJ. 2007. Annu. Rev. Astron. Astrophys. 45: 117-175\\
Meier DL. 1999. Ap. J. 522:753-766\\
Meier DL. 2001. Ap. J. 548:L9-L12\\
Merloni A, Heinz S, Di Matteo T. 2003. MNRAS 345:1057-76\\
Merloni A, Heinz S. 2007. MNRAS 381: 589-601\\
Merloni A, Heinz S. 2008. MNRAS 388: 1011-1030\\
Miley G. 1980. Annu. Rev. Astron. Astrophys. 18: 165-218\\
Miley GK, Neugebauer G, Soifer BT. 1985. ApJ 293: L11-14\\
Miller CJ, Nichol RC, G{\'o}mez PL, Hopkins AM, Bernardi M. 2003. Ap.J. 597: 142-156\\
Miller JM, Raymond J, Reynolds CS, Fabian AC, Kallman TR, Homan J. 2008. Ap. J. 680:1359-77\\
Mirabel IF, Wilson AS. 1984. ApJ 277: 92-105\\
Moe M, Arav N, Bautista MA, Korista KT. 2009. ApJ 706: 525-534\\
Mullaney JR, Pannella M, Daddi E et~al. 2012a. MNRAS 419:95-115\\
Mullaney JR, Daddi E, B{\'e}thermin M et~al. 2012b. ApJ 753: L30-34\\
Murphy KD, Yaqoob T. 2009. ApJ 701: 635-641\\
Murray N, Menard B, Thompson TA. 2011. ApJ 735:66-77\\
Mutch SJ, Croton DJ, Poole GB. 2013. MNRAS 435: 2445-2459\\
Nagar NM, Falcke H, Wilson AS. 2005. Ap. J. 435:521-543\\
Narayan R. 2002. `Lighthouses of the Universe' eds. M. Gilfanov, R. Sunyaev, E Churazov, pp. 405-429, Berlin,Heidelberg: Springer-Verlag\\
Narayan R. 2005. Astrophys.Space Sci. 300: 177-188\\
Narayan R, Yi I. 1994. Ap. J. 428: L13-L16 \\
Narayan R, Yi I. 1995. Ap. J. 444: 710-735\\
Neilsen J, Lee JC. 2009. Nature 458:481-84\\
Nemmen RS, Bower RG, Babul A, Storchi-Bergmann T. 2007. MNRAS 377: 1652-1662\\
Netzer H, Mainieri V, Rosati, P, Trakhtenbrot, B. 2006. A\&A 453: 525-533\\
Netzer H. 2009. MNRAS 399: 1907-1920\\
Netzer H. 2013. `The Physics and Evolution of Active Galactic Nuclei'. Cambridge UK: Cambridge University Press\\
Nesvadba NPH, Lehnert MD, Eisenhauer F et~al. 2006. ApJ 650: 661-668\\
Nesvadba NPH. Lehnert MD, De Breuck C, Gilbert AM, van Breugel W. 2008. A\&A 491: 407-424\\
Norman CA, Scoville NZ. 1988.ApJ 332: 124-134\\
Nulsen PEJ, Jones C, Forman WR et~al. 2007. In ESO Astrophys. Symp., ed. H B{\"o}hringer, GW Pratt, A Finoguenov, P Schuecker, pp. 210-15. Berlin,Heidelberg: Springer-Verlag\\
Nusser A, Silk J, Babul A. 2006. MNRAS 373: 739-746\\
O'Dea CP, Baum SA, Privon G, Noel-Storr J, Quillen AC, et~al. 2008. Ap. J. 681:1035-45\\
O'Dea CP, Daly RA, Kharb P, Freeman KA, Baum SA. 2009. Astron. Astrophys. 494: 471-488\\
Osterbrock DE, Ferland G. 2005. `Astrophysics of Gaseous Nebulae and Active Galactic Nuclei'. Mill Valley CA: University Science Books\\
Padovani P, Miller N, Kellermann KI et~al. 2011. Ap.J. 740: 20-37\\
Panter B, Heavens AF, Jimenez R. 2003. MNRAS 343: 1145-1154\\
Panter B, Jimenez, R, Heavens AF, Charlot S. 2007. MNRAS 378: 1550-1564\\
Pasquali A, Kauffmann G, Heckman TM. 2005. MNRAS 361: 1121-1130\\
Peterson BM. 1997. `An Introduction to Active Galactic Nuclei'. Cambridge UK: Cambridge University Press\\
Peterson BM. 2013. Space Sci Rev in press\\
Peterson JR, Bleeker JAM, Ferrigno C et~al. 2001. Astron. Astrophys. 365:L104-9\\
Peterson JR, Kahn SM, Paerels FBS et~al. 2003. Ap. J. 590:207-24\\
Pettini M, Pagel BEJ. 2004. MNRAS 348: L59-63 \\
Pierce CM, Lotz JM, Laird ES et~al. 2007. ApJ 660: L19-22\\
Pizzolato F, Soker N. 2010. MNRAS 408:961-74\\
Ponti G, Fender RP, Begelman MC et~al. 2012. MNRAS 422:L11-15\\
Porciani C, Magliocchetti M, Norberg P. 2004. MNRAS 355: 1010-1030\\
Punsly B, Coroniti FV. 1990. Ap.J. 350: 518-535\\
Qiao E, Liu BF. 2009. Pub. Ast. Soc. Japan. 61: 403-410\\
Quataert E. 2001. ASP Conf Ser. 224: 71-85\\
Rafferty DA, McNamara BR, Nulsen PEJ, Wise MW. 2006. Ap. J. 652:216-31\\
Reviglio P, Hefland DJ. 2006. Ap.J. 650: 717-726\\
Rawlings S, Saunders R. 1991. Nature 349: 138-140\\
Reichard TA, Heckman TM, Rudnick G et~al. 2009. ApJ 691: 1005-1020\\
Remillard RA, McClintock JE. 2006. Annu. Rev. Astron. Astrophys. 44:49-92\\
Rengelink RB, Tang Y, de Bruyn AG, et~al. 1997. Astron. Astrophys. Supp. 124: 259-280\\
Richards G, Fan X, Newberg H et~al. 2002. AJ 123: 2945-2975\\
Richards G, Lacy M, Storrie-Lombardi LJ et~al. 2006. ApJS 166: 470-497\\
Rigby EE, Best PN, Brookes MH et~al. 2011. MNRAS 416: 1900-1915\\
Rigby JR, Diamond-Stanic AM, Aniano G. 2009. ApJ 700: 1878-1883\\
Rosario DJ, Shields GS, Taylor GB, Salviander S, Smith KL. 2010a. ApJ 716: 131-143\\
Rosario DJ, Whittle, M, Nelson CH, Wilson AS. 2010b. ApJ 711: L94-98\\
Rosario DJ, Trakhtenbrot B, Lutz D et~al. 2013. arXiv 1310.1922\\
Rosenfield P, Johnson L, Clifton G et~al. 2012. ApJ 755: 131-143\\
Ross NP, Shen Y, Strauss MA et~al. 2009. Ap.J. 697: 1634-1655\\
Roy AL, Norris RP, Kesteven MJ, Troup ER, Reynolds JE. 1998. MNRAS 301: 1019-1030\\
Russell HR, McNamara BR, Edge AC et~al. 2013a. Ap.J. submitted, arXiv:1309.0014\\
Russell DM, Gallo E, Fender RP. 2013b. MNRAS 431: 405-414\\
Sabater J, Best PN, Argudo-Fern{\'a}ndez M. 2013. MNRAS 430: 638-651\\
Sadler EM, Jenkins CR, Kotanyi CG. 1989. MNRAS 240:591-635\\
Sadler EM, McIntyre VJ, Jackson CA, Cannon RD. 1999. Pub. Astron. Soc. Aust. 16: 247-256\\
Sadler EM, Jackson CA, Canoon RD et~al. 2002. MNRAS 329: 227-245\\
Sadler EM, Cannon RD, Mauch T et~al. 2007. MNRAS 381: 211-227\\
Sadler EM, Ekers RD, Mahony E, Mauch T, Murphy T. 2013. MNRAS in press, arXiv:1304.0268\\
Saintonge A, Kauffmann G, Cramer C et~al. 2011. MNRAS 415: 32-60\\
Saintonge A, Tacconi LJ, Fabello S et~al. 2012. ApJ 758: 73-89\\
Salim S, Charlot S, Rich RM et~al. 2005. ApJ 619: L39-42\\
Sandage A. 1965. ApJ 141: 1560-1580\\
Sanders DB, Soifer, BT, Elias JH, Neugebauer G, Matthews K. 1988a. ApJ 328: L35-39\\
Sanders DB, Soifer BT, Elias JH et~al. 1988b. ApJ 325: 74-91\\
Santini P, Rosario DJ, Shao L et~al. 2012. A\&A 540: 109-128\\
Sargent MT, Schinnerer E, Murphy E et~al. 2010. Ap.J.Supp. 186: 341-377\\
Sarzi M, Shields JC, Schawinski K et~al. 2010. MNRAS 402: 2187-2210\\
Satyapal S, Vega D, Dudik RP, Abel NP, Heckman T. 2008. ApJ 677: 926-942\\
Satyapal S, B{\"o}ker T, McAlpine W et~al. 2009. ApJ 704: 439-452\\
Schawinski K, Thomas D, Sarzi M et~al. 2007. MNRAS 382: 1415-1431\\
Schawinski K, Urry CM, Virani S et~al. 2010. Ap.J. 711: 284-302\\
Schawinski K, Treister E, Urry CM et~al. 2011. ApJ 727: L31-36\\
Schawinski K, Simmons BD, Urry CM, Treister E, Glikman E. 2012. MNRAS 425: L61-65\\
Scheuer P. 1974. MNRAS 166: 513-528\\
Schiminovich D, Wyder T, Martin DC et~al. 2007. ApJS 173: 315-341\\
Schmidt M 1963. Nature 197: 1040\\
Seljak U. 2000. MNRAS 318: 203-213\\
Selwood J. 2014. arXiv 1310.0403\
Seyfert CK. 1943. ApJ 97: 28-40\\
Shabala SS, Godfrey LEH. 2013. Ap.J. 769: 129-136\\
Shankar F, Weinberg D, Miralda-Escude, J. 2009. ApJ 690: 20-41\\
Shankar F, Marulli F, Mathur S et al. 2012. A\&A 540: 23-31\\
Shankar F, Weinberg D, Miralda-Escude J. 2013. MNRAS 428: 421-446\\
Shao L, Kauffmann G, Cheng L, Wang, J, Heckman TM. 2013. arXiv 1304.7175\\
Shapley A. 2011. ARA\&A 49: 525-580\\
Shen S et~al. 2006. MNRAS 369: 1639-1653\\
Shen S, Kauffmann G, von der Linden A, White SDM, Best PN. 2008. MNRAS 389: 1074-1086\\
Shen Y, Strauss MA, Ross NP et~al. 2009. Ap.J. 697: 1656-1673\\
Sheth RK, Mo HJ, Tormen G. 2001. MNRAS 323: 1-12\\
Shimasaku K, Fukugita M, Doi M et~al. 2001. AJ 122: 1238-1250\\
Shlosman I, Begelman MC, Frank J. 1990. Nature 345: 679-686\\
Silverman JD, Lamareille F, Maier C et~al. 2009. ApJ 696: 396-410\\
Silverman JD, Kampczyk P, Jahnke K et~al. 2011. ApJ 743: 2-11\\
Silk J. 2013. arXiv 1305.5840\\
Simoes Lopes R, Ramiro D, Storchi-Bergmann T, de Fatima Saraiva, M, Martini P. 2007. ApJ 655: 718-734\\
Simmons BD, Lintott C, Schawinski K et~al. 2013. MNRAS 429: 2199-2211\\
Simpson C, Rawlings S, Ivison RJ et~al. 2012. MNRAS 421: 3060-3083\\
Simpson C, Westoby P, Arumugam V et~al. 2013. MNRAS 433: 2647-2656\\
Soker N. 2006. New. Astron. 12: 38-46 \\
Soltan A. 1982. MNRAS 200: 115-122\\
Spinoglio L, Malkan MA. 1989. ApJ 342: 83-99\\
Spoon HWW, Farrah D, Lebouteiller V et~al. 2013. ApJ 775: 127-146\\
Springel V, Di Matteo, T, Hernquist L. 2005. MNRAS 361: 776-794 \\
Stasinska G, Cid Fernandes R, Mateus A, Sodre L. Asari NV. 2006. MNRAS 371: 972-982\\
Steidel CC, Erb DK, Shapley AE et~al. 2010. ApJ 717: 289-322\\
Steiner JF, McClintock JE, Narayan R. 2013. Ap.J. 762: 104-113\\
Stern D, Eisenhardt P, Gorjian V et~al. 2005. ApJ 631: 163-168\\
Stern D, Assef RJ, Benford DJ et~al. 2012. ApJ 753: 30-47\\
Stern J, Laor A. 2012a. MNRAS 426: 2703-2718\\
Stern J, Laor A. 2012b. MNRAS 423: 600-631\\
Strauss MA, Weinberg DH, Lupton RH et~al. 2002. AJ 124: 1810-1824\\
Sturm E, Gonz{\'a}lez-AlfonsonE, Veilleux S et~al. 2011. ApJ 733: L16-20\\
Tamura T, Bleeker JAM, Peterson JR, et~al. 2001. Astron. Astrophys. 365:L87-92\\
Tasse C, Best PN, R{\"o}ttgering HJA, Le Borgne D. 2008. Astron. Astrophys. 490: 893-904\\
Terlevich R, Melnick J. 1985. MNRAS 213: 841-856\\
Terlevich R, Tenorio-Tagle G, Franco, J, Melnick J. 1992. MNRAS 255: 713-728\\
Toffolatti L, Franceschini A, Danese L, de Zotti G. 1987. Astron. Astrophys. 184: 7-15\\
Treister E, Schawinski K, Volonteri M, Natarajan P. 2013. arXiv 1310.2249\\
Tremaine S, Gebhardt K, Bender R et~al. 2002. ApJ 574: 740-753\\
Trump JR, Impey CD, Kelly BC et~al. 2009. Ap.J. 700: 49-55\\
Ueda Y, Hiroi K, Isobe N et~al. 2011. PASJ 63: 937-945\\
Veilleux S, Osterbrock DE. 1987. ApJS 63: 295-310\\
Veilleux S, Cecil G, Bland-Hawthorn J. 2005. ARA\&A 43: 769-826\\
Veilleux S, Rupke DSN, Kim D-C et~al. 2009. ApJS 182: 628-666\\
Veilleux S, Mel{\'e}ndez M, Sturm E et~al. 2013. ApJ 776: 27-47\\
Vitale M, Zuther J, Garc{\'{\i}}a-Mar{\'{\i}n} M, Eckart A, Bremer M et~al. 2012. Astron. Astrophys. 546: A17-A33\\
Voges W, Aschenbach B, Boller T et~al. 1999. A\&A 349: 389-405\\
von der Linden A, Best PN, Kauffmann G, White SDM. 2007. MNRAS 379: 867-893\\
Whitaker K, van Dokkum PG, Brammer G, Franx M. 2012. ApJ 754: L29-34\\
Willott CJ, Rawlings S, Blundell KM, Lacy M. 1999. MNRAS 309: 1017-1033\\
Wilson AS, Colbert EJM. 1995. Ap.J. 438: 62-71\\
Wilson AS, Heckman TM. 1985. Astrophysics of active galaxies and quasi-stellar objects: 39-109\\
Wild V, Kauffmann G, Heckman T et~al. 2007. MNRAS 381: 543-572\\
Wild V, Heckman T, Charlot C. 2010. MNRAS 405: 933-947\\
Winter L, Lewis KT, Koss M et~al. 2010. ApJ 710: 503-539\\
Woo J-H, Schulze A, Park D et~al. 2013. ApJ 772: 49-57\\
Worthey G, Ottaviani DL. 1997. ApJS 111: 377-386\\
Wright EL, Eisenhardt PPM, Mainser AK et~al. 2010. AJ 140: 1868-1881\\
Wuyts S, F{\"o}rster Schreiber NM, Lutz D et~al. 2011a. ApJ 738: 106-123\\
Wuyts S, F{\"o}rster Schrieber NM, van der Wel A et~al. 2011b. ApJ 742: 96-15\\
Xu C, Livio M, Baum S. 1999. AJ 118: 1169-1176\\
Yan R, Newman JA, Faber SM et~al. 2006. ApJ 648: 281-298\\
Yan R, Blanton MR. 2012. Ap.J. 747: 61-85\\
Yee HKC, Green RF. 1984. Ap.J. 280: 79-90 \\
Yip CW, Connolly AJ, Szalay AS et~al. 2004. AJ 128: 585-609\\
Yu Q, Tremaine S. 2002. MNRAS 335: 965-976\\
Yuan F, Narayan R. 2004. Ap.J. 612: 724-728\\
Zakamska N, Strauss MA, Heckman T, Ivezic Z, Krolik J. 2004. AJ 128: 1002-1016\\
Zinn PC, Middelberg E, Norris RP, Dettmar RJ. 2013. ApJ 774: 66-75
\end{document}